\documentclass[twocolumn,showpacs,preprintnumbers,amsmath,amssymb,seceq]{revtex4}%
\usepackage[dvips]{graphicx}
\usepackage{dcolumn}
\usepackage{bm}
\usepackage{amsmath}
\usepackage{amsfonts}
\usepackage{amssymb}%
\providecommand{\U}[1]{\protect\rule{.1in}{.1in}}

\newcommand{\ignore}[1]{}
\newcommand{\beq}{\begin{equation}}
\newcommand{\eeq}{\end{equation}}

\def\bwt{\begin{widetext}}
\def\ewt{\end{widetext}}
\def\beq{\begin{equation}}
\def\eeq{\end{equation}}
\def\bal{\begin{align}}
\def\eal{\end{align}}
\def\H{{\mathcal{H}}}
\def\P{{\mathcal{P}}}
\def\Q{{\mathcal{Q}}}
\def\T{{\mathtt{T}}}
\def\HH{{\mathtt{H}}}
\def\SS{{\mathtt{S}}}
\def\B{{\mathtt{B}}}
\def\SR{{\mathtt{SR}}}
\def\R{{\mathtt{R}}}
\def\TC{{\mathtt{TC}}}
\def\TCL{{\mathtt{TCL}}}
\def\TCE{{\mathtt{TCE}}}
\def\TCLE{{\mathtt{TCLE}}}
\def\tr{{\mathrm{tr}}}
\def\Tr{{\mathrm{Tr}}}
\def\HOT{{\mathrm{HOT}}}
\def\v{{\mathrm{v}}}
\def\oc{{\mathrm{oc}}}
\def\pc{{\mathrm{pc}}}
\def\ed{{\mathtt{ed}}}
\def\TE{{\mathtt{TE}}}

\begin{document}

\title{Comparison among Various Expressions of Complex Admittance\\for Quantum System in Contact with Heat Reservoir}%

\author{Mizuhiko Saeki$^{1,3}$, Chikako Uchiyama$^{2}$, Takashi Mori$^{1,3}$ and Seiji Miyashita$^{1,3}$}
\affiliation{
$^1${\it Department of Physics, The University of Tokyo, 7-3-1, Hongo, Bunkyo-ku, Tokyo 113-0033, Japan} \\
$^2${\it Faculty of Engineering, University of Yamanashi, 4-3-11, Takeda, Kofu, Yamanashi 400-8511, Japan} \\
$^3${\it CREST, JST, 4-1-8, Honcho, Kawaguchi, Saitama 332-0012, Japan}
}

\date{}%

\begin{abstract}%

Relation among various expressions of the complex admittance for quantum systems in contact with heat reservoir is studied. 
Exact expressions of the complex admittance are derived in various types of formulations of equations of motion under contact with heat reservoir. 
Namely, the complex admittance is studied in the relaxation method and the external-field method. 
In the former method, the admittance is calculated using the Kubo formula for quantum systems in contact with heat reservoir in no external driving fields, 
while in the latter method the admittance is directly calculated from equations of motion with external driving terms. 
In each method, two types of equation of motions are considered, i.e., the time-convolution (TC) equation and time-convolutionless (TCL) equation. 
That is, the full of the four cases are studied. 
It is turned out that the expression of the complex admittance obtained by using the relaxation method with the TC equation exactly coincides with that obtained by using the external-field method with the TC equation, while other two methods give different forms. 
It is also explicitly demonstrated that all the expressions of the complex admittance coincide with each other in the lowest Born approximation for the system-reservoir interaction. 
The formulae necessary for the higher order expansions in powers of the system-reservoir interaction are derived, and also the expressions of the admittance in the $n$-th order approximation are given. 
By transforming inverse-temperature-integrals into time-integrals, the admittances are also given in the formulae with time-integrals alone. 
To characterize the TC and TCL methods, we study the expressions of the admittances of two exactly solvable models. 
Each exact form of admittance is compared with the results of the two methods in the lowest Born approximation. It is found that depending on the model, either of TC and TCL would be the better method.
\end{abstract}%


\maketitle

\section{Introduction} 

The complex admittance is a fundamental quantity to describe response of systems to an external field. 
Since the magnetic resonance has been discovered \cite{MR}, 
experimental and theoretical studies have been developed simultaneously \cite{theory,Bloch,book}. 
For example, the magnetic resonance (MR) has been one of the most active fields to study the quantum response. 
Kubo and Tomita gave a unified view point to express the response function by the autocorrelation function \cite{kubo-Tomita}, 
which was extended to the general formula of linear response \cite{Kubo}.

Although the Kubo formula \cite{Kubo} gives a general form of the complex admittance, 
there are several types of formulation to take into account effects of the interaction of a quantum system with its heat reservoir. 
When we study the time dependence of a physical quantity $A$ under a time-dependent external field $F$ conjugate to a physical quantity $B$, 
the response function can be expressed in terms of a time correlation function of quantities $A$ and $B$, i.e., $\langle A(t)B\rangle$. 
In the simplest treatment, we give the time evolution by the pure quantum dynamics given by Hamiltonian $\H_\SS$ of the system. 
In this case, the complex admittance $\chi_{AB}(\omega)$ is given by an ensemble of delta function $\delta(E_i - E_j - \hbar \omega)$, 
which denotes a resonance between energy levels $E_i$ and $E_j$ of the system.

In many-body systems, the energy levels form a continuous spectrum, and also the complex admittance has continuous forms. 
There, line shape of the spectrum (the imaginary part of the complex susceptibility) has an intrinsic line width which is attributed to the energy structure of the system. 
In such systems, effects of interactions on the spectrum line shape have been developed \cite{Kanamori-Tachiki}, and various properties have been clarified. 
For example, Nagata and Tazuke found the shift of resonance peak due to dipole-dipole interaction as a function of relative angle between the external field and 
the lattice axis of one-dimensional magnets \cite{Nagata-Tazuke,ESR3}. 
Recently, studies to evaluate the susceptibility from microscopic Hamiltonian have been proposed, and new aspects of the resonance have been discussed \cite{MiyashitaESR,Oshikawa-Affleck}.

On the other hand, the width of the line shape also appears due to the interaction of system with its heat reservoir. 
Because of the developments of experimental methods, real time quantum dynamics has become accessible in microscopic quantum systems, e.g., the magnetization dynamics of single molecule magnets \cite{SMM1,SMM2,SMM3,SMM4,SMM5,SMM6,SMM7,SMM8,SMM9} and also microscopic circuits which manipulate the degree of freedom of qubit \cite{qubit}. 
In these systems, it has been pointed out that the dissipation plays important roles \cite{Fe2,SMM6,SM}, and also the nature of dissipation has been discussed in detail \cite{noise}. 
It is interesting problems to study effects of the dissipation on the complex admittance for those systems.

If we consider the total system which consists of a system of interest and of its heat reservoir, 
the formulation of a pure quantum dynamics works in principle. 
However, it is difficult in practice, and we introduce a kind of dissipative dynamics instead of the quantum dynamics for the total system. 
Usually we introduce the dissipative dynamics by coarse-graining of the heat reservoir. 
The dissipative dynamics is expressed by the so-called quantum master equation \cite{Nakajima,Zwanzig,Louisell,HSS,KTH,QME5,Redfield,STM2000}. 

In order to take into account the dissipative dynamics, we need to obtain the time evolution function for dissipative systems. 
The phenomenological treatment of the dissipation effects had been done by using the Bloch equation including disipation terms with the relaxation times $T_1$ and $T_2$ \cite{Bloch}. 
The microscopic treatment was proposed using the projection operator techniques by Nakajima \cite{Nakajima}, Zwanzig \cite{Zwanzig} and Mori \cite{Mori}. 
They obtained the equations of motion in the non-Markovian or ``time-convolution (TC)" integral forms with memory. 
However, in order to treat the memory term in a compact way, the so-called ``time-convolutionless (TCL)" formalisim has been introduced \cite{TM,HSS,STH}. 
Recently, we have studied a complete expression of the complex admittance in the dissipative environment in the TC formalism \cite{MoriMiyashita,Uchiyama}. 
There, we have used a standard equation of the reduced density operator, but we have pointed out that the initial correlation between the system and the heat reservoir can not be ignored in general when we treat the non-Markovian case.

Moreover, one of the authors has proposed the formalism called the ``TCLE method" 
in which the complex admittance is directly calculated from time-convolutionless (TCL) equations of motion with external driving terms \cite{S1,S2,S3,S4,S5,S6,S7}, 
by generalizing the method of Argyres and Kelley in which the admittance is calculated 
from time-convolution (TC) equations of motion with external driving terms in the Markovian approximation \cite{AK}. 
Hereafter, we call the method in which the complex admittance is directly calculated from equations of motion with external driving terms, the ``external-field method", 
while we call the method in which the Kubo formula is calculated for systems with no external driving fields, the ``relaxation method". 
Thus, we have the four types of formalism, that is, the relaxation method or the external-field method with TC or TCL equation of motion. 
The external-field method is considered to give the same results as those obtained using the relaxation method in principle. 
They must also give the same results because the original equation is the same. 
However, the treatment of the contact with the heat reservoir can be simplified in the external-field method. 
In practical calculations, we need to approximate perturbatively in order of the strength of the interaction with the heat reservoir. 
If we truncate the perturbation, the higher order corrections may be different in the four formalisms. 

In the present paper, we consider a system interacting with a heat reservoir in an external driving field, 
and derive forms of the complex admittance using the four types of formalism. Then, we investigate relations among them. 
Namely, we examine the forms of admittance obtained by using the following four methods. \\
(1) The relaxation TC method (the relaxation method with the TC equation) \\
(2) The relaxation TCL method (the relaxation method with the TCL equation) \\
(3) The TCE method (the external-field method with the TC equation) \\
(4) The TCLE method (the external-field method with the TCL equation) \\
We also derive the formulae necessary for the higher order expansions in powers of the system-reservoir interaction, 
and give the expressions of the admittance in the $n$-th order approximation. 
Moreover, in order to discuss the truncation effect in the four methods, 
we study the two exactly solvable models, and compare each exact form of admittance 
with the results obtained by using the above methods in the lowest Born approximation for the system-reservoir interaction.

In Section II, we survey the derivation of the Kubo formula, 
and explain the basic difference between the relaxation method and external-field method. 
In Section III, we derive forms of the complex admittance obtained by using the above four methods. 
In Section IV, we investigate relations among forms of the admittance obtained by using the four methods in the lowest Born approximation for the system-reservoir interaction. 
In Section V, we derive the formulae necessary for the higher order expansions in powers of the system-reservoir interaction. 
In Section VI, we study the two exactly solvable models to discuss the truncation effects in different methods. 
In Section VII, we give a short summary and some concluding remarks.
%
\section{Model}

We consider a quantum system interacting with a quantum heat reservoir in an external driving field. We take the Hamiltonian $\H_\T(t)$ of the total system as
\begin{align}
\H_\T(t) &= \H_\SS + \H_\R + \H_\SR + \H_\ed(t) \nonumber \\
         &= \H_0 + \H_\SR + \H_\ed(t) = \H + \H_\ed(t), 
\label{Htotal}
\end{align}
where $\H_\SS$ is the Hamiltonian of the quantum system, $\H_\R$ is the Hamiltonian of the heat reservoir, $\H_\SR$ is the interaction Hamiltonian between the system and heat reservoir, and $\H_\ed(t)$ is the interaction Hamiltonian of the quantum system with the external driving field. 
Here, we express the Hamiltonian of the system and heat reservoir with interaction by $ \H= \H_\SS + \H_\R + \H_\SR $, and the unperturbed Hamiltonian, i.e., the Hamiltonian of the system and heat reservoir without interaction by $ \H_0 = \H_\SS + \H_\R $. 
The denisity operator $\rho_\T(t)$ of the total system satisfies the Liouville equation
\begin{equation}
(d / dt) \, \rho_\T(t) = - (i / \hbar)\, [\, \H_\T(t) \,,\, \rho_\T(t) \,] \equiv - \, i \, L_\T(t) \, \rho_\T(t),
\label{Leq}
\end{equation}
which decides the dynamics corresponding to the Hamiltonian $\H_\T(t)$, where the Liouvillian $L_\T(t)$ corresponds to the Hamiltonian $\H_\T(t)$. 
Hereafter, a Liouvillian corresponding to a Hamiltonian, say $\H$, is denoted as $L$, i.e., $ L A $\,=\,$ [\, \H, A \,]/ \hbar $. 

We assume that the system and heat reservoir are in the thermal equilibrium state $\rho_\TE$ at temperature $T$ initially, i.e., before the external driving field is turned on, where $\rho_\TE$ is the thermal equilibrium density operator for the system and reservoir and is given by
\begin{equation}
\rho_\TE = \exp(- \, \beta \, \H) / \, \Tr \exp(- \, \beta \, \H), 
\label{rhoTE}
\end{equation}
with $ \beta $\,=\,$ 1 / (k_\B T) $. Here, notation $\Tr$ denotes the trace operations in the system and reservoir spaces, i.e., $ \Tr $\,=\,$ \tr \, \tr_\R $, where notations $\tr$ and $\tr_\R$ denote the trace operations in the system and reservoir spaces, respectively. 
If we consider only the system, the density operator for the system alone is given by
\begin{equation}
\rho_\SS = \exp(- \, \beta \, \H_\SS) / \, \tr \exp(- \, \beta \, \H_\SS). 
\label{rhoS}
\end{equation}
The density operator for the heat reservoir alone is given by
\begin{equation}
\rho_\R = \exp(- \, \beta \, \H_\R) / \, \tr_\R \exp(- \, \beta \, \H_\R). 
\label{rhoR}
\end{equation}
It should be noted that the equlibrium state of the system is given by
\begin{equation}
\rho_0 = \tr_\R \, \rho_\TE = \tr_\R \exp(- \, \beta \, \H) / \, \Tr \exp(- \, \beta \, \H), 
\label{rho}
\end{equation}
which includes the interaction between the system and reservoir and is different from $\rho_\SS$.

For the convenience of expressions, we define the notation $ \langle \cdots \rangle_\R $\,=\,$ \tr_\R(\cdots \rho_\R) $, and renormalize the Hamiltonian $\H_\SS$ of the system and the system-reservoir interaction $\H_\SR$ as
\begin{subequations}
\label{renormalize}
\begin{align}
& \H_\SS \longrightarrow \H_\SS + \langle \H_\SR \rangle_\R \Longrightarrow \H_\SS, 
\label{renormalize1} \\
& \H_\SR \longrightarrow \H_\SR - \langle \H_\SR\rangle_\R \Longrightarrow \H_\SR . \qquad \label{renormalize2}
\end{align}
\end{subequations}
Hereafter, we use $\H_\SS$ and $\H_\SR$ for the Hamiltonians renormalized by (\ref{renormalize}), and then, we have $ \langle \H_\SR \rangle_\R $\,=\,0.

We now take the interaction of the system with the external driving field as \cite{Kubo,AK}
\begin{align}
\H_\ed(t) & = - \sum_j A_j \, F_j(t) = - \sum_j \sum_\omega A_j \, F_j[\omega] \, e^{- \, i \, \omega \, t}, \nonumber \\
    & = \sum_\omega \, \H_\ed[\omega] \, e^{- \, i \, \omega \, t},
\label{edf}
\end{align}
where $A_j$ is the physical quantity conjugate to the force $F_j(t)$ which is a $c$-number function of the time $t$, and the summation $\sum_\omega$ is over $\omega$ and $- \omega$ for each $\omega$.
%
\subsection{The Kubo formula}

We first survey the derivation of the Kubo formula briefly \cite{Kubo}. 
The time evolution for the density operator of the total system is given by Eq. (\ref{Leq}). The external driving field is assumed to be turned on adiabatically at the initial time $t$\,=\,$t_0$ which is infinite past ($ t_0 $\,=\,$ - \, \infty $) in the Kubo theory \cite{Kubo}. 
Expanding the density operator $\rho_\T(t)$ in powers of the external driving field as
\begin{equation}
\rho_\T(t) = \rho_{\T0}(t) + \rho_{\T1}(t) + \rho_{\T2}(t) + \cdots \,, \quad
\label{rhoT-expand}
\end{equation}
the zeroth-order part $\rho_{\T0}(t)$ and first-order part $\rho_{\T1}(t)$ satisfy the following equations and initial conditions\,:
\begin{align}
& \frac{d}{dt}\, \rho_{\T0}(t) = - \, i L \, \rho_{\T0}(t) \,; \qquad \quad \rho_{\T0}(t_0) = \rho_\TE \,,
\label{order0-1} \\
& \frac{d}{dt}\, \rho_{\T1}(t) = - \, i L \, \rho_{\T1}(t) - i L_\ed(t) \, \rho_{\T0}(t) \,; \quad \ \rho_{\T1}(t_0) = 0 \,. \label{order1-1}
\end{align}
Equation (\ref{order0-1}) has, by virtue of the thermal equilibrium density operator (\ref{rhoTE}), the solution
\begin{equation}
\rho_{\T0}(t) = \exp\{- \, i \, L \, (t - t_0) \}\, \rho_{\T0}(t_0) = \rho_\TE \,,
\end{equation}
by which Eq. (\ref{order1-1}) can be formally solved as
\begin{equation}
\rho_{\T1}(t) = - \, i \int_{t_0}^t d\tau \exp\{- \, i \, L \cdot (t - \tau)\}\, L_\ed(\tau)\, \rho_\TE \,.
\end{equation}
Then, the first-order part in powers of the external driving field for the expectation value of a physical quantity $A_i$ of the system, can be described as
\begin{align}
& \Tr \, A_i \, \rho_{\T1}(t) \nonumber \\
    = & \frac{i}{\hbar} \sum_j \sum_\omega \int_{t_0}^t d\tau \, \Tr \, A_i \, e^{- \, i \, L \cdot (t - \tau)}\, [ A_j , \rho_\TE ] \, F_j[\omega] \, e^{- \, i \, \omega \, \tau}, \nonumber \\
    = & \frac{i}{\hbar} \sum_j \sum_\omega \int_0^{t - t_0} d\tau \, \Tr \, A_i \, e^{- \, i \, L \, \tau}\, [\, A_j \,,\, \rho_\TE \,] \, F_j[\omega] \nonumber \\
& \qquad \qquad \qquad \qquad \qquad { } \times e^{i \, \omega \, \tau \, - \, i \, \omega \, t }.
\end{align}
The complex admittance $\chi_{ij}(\omega)$ is defined in the limit $t_0$\,$\to$\,$ - \,\infty $, as \cite{Kubo}
\begin{align}
\Tr \, A_i \, \rho_{\T1}(t) & = \sum_j \, \sum_\omega \, \chi_{ij}(\omega) \, F_j[\omega] \, e^{- \, i \, \omega \, t}, \label{Kadmit}\\
    & \qquad \qquad \qquad (t_0 \to - \, \infty) \nonumber 
\end{align}
and takes the expressions \cite{Kubo}
\begin{subequations}
\label{KF}
\begin{align}
\chi_{ij}(\omega) & = \frac{i}{\hbar} \int_0^\infty \! dt \, \Tr \, A_i \, e^{- \, i \, L \, t} \big[ A_j , \, \rho_\TE \big] \, e^{i \, \omega \, t \, - \, \epsilon \, t}, \label{KF1}\\
    & = \frac{i}{\hbar} \int_0^\infty dt \, \Tr \, A_i \, [\, A_j^\HH(- \, t) ,\, \rho_\TE \,] \, e^{i \, \omega \, t \, - \, \epsilon \, t}, \label{KF2}\\
    & = \frac{i}{\hbar} \int_0^\infty dt \, \Tr \, A_i^\HH(t) \, [\, A_j ,\, \rho_\TE \,] \, e^{i \, \omega \, t \, - \, \epsilon \, t}, \label{KF3}\\
    & = \frac{i}{\hbar} \int_0^\infty dt \, \Tr \, [\, A_i^\HH(t) ,\, A_j \,]\, \rho_\TE \, e^{i \, \omega \, t \, - \, \epsilon \, t}, \label{KF4}
\end{align}
\end{subequations}
which are called the Kubo formula, where $\epsilon $$ \to $$ +0$. Here, $A_i^\HH(t)$ is the Heisenberg operator defined by
\begin{equation}
A^\HH(t) = \exp(i \, L \, t) A = \exp(i \, \H \, t / \,\hbar) A \exp(- \, i \, \H \, t / \, \hbar).
\end{equation}
In the definition (\ref{Kadmit}) of the complex admittance, the summation $\sum_\omega$ for $\omega$ is over the same frequencies as those in (\ref{edf}). 
In the Kubo formula, $A_i^\HH(t)$ and $A_j^\HH(t)$ denote the time evolutions of the physical quantities $A_i$ and $A_j$. Suzuki and Kubo discussed the linear response by deriving the time evolution of the physical quantity $A_i$ using the stochastic dynamics \cite{Suzuki-Kubo}. 
In the Kubo theory \cite{Kubo}, the complex admittance is given by the time correlation function for the system with no external driving fields. We call the method that evaluates the complex admittance for a quantum system in contact with its heat reservoir using the Kubo formula, the ``relaxation method".

We next mention the basic difference between the relaxation method and external-field method in the following subsections. 
%
\subsection{The relaxation method}

The relaxation method is the method in which complex admittances are calculated using the Kubo formula for a quantum system in contact with its heat reservoir. For a quantum system in contact with its heat reservoir, the operator with a physical quantity $A$ of the quantum system\,:
\begin{equation}
\tilde{A}(t) = e^{- \, i \, L \, t}\, [ A \,, \rho_\TE \,] = e^{- \, i \, \H \, t / \hbar}\, [ A \,, \rho_\TE \,] \, e^{i \, \H \, t / \hbar},
\end{equation}
satisfies the equation of motion
\begin{equation}
(d/dt) \, \tilde{A}(t) = - \, (i / \hbar)\, [\, \H \,, \, \tilde{A}(t) \,] = - \, i \, L \, \tilde{A}(t).
\label{Atilde-eq}
\end{equation}
In order to eliminate irrelevant variables associated with the heat reservoir, we introduce the time-independent projection operators $\P$ and $\Q$, which are taken as 
\begin{equation}
\P = \rho_\R \, \tr_\R\,, \qquad \quad \Q = 1 - \P \ ; \qquad \quad \tr_\R\, \rho_\R = 1,
\label{projectionPQ}
\end{equation}
where $\rho_\R$ is the equilibrium density operator (\ref{rhoR}) of the heat reservoir alone. Applying the projection operators $\P$ and $\Q$ to Eq. (\ref{Atilde-eq}), we have the coupled equations\,:
\begin{subequations}
\label{PQA}
\begin{align}
(d/dt)\, \P \, \tilde{A}(t) & = - \, i \, \P \, L\, \P \, \tilde{A}(t) - i \, \P \, L \, \Q \, \tilde{A}(t),
\label{PA} \\
(d/dt)\, \Q\, \tilde{A}(t) & = - \, i \, \Q \, L\, \P \, \tilde{A}(t) - i \, \Q \, L \, \Q \, \tilde{A}(t).
\label{QA}
\end{align}
\end{subequations}
In the relaxation method, complex admittances are calculated by deriving the equation of motion for $\tr_\R \tilde{A}(t)$ from the above two equations (\ref{PQA}) and by substituting its solution into the Kubo formula (\ref{KF}).
%
\subsection{The external-field method}

The external-field method is the method in which complex admittances are directly calculated by solving the equation of motion including external driving terms for the total system, which includes the external driving field. 
The density operator $\rho_\T(t)$ of the total system satisfies the Liouville equation 
(\ref{Leq}). 
We assume that the external driving field is turned on adiabatically at the initial time $t_0$\,=\,0, though the initial time is infinite past ($t_0$\,=\,$ - \, \infty $) in the Kubo theory \cite{Kubo}. 
Expanding the density operator $\rho_\T(t)$ in powers of the external driving field as in (\ref{rhoT-expand}), 
the zeroth-order part $\rho_{\T0}(t)$ and first-order part $\rho_{\T1}(t)$ satisfy the following equations and initial conditions similar to (\ref{order0-1}) and (\ref{order1-1})\,:
\begin{align}
& \frac{d}{dt}\, \rho_{\T0}(t) = - \, i L \, \rho_{\T0}(t) \,; \qquad \quad \rho_{\T0}(0) = \rho_\TE \,, \label{order0-2} \\
& \frac{d}{dt}\, \rho_{\T1}(t) = - \, i L \, \rho_{\T1}(t) - i L_\ed(t) \, \rho_{\T0}(t) \,; \quad \ \rho_{\T1}(0) = 0 \,. \label{order1-2}
\end{align}
Applying the projection operators $\P$ and $\Q$ defined by (\ref{projectionPQ}) to Eq. (\ref{order1-2}), we have the coupled equations\,:
\begin{subequations}
\label{PQrhoE}
\begin{align}
(d/dt)\, \P \, \rho_{\T1}(t) = & - \, i \, \P \, L \, \P \, \rho_{\T1}(t) - \, i \, \P \, L \, \Q \, \rho_{\T1}(t) \qquad \quad \nonumber \\
    & - i \, \P \, L_\ed(t) \, \rho_\TE \,, \label{PrhoE} \\
(d/dt)\, \Q \, \rho_{\T1}(t) = & - \, i \, \Q \, L \, \P \, \rho_{\T1}(t) - \, i \, \Q \, L \, \Q \, \rho_{\T1}(t) \qquad \nonumber \\
    & - i \, \Q \, L_\ed(t) \, \rho_\TE \,, \label{QrhoE}
\end{align}
\end{subequations}
where we have used the solution of Eq. (\ref{order0-2})\,:
\begin{equation}
\rho_{\T0}(t) = \exp(- \, i \, L \, t) \, \rho_{\T0}(0) = \rho_\TE \,.
\label{rhoT0}
\end{equation}
In the external-field method, complex admittances are calculated by deriving the equation of motion for $\tr_\R \rho_{\T1}(t)$ from the above two equations (\ref{PQrhoE}) and by solving it in the limit $t$\,$ \to $\,$ \infty $, which corresponds to the assumption that the initial time is infinite past ($ t_0 $\,=\,$ - \, \infty $) in the Kubo theory \cite{Kubo}. 

In the external-field method, it should be noticed that the time evolution of the external driving force is included in the projection procedure. Thus, the projection procedure is performed for the equation (\ref{order1-2}) of motion including external driving terms in the external-field method, while the projection procedure is performed for the equation (\ref{Atilde-eq}) of motion including no external driving terms in relaxation method.
%
\section{Complex admittance}

In this section, we derive forms of complex admittance $\chi_{ij}(\omega)$ for the two methods mentioned in the previous section, i.e., the relaxation method and external-field method, and for the two types of equation of motion, i.e., the time-convolution (TC) equation and time-convolutionless (TCL) equation.
\subsection{The Relaxation TC method}

We first derive the form of complex admittance by the ``relaxation TC method" in which the Kubo formula is calculated by solving the ``time-convolution" (TC) equation of motion for systems with no external driving fields. 
\begin{widetext}
Equation (\ref{QA}) has the formal solution
\begin{equation}
\Q \, \tilde{A}(t) = - \, i \int_0^t d \tau \exp\{- \, i \, \Q \, L \, \Q \, (t - \tau) \}\, \Q \, L \, \P \, \tilde{A}(\tau) + \exp(- \, i \, \Q \, L \, \Q \, t ) \, \Q \, \tilde{A}(0), \qquad
\label{QATC}
\end{equation}
which has a form of the time-convolution (TC). By substituting (\ref{QATC}) into (\ref{PA}), the ``time-convolution" (TC) equation of motion for $\P \tilde{A}(t)$ can be obtained as
\begin{equation}
(d/dt)\, \P \tilde{A}(t) = - \, i \, \P L \, \P \tilde{A}(t) - \int_0^t d \tau \, \P L \exp\{- \, i \, \Q \, L \, \Q \, (t - \tau) \}\, \Q \, L \, \P \tilde{A}(\tau) - i \, \P L \exp(- \, i \, \Q \, L \, \Q \, t )\, \Q \, \tilde{A}(0).
\label{PATC}
\end{equation}
By virtue of the projection operators (\ref{projectionPQ}), the above equation is reduced to the TC equation of motion
\begin{equation}
(d/dt) \, \tilde{a}(t) = - \, i \, L_\SS \, \tilde{a}(t) + \bar{C}(t \,,\, \{\tilde{a}\}) + \bar{I}(t), \qquad \quad
\label{tilde-a-TC}
\end{equation}
for the reduced operator $\tilde{a}(t)$ defined by
\begin{equation}
\tilde{a}(t) = \tr_\R \, \tilde{A}(t) = \tr_\R \exp(- \, i \, L \, t )\, [A \, , \, \rho_\TE \,], \qquad \quad
\label{Def-tilde-a}
\end{equation}
where the collision term $\bar{C}(t , \{\tilde{a}\})$ and the inhomogeneous term $\bar{I}(t)$ are, respectively, given by
\begin{align}
    & \bar{C}(t \,,\, \{\tilde{a}\}) = - \int_0^t d \tau \, \tr_\R \, L_\SR \exp\{- \, i \, \Q \, L \, \Q \, (t - \tau) \} \, L_\SR \, \rho_\R \, \tilde{a}(\tau), \qquad
\label{bar-C-t} \\
    & \bar{I}(t) = - \, i \!\cdot\! \tr_\R \, L_\SR \exp(- \, i \, \Q \,  L \, \Q \, t) \, \Q \, [\, A \,,\, \rho_\TE \,]\,.
\label{bar-I-t}
\end{align}
The inhomogeneous term $\bar{I}(t)$ represents the effects of initial correlation of the system and reservoir, because if the initial state $\rho_\TE$ given by (\ref{rhoTE}) is approximated to the decoupled one $\rho_0 \rho_\R$, i.e., $ \rho_\TE $\,=\,$ \rho_0 \, \rho_\R $, the term $\bar{I}(t)$ vanishes since $\Q \, [A , \rho_\TE ] $\,=\,$ \Q \, [A, \rho_0] \, \rho_\R $\,=\,0. 
Performing the Fourier-Laplace transformation for Eq. (\ref{tilde-a-TC}), we have 
\begin{equation}
- \, \tilde{a}(0) - i \, \omega \, \tilde{a}[\omega] = - \, i \, L_\SS \, \tilde{a}[\omega] + \bar{C}[\omega] \, \tilde{a}[\omega] + \bar{I}[\omega], \qquad \quad
\label{tilde-a-TComega}
\end{equation}
where $\tilde{a}[\omega]$, $\bar{C}[\omega]$ and $\bar{I}[\omega]$ are given by
\begin{align}
    & \tilde{a}[\omega] = \int_0^\infty dt \, \tilde{a}(t) \exp(i \, \omega \, t - \epsilon \, t) = \int_0^\infty dt \, \tr_\R \exp(- \, i \, L \, t) \big[ A \,,\, \rho_\TE \,\big] \exp(i \, \omega \, t - \epsilon \, t) \big|_{\epsilon \to +0}\,, \quad \label{tilde-a-omega} \\
    & \bar{C}[\omega] = - \int_0^\infty d t \, \big\langle L_\SR \exp(- \, i \, \Q \, L \, \Q \, t) \, L_\SR \, \big\rangle_\R \exp(i \, \omega \, t - \epsilon \, t) \big|_{\epsilon \to +0}\,, \label{bar-C-omega} \\
    & \bar{I}[\omega] = - \, i \int_0^\infty dt \, \tr_\R \, L_\SR \exp(- \, i \, \Q \,  L \, \Q \, t)\, \Q \, \big[ A \,,\, \rho_\TE \,\big] \exp(i \, \omega \, t - \epsilon \, t) \big|_{\epsilon \to +0}\,. \label{bar-I-omega}
\end{align}
Considering that 
\begin{equation}
\tilde{a}(0) = \tr_\R \,\tilde{A}(0) = \tr_\R \,[\, A \,,\, \rho_\TE \,] = [\, A \,,\, \rho_0 \,]\,, \qquad \qquad
\end{equation}
Eq. (\ref{tilde-a-TComega}) can be formally solved as
\begin{equation}
\tilde{a}[\omega] = \big( i \, (L_\SS - \omega) - \bar{C}[\omega] \big)^{-1} \big\{ [\, A \,,\, \rho_0 \,] + \bar{I}[\omega] \big\}. \qquad \quad
\label{tilde-a-solution}
\end{equation}
Substituting (\ref{tilde-a-solution}) into the Kubo formula (\ref{KF}), the admittance takes the form
\begin{align}
\chi^{\R\TC}_{ij}(\omega) & = \frac{i}{\hbar} \int_0^\infty dt \, \Tr \, A_i \exp(- \, i \, L \, t) \, [\, A_j \,,\, \rho_\TE \,] \exp(i \,\omega \, t) = \frac{i}{\hbar} \int_0^\infty dt \, \tr \, A_i \, \tilde{a}_j(t) \exp(i \,\omega \, t) = \frac{i}{\hbar}\, \tr \, A_i \, \tilde{a}_j[\omega], \quad \nonumber \\
    & = \tr \, A_i \, \frac{1}{\, i \, (L_\SS - \omega) - \bar{C}[\omega] \,} \Big\{ \frac{i}{\hbar}\, \big[ A_j \,,\, \rho_0 \,\big] + \bar{D}_j[\omega] \Big\}, 
\label{RTCadmit}
\end{align}
which is a general form of admittance derived using the relaxation TC method under thermal equilibrium initial conditions, where $\bar{D}_j[\omega]$ comes from the inhomogeneous term $\bar{I}(t)$ in Eq. (\ref{tilde-a-TC}), represents the effects of initial correlation of the system and reservoir and is given by
\begin{equation}
\bar{D}_j[\omega] = \frac{i}{\hbar} \, \bar{I}_j[\omega] = \frac{1}{\hbar} \int_0^\infty dt \, \tr_\R \, L_\SR \, \exp(- \, i \, \Q\,  L\, \Q\, t)\, \Q \, \big[ A_j \,,\, \rho_\TE \,\big] \exp(i \, \omega \, t - \epsilon \, t) \big|_{\epsilon \to +0} \, . \quad
\label{barDj-omega}
\end{equation}

Expanding Eq. (\ref{tilde-a-TC}) up to second order in powers of the system-reservoir interaction $\H_\SR$, it reduces to
\begin{equation}
(d/dt)\, \tilde{a}(t) = - \, i \, L_\SS \, \tilde{a}(t) + \bar{C}^{(2)}(t \,,\, \{\tilde{a}\}) + \bar{I}^{(2)}(t), \qquad \quad \label{tilde-a-TC2}
\end{equation}
where the collision term $\bar{C}^{(2)}(t \,,\, \{\tilde{a}\})$ and the inhomogeneous term $\bar{I}^{(2)}(t)$ are, respectively, given by
\begin{align}
    & \bar{C}^{(2)}(t , \{\tilde{a}\}) = - \int_0^t d\tau \, \big\langle L_\SR \exp\{- \, i \, L_0 \, (t - \tau) \} \, L_\SR \, \big\rangle_\R \, \tilde{a}(\tau), 
\label{bar-C2-t} \\
    & \bar{I}^{(2)}(t) = i \int_0^\beta d\beta' \, \tr_\R \, L_\SR \exp(- \, i \, L_0 \, t) \, \big[ A\,,\, \rho_\SS \, \rho_\R \, \H_\SR(- \, i \, \hbar \, \beta') \, \big]. 
\label{bar-I2-t}
\end{align}
Here, $\H_\SR(t)$ is defined by
\begin{equation}
\H_\SR(t) = \exp(i \, L_0 \, t) \, \H_\SR = \exp( i \, \H_0 \, t / \hbar ) \, \H_\SR \exp(- \, i \, \H_0 \, t / \hbar ). \qquad \qquad \label{H-SR-t}
\end{equation}
Performing the Fourier-Laplace transformation for Eq. (\ref{tilde-a-TC2}) and substituting its formal solution into the Kubo formula (\ref{KF}), the admittance takes the form
\begin{equation}
\chi^{\R\TC}_{ij}(\omega) = \tr \, A_i \, \frac{1}{\, i \, (L_\SS - \omega) - \bar{C}^{(2)}[\omega] \,} \Big\{ \frac{i}{\hbar}\, \big[ A_j \,,\, \rho_\SS + \rho_0^{(2)} \,\big] + \bar{D}_j^{(2)}[\omega] \Big\}, \qquad 
\label{RTC2admit}
\end{equation}
which is a general form of admittance derived using the relaxation TC method 
in the lowest Born approximation for the system-reservoir interaction, 
where $\rho_0^{(2)}$, $\bar{C}^{(2)}[\omega]$ and $\bar{D}_j^{(2)}[\omega]$ are, respectively, given by
\begin{align}
    & \rho_0^{(2)} = \rho_\SS \int_0^\beta d\beta_1 \int_0^{\beta_1} d\beta_2 \, \big\{ \tr_\R \, \rho_\R \, \H_\SR(- \, i \, \hbar \, \beta_1) \, \H_\SR(- \, i \, \hbar \, \beta_2) - \Tr \, \rho_\SS \, \rho_\R \, \H_\SR(- \, i \, \hbar \, \beta_1) \, \H_\SR(- \, i \, \hbar \, \beta_2) \, \big\}, \quad
\label{rho-0-2} \\
    & \bar{C}^{(2)}[\omega] = - \int_0^\infty d \tau \, \big\langle L_\SR \exp(- \, i \, L_0 \, \tau ) \, L_\SR \big\rangle_\R \exp(i \, \omega \, \tau ) = - \int_0^\infty d \tau \, \big\langle L_\SR \, L_\SR(- \, \tau) \big\rangle_\R \exp\{i \, (\omega - L_\SS) \, \tau \}, \label{bar-C2-omega} \\
    & \bar{D}_j^{(2)}[\omega] = \frac{i}{\hbar} \, \bar{I}_j^{(2)}[\omega] = \frac{- 1}{\hbar} \int_0^\infty d \tau \int_0^\beta d \beta' \, \tr_\R \, L_\SR \exp(- \, i \, L_0 \, \tau) \big[\, A_j \,,\, \rho_\SS \, \rho_\R \, \H_\SR(- \, i \, \hbar \, \beta') \,\big] \exp(i \, \omega \, \tau).
\label{barDj2b-omega}
\end{align}
The term $\bar{D}_j^{(2)}[\omega]$ represents the effects of initial correlation of the system and reservoir. The above result (\ref{RTC2admit}) can be obtained by expanding Eq. (\ref{tilde-a-TComega}) up to second order in powers of the system-reservoir interaction too. By using this formula, we can make a unified numerical method to treat systems with large number of spins, and can study relations between the spin-spin interaction and the noise effects on the line shape of magnetic resonace \cite{Uchiyama}.
%
\subsection{The Relaxation TCL method}

One of the authors has derived the form of complex admittance by the ``relaxation TCL method" in which the Kubo formula is calculated by solving the ``time-convolutionless" (TCL) equation of motion for systems with no external driving fields \cite{S6,S8}. We here survey the derivation of the form of complex admittance by the relaxation TCL method for comparison. In order to renormalize the time-convolution (TC) in the formal solution (\ref{QATC}) of Eq. (\ref{QA}), we write the formal solution of Eq. (\ref{Atilde-eq}) in the form
\begin{equation}
\tilde{A}(t) = \exp\{- \, i \, L \cdot (t - \tau)\} \tilde{A}(\tau) \qquad \qquad \qquad \quad \mathrm{or} \qquad \qquad \qquad \quad \tilde{A}(\tau) = \exp\{i \, L \cdot (t - \tau)\} \tilde{A}(t), \quad
\label{Atilde-sol}
\end{equation}
which is substituted into (\ref{QATC}) to give
\begin{equation}
\Q \, \tilde{A}(t) = \{\theta(t) - 1 \} \, \P \tilde{A}(t) + \theta(t) \exp(- \, i \, \Q \, L \, \Q \, t ) \, \Q \, \tilde{A}(0), \qquad
\label{QATCL}
\end{equation}
where $\theta(t)$ is defined by
\begin{equation}
\theta(t) = \Big\{ 1 + i \int_0^t d\tau \exp(- \, i \, \Q \, L\, \Q\, \tau)\, \Q \, L_\SR \, \P \exp(i \, L \, \tau) \Big\}^{-1}. \qquad
\label{theta-t}
\end{equation}
By substituting the expression (\ref{QATCL}) for $\Q \tilde{A}(t)$ into (\ref{PA}), 
the ``time-convolutionless" (TCL) equation of motion for $\P \tilde{A}(t)$ can be obtained as
\begin{equation}
(d/dt)\, \P \tilde{A}(t) = - \, i \, \P L \, \P \tilde{A}(t) - \, i \, \P  L \, \{\theta(t) - 1 \}\,  \P \tilde{A}(t) - i \, \P L \, \theta(t) \exp(- \, i \, \Q \, L \, \Q \, t)\, \Q \, \tilde{A}(0). \quad
\end{equation}
By virtue of the projection operators (\ref{projectionPQ}), the above equation is reduced to the TCL equation of motion for the reduced operator $\tilde{a}(t)$ as
\begin{equation}
(d/dt)\, \tilde{a}(t) = - \, i \, L_\SS \, \tilde{a}(t) + C(t) \, \tilde{a}(t) + I(t), \qquad \qquad 
\label{tilde-a-TCL}
\end{equation}
where the collision operator $C(t)$ and the inhomogeneous term $I(t)$ are, respectively, given by
\begin{align}
& C(t) = - \, i \cdot \tr_\R \, L_\SR \, \{\theta(t) - 1 \} \, \rho_\R\, \equiv - \, i \, \langle \, L_\SR \, \{\theta(t) - 1 \} \, \rangle _\R, \qquad \quad
\label{C-t} \\
& I(t) = - \, i \cdot \tr_\R\, L_\SR\, \theta(t) \exp(- \, i \, \Q \, L \, \Q \, t) \, \Q \, [\, A \,,\, \rho_\TE \,]\,.
\label{I-t}
\end{align}
Equation (\ref{tilde-a-TCL}) corresponds to the TCL equation of motion for the reduced density operator of the system \cite{HSS}. The inhomogeneous term $I(t)$ represents the effects of initial correlation of the system and reservoir, because if the initial state $\rho_\TE$ [(\ref{rhoTE})] is approximated to the decoupled one $\rho_0 \rho_\R$, the term $I(t)$ vanishes. Equation (\ref{tilde-a-TCL}) has the formal solution 
\begin{equation}
\tilde{a}(t) = \exp_\gets \!\Big\{\! - i \, L_\SS \, t + \int_0^t d\tau \, C(\tau) \Big\}\, \tilde{a}(0) + \int_0^t d\tau \exp_\gets \!\Big\{\! - i \, L_\SS \cdot (t - \tau) + \int_\tau^t ds \, C(s) \Big\} \, I(\tau). \quad
\label{tilde-a-solTCL}
\end{equation}
Substituting (\ref{tilde-a-solTCL}) into the Kubo formula (\ref{KF}), 
the admittance takes the form \cite{S6}
\begin{align}
\chi^{\R\TCL}_{ij}(\omega) & = \frac{i}{\hbar} \int_0^\infty dt \,\, \Tr \, A_i \exp(- \, i \, L \, t ) \, [\, A_j \,,\, \rho_\TE \,] \exp(i \, \omega \, t) = \frac{i}{\hbar} \int_0^\infty dt \, \tr \, A_i \, \tilde{a}_j(t) \exp(i \, \omega \, t), \nonumber \\
    & = \frac{i}{\hbar} \int_0^\infty dt \, \tr \, A_i \, \exp_\gets \! \Big\{i \, (\omega - L_\SS) \, t + \int_0^t d\tau \, C(\tau) \Big\} \, [\, A_j \,,\, \rho_0 \,] \nonumber \\
    & \quad + \, \frac{i}{\hbar} \int_0^\infty dt \int_0^t d\tau \, \tr \, A_i \exp_\gets \! \Big\{i \, (\omega - L_\SS) \, (t - \tau) + \int_\tau^t ds \, C(s) \Big\}\, I_j(\tau) \exp(i \, \omega \, \tau), \qquad \label{RTCLadmit}
\end{align}
which is a general form of admittance derived using the relaxation TCL method under thermal equilibrium initial conditions, where $I_j(t)$ is equal to $I(t)$ with $A_j$ in place of $A$. 
The first term of the above admittance (\ref{RTCLadmit}) is the admittance obtained using the conventional relaxation method under the decoupled initial condition 
$\rho_\T(t_0) $\,=\,$ \rho_\TE \Rightarrow \rho_0 \, \rho_\R$, 
and its second term comes from the inhomogeneous term $I(t)$ in Eq. (\ref{tilde-a-TCL}) and represents the effects of initial correlation of the system and reservoir. 

Expanding Eq. (\ref{tilde-a-TCL}) up to second order in powers of the system-reservoir interaction $\H_\SR$, it reduces to
\begin{equation}
(d/dt)\, \tilde{a}(t) = - \, i \, L_\SS \, \tilde{a}(t) + C^{(2)}(t)\, \tilde{a}(t) + I^{(2)}(t), \qquad \quad
\label{tilde-a-TCL2}
\end{equation}
with
\begin{align}
& C^{(2)}(t) = - \int_0^t d\tau \, \big\langle L_\SR \exp(- \, i \, L_0\, \tau) \, L_\SR \exp(i \, L_0 \, \tau) \big\rangle_\R = - \int_0^t d\tau \, \big\langle L_\SR \, L_\SR(- \, \tau) \big\rangle_\R \,, \qquad
\label{C2-t} \\
& I^{(2)}(t) = \bar{I}^{(2)}(t) = i \int_0^\beta d\beta' \, \tr_\R \, L_\SR \exp(- \, i \, L_0 \, t) \, \big[ A\,,\, \rho_\SS \, \rho_\R \, \H_\SR(- \, i \, \hbar \, \beta') \, \big].
\label{I2-t}
\end{align}
Solving Eq. (\ref{tilde-a-TCL2}) formally and substituting its solution into the Kubo formula (\ref{KF}), the admittance takes the form \cite{S5,S6}
\begin{align}
\chi^{\R\TCL}_{ij}(\omega) = & \, \frac{i}{\hbar} \int_0^\infty dt \, \tr \, A_i \exp_\gets \! \Big\{ i \, (\omega - L_\SS)\, t + \int_0^t d\tau \, C^{(2)}(\tau) \Big\} \, \big[ A_j \,,\, \rho_\SS + \rho_0^{(2)} \,\big] \nonumber \\
    & + \, \frac{i}{\hbar} \int_0^\infty dt \int_0^t d\tau \, \tr \, A_i \exp_\gets \! \Big\{ i \, (\omega - L_\SS) \, (t - \tau) + \int_\tau^t ds \, C^{(2)}(s) \Big\}\, I_j^{(2)}(\tau) \exp(i \, \omega \, \tau), \quad \label{RTCL2admit}
\end{align}
which is a general form of the admittance derived using the relaxation TCL method in the lowest Born approximation for the system-reservoir interaction, where $I_j^{(2)}(t)$ is equal to $I^{(2)}(t)$ with $A_j$ in place of $A$. The second term of the above admittance comes from the inhomogeneous term $I^{(2)}(t)$ in Eq. (\ref{tilde-a-TCL2}) and represents the effects of initial correlation of the system and reservoir.
%
\subsection{The TCE method}

We next derive the form of complex admittance by the ``TCE method" in which complex admittances are directly calculated by solving the \underline{t}ime-\underline{c}onvolution (TC) equation of motion including \underline{e}xternal driving terms. 
Equation (\ref{QrhoE}) has the formal solution
\begin{equation}
\Q \, \rho_{\T1}(t) = - \, i \int_0^t d\tau \exp\{- \, i \, \Q \, L \, \Q \,(t - \tau)\}\, \big\{ \Q \, L \, \P \rho_{\T1}(\tau) + \Q \, L_\ed(\tau) \, \rho_\TE \big\}, \qquad
\label{QrhoTC}
\end{equation}
which has a form of the time-convolution (TC). By substituting the time-convolution expression (\ref{QrhoTC}) for $\Q \, \rho_{\T1}(t)$ into (\ref{PrhoE}) directly, the time-convolution (TC) equation of motion for $\P \rho_{\T1}(t)$ can be obtained as
\begin{align}
(d/dt)\, \P \rho_{\T1}(t) = & - \, i \, \P L \, \P \rho_{\T1}(t) - \P L \int_0^t d\tau \exp\{- \, i \, \Q \, L \, \Q \, (t - \tau) \} \, \Q \, L \, \P \rho_{\T1}(\tau) \nonumber \\
    & - \, i \, \P L_\ed(t) \, \rho_\TE - \P L \int_0^t d\tau \exp\{- \, i \, \Q \, L \, \Q \, (t - \tau) \} \, \Q \, L_\ed(\tau) \, \rho_\TE \,. \quad
\label{PrhoTC}
\end{align}
By virtue of the projection operators (\ref{projectionPQ}), Eq. (\ref{PrhoTC}) is reduced to the TC equation of motion for $\rho_1(t)$ [=\,$ \tr_\R \, \rho_{\T1}(t)]$ \cite{AK}\,:
\begin{equation}
(d/dt) \, \rho_1(t) = - \, i \, L_\SS \, \rho_1(t) + \bar{C}(t , \{\rho_1 \}) - i \, L_\ed(t) \, \rho_0 + \bar{D}(t). \qquad 
\label{rhoTC}
\end{equation}
On the right-hand side of Eq. (\ref{rhoTC}), the collision term $\bar{C}(t , \{\rho_1 \})$ and the inhomogeneous term $\bar{D}(t)$ are given by
\begin{align}
& \bar{C}(t \,,\, \{\rho_1 \}) = - \int_0^t d \tau \, \tr_\R \, L_\SR \exp\{- \, i \, \Q \, L \, \Q \, (t - \tau) \} \, L_\SR \, \rho_\R \, \rho_1 (\tau) = - \int_0^t d \tau \, \big\langle \, L_\SR \exp(- \, i \, \Q \, L \, \Q \, \tau) \, L_\SR \, \big\rangle_\R \, \rho_1 (t - \tau),
\label{CbarTC} \\
& \bar{D}(t) = - \int_0^t d \tau \, \tr_\R \, L_\SR \exp\{- \, i \, \Q \, L \, \Q \, (t - \tau) \} \, L_\ed(\tau) \, \Q \, \rho_\TE = - \int_0^t d \tau \, \tr_\R \, L_\SR \exp(- \, i \, \Q \, L \, \Q \, \tau) \, L_\ed(t - \tau) \, \Q \, \rho_\TE \,.
\label{DbarTC}
\end{align}
The inhomogeneous term $\bar{D}(t)$ is called the ``interference term" in the TC equation, and represents the effects of initial correlation of the system and reservoir, because if the initial state $\rho_\TE$ [(\ref{rhoTE})] is approximated to the decoupled one $\rho_0 \rho_\R$, the interference term $\bar{D}(t)$ vanishes. 
We now take into account the interaction of the system with the external driving field, which is turned on adiabatically at the initial time $t$\,=\,0, to be given by the Hamiltonian (\ref{edf}). 
The admittance $\chi_{ij}(\omega)$ for the physical quantities $A_i$ and $A_j$ of the system is defined in the limit $t$\,$ \to $\,$ \infty $, as
\begin{equation}
\Tr \, A_i \, \rho_{\T1}(t) = \tr \, A_i \, \rho_1(t) = \sum_j \, \sum_\omega \, \chi_{ij}(\omega) \, F_j[\omega] \, e^{- \, i \, \omega \, t}, \qquad \qquad \qquad (t \to \infty) \quad \ \label{Eadmit}
\end{equation}
which corresponds to (\ref{Kadmit}). 
We assume that the correlation times of the heat reservoir are finite, that the integrands of (\ref{CbarTC}) and (\ref{DbarTC}) vanish for finite $\tau$ in the limit $t$\,$ \to $\,$ \infty $, and that Eq. (\ref{rhoTC}) has a stationary solution of the form 
\begin{equation}
\rho_1(t) = \sum_\omega \, \rho_1[\omega] \, e^{- \, i \, \omega \, t}. \qquad \qquad \qquad \qquad (t \to \infty) \qquad 
\label{rho1-inf}
\end{equation}
The summations $\sum_\omega$ for $\omega$ in (\ref{Eadmit}) and (\ref{rho1-inf}) are over the same frequencies as those in (\ref{edf}). Then, $\rho_1[\omega]$ satisfies the equation
\begin{equation}
- \, i \, \omega \, \rho_1[\omega] = - \, i \, L_\SS\, \rho_1[\omega] + \bar{C}[\omega] 
\, \rho_1[\omega] - i \, L_\ed[\omega] \, \rho_0 + \bar{D}[\omega] \,, \qquad
\label{rho1TCeq}
\end{equation}
which has the formal solution
\begin{equation}
\rho_1[\omega] = \big(\, i \, (L_\SS - \omega) - \bar{C}[\omega] \,\big)^{-1} \big\{\! - i \, L_\ed[\omega]\, \rho_0 + \bar{D}[\omega] \, \big\}\,, \qquad
\label{rho1TC}
\end{equation}
where $\bar{C}[\omega]$ is the collision operator given by (\ref{bar-C-omega}), and $\bar{D}[\omega]$ is the interference term defined through the relation
\begin{equation}
\bar{D}(t) = \sum_\omega \bar{D}[\omega] \, e^{- \, i \,\omega \, t}, \qquad \qquad \qquad \qquad (t \to \infty) \qquad
\label{bar-D}
\end{equation}
and takes the expression
\begin{equation}
\bar{D}[\omega] = - \int_0^\infty d\tau \, \tr_\R \, L_\SR \exp(- \, i \, \Q \, L \, \Q \, \tau) \, L_\ed[\omega]\, \Q \, \rho_\TE \exp(i \, \omega \, \tau). \qquad 
\label{bar-D-omega}
\end{equation}
In the relation (\ref{bar-D}), the summation $\sum_\omega$ for $\omega$ is over the same frequencies as those in (\ref{edf}). By substituting the formal solution (\ref{rho1TC}) into (\ref{rho1-inf}) and by using the definition (\ref{Eadmit}) of the admittance, we can obtain
\begin{equation}
\chi^{\TCE}_{ij}(\omega) = \tr \, A_i \, \frac{1}{\, i \, (L_\SS - \omega) 
- \bar{C}[\omega] \,} \, \Big\{ \frac{i}{\hbar} \, [\, A_j \,,\, \rho_0 \,] 
+ \bar{D}_j[\omega] \Big\}, \qquad
\label{TCEadmit}
\end{equation}
which is a general form of the admittance obtained using the TCE method, 
where $\bar{D}_j[\omega]$ is the interference term given by (\ref{barDj-omega}) and represents the effects of initial correlation of the system and reservoir. 
We notice that the admittance (\ref{TCEadmit}) obtained using the TCE method 
has exactly the same form as the admittance (\ref{RTCadmit}) obtained using the relaxation TC method.

Expanding Eq. (\ref{rho1TCeq}) up to second order in powers of the system-reservoir 
interaction $\H_\SR$, it reduces to
\begin{equation}
- \, i \, \omega \, \rho_1[\omega] = - \, i \, L_\SS\, \rho_1[\omega] + \bar{C}^{(2)}[\omega] \, \rho_1[\omega] - i \, L_\ed[\omega] \, \big( \rho_\SS + \rho_0^{(2)} \big) + \bar{D}^{(2)}[\omega] \,, \qquad \label{rho1TC2eq}
\end{equation}
where the collision operator $\bar{C}^{(2)}[\omega]$ is given by (\ref{bar-C2-omega}) 
and the interference term $\bar{D}^{(2)}[\omega]$ is given by
\begin{equation}
\bar{D}^{(2)}[\omega] = \int_0^\infty d\tau \int_0^\beta d\beta' \, \tr_\R \, L_\SR \exp(- \, i \, L_0 \, \tau) \, L_\ed[\omega] \, \rho_\SS \, \rho_\R \, \H_\SR(- \, i \, \hbar \, \beta' \,) \exp(i \, \omega \, \tau), \qquad
\label{bar-D2b-omega}
\end{equation}
which is shown in Appendix A to coincide with
\begin{equation}
\bar{D}^{(2)}[\omega] = i \int_0^\infty d\tau \int_0^\tau ds \, \tr_\R \, L_\SR \exp(- \, i \, L_0 \, s) \, L_\ed[\omega] \exp\{i \, L_0 \, (s - \tau) \}\, L_\SR \, \rho_\SS \, \rho_\R \exp(i \, \omega \, s). \qquad
\label{bar-D2t-omega}
\end{equation}
Substituting the formal solution of Eq. (\ref{rho1TC2eq})\,:
\begin{equation}
\rho_1[\omega] = \big(\, i \,(L_\SS - \omega) - \bar{C}^{(2)}[\omega] \,\big)^{-1} \big\{\! - i \, L_\ed[\omega]\, \big( \rho_\SS + \rho_0^{(2)} \big) + \bar{D}^{(2)}[\omega] \,\big\} \qquad 
\label{rho1TC2}
\end{equation}
into (\ref{rho1-inf}) and using the definition (\ref{Eadmit}) of the admittance, 
the admittance takes the form
\begin{equation}
\chi_{ij}^{\TCE}(\omega) = \tr \, A_i \, \frac{1}{\, i \, (L_\SS - \omega) - \bar{C}^{(2)}[\omega] \,} \Big\{ \frac{i}{\hbar} \, \big[ A_j \,,\, \rho_\SS + \rho_0^{(2)} \,\big] + \bar{D}_j^{(2)}[\omega] \Big\}, \qquad 
\label{TCE2admit}
\end{equation}
which has the same form as the admittance (\ref{RTC2admit}) obtained using the relaxation TC method in the lowest Born approximation for the system-reservoir interaction. Here, the interference term $\bar{D}_j^{(2)}[\omega]$ is given by (\ref{barDj2b-omega}), is shown in Appendix A to coincide with
\begin{equation}
\bar{D}_j^{(2)}[\omega] = - \, \frac{i}{\hbar} \int_0^\infty d\tau \int_0^\tau ds \, \tr_\R\, L_\SR \exp(- \, i \, L_0 \, s) \, \big[ A_j \,,\, \exp\{i \, L_0 \, (s - \tau) \} \, L_\SR \, \rho_\SS \, \rho_\R \,\big] \exp(i \, \omega \, s), \quad
\label{barDj2t-omega}
\end{equation}
and represents the effects of initial correlation of the system and reservoir. The above result (\ref{TCE2admit}) can be obtained by expanding the TC equation (\ref{rhoTC}) up to second order in powers of the system-reservoir interaction too. 
In the TCE method, the admittance (\ref{TCE2admit}) is obtained by the lowest Born approximation of the equation (\ref{rhoTC}) or (\ref{rho1TCeq}) including external driving terms, while in the relaxation TC method, the admittance (\ref{RTC2admit}) is obtained by the lowest Born approximation of the equation (\ref{tilde-a-TC}) or (\ref{tilde-a-TComega}) including no external driving terms. It should be noticed that although the equations in which the perturbation is truncated are different from each other in the two methods, the expressions of the obtained admittances coincides with each other.
%
\subsection{The TCLE method}

One of the authors has proposed and has studied the ``TCLE method" in which the complex admittance is directly calculated by solving the \underline{t}ime-\underline{c}onvolution\underline{l}ess (TCL) equations of motion including \underline{e}xternal driving terms \cite{S1,S2,S3,S4,S5,S6,S7}. 
However, the exact expression of complex admittance obtained using the TCLE method has not been derived. 
We here derive the exact form of complex admittance by the TCLE method and survey that method. 
In order to renormalize the time-convolution (TC) in (\ref{QrhoTC}), we write the formal solution of Eq. (\ref{order1-2}) in the form
\begin{equation}
\rho_{\T1}(\tau) = \exp\{ i \, L \, (t - \tau) \} \, \rho_{\T1}(t) + i \int_\tau^t ds 
\exp\{ i \, L \,(s - \tau) \} \, L_\ed (s)\, \rho_\TE \,, \qquad 
\label{rhoT1-tau}
\end{equation}
where we have used the solution (\ref{rhoT0}) of Eq. (\ref{order0-2}). Substituting (\ref{rhoT1-tau}) into (\ref{QrhoTC}) and solving it for $\Q \rho_{\T1}(t)$, we have
\begin{equation}
\Q\, \rho_{\T1}(t) = \{\theta(t) - 1 \}\, \P \rho_{\T1}(t) - i \, \theta(t) \int_0^t  d\tau \exp\{- \, i \, \Q \, L \, \Q \, (t - \tau) \} \, \Q \, \theta^{-1}(\tau) \, L_\ed(\tau) \, \rho_\TE \,, \quad 
\label{QrhoTCL}
\end{equation}
where $\theta(t)$ is given by (\ref{theta-t}). 
By substituting the above expression for $\Q \rho_{\T1}(t)$ into (\ref{PrhoE}), 
the time-convolutionless (TCL) equation of motion for $\P \rho_{\T1}(t)$ can be obtained as
\begin{align}
(d/dt)\, \P \rho_{\T1}(t) = & - \, i \, \P L \, \P \rho_{\T1}(t) - i \, \P L \, \{\theta(t) - 1\}\, \P \rho_{\T1}(t) - i \, \P L_\ed(t)\, \rho_\TE \nonumber \\
    & - \int_0^t d\tau \, \P L \, \theta(t) \exp\{- \, i \, \Q \, L \, \Q \, (t - \tau) \} \, \Q \, \theta^{-1}(\tau) L_\ed(\tau)\, \rho_\TE \,. \qquad 
\label{PrhoTCL}
\end{align}
By virtue of the projection operators (\ref{projectionPQ}), Eq. (\ref{PrhoTCL}) is reduced to the TCL equation of motion for $ \rho_1(t) $ [=\,$ \mathrm{tr}_\R \,\rho_{\T1}(t)]$ \cite{S1,S2,S6}\,:
\begin{equation}
(d/dt)\, \rho_1(t) = - \, i \, L_\SS \, \rho_1(t) + C(t) \, \rho_1(t) - i \, L_\ed(t)\, \rho_0 + D(t). \qquad \quad
\label{rhoTCL}
\end{equation}
On the right-hand side of Eq. (\ref{rhoTCL}), $C(t)$ is the collision operator given by (\ref{C-t}), and $D(t)$ describes the effects of interference between the external driving field and heat reservoir, is called the ``interference term" and take the form 
\begin{align}
D(t) = & - \int_0^t d\tau \, \tr_\R \, L_\SR\, \frac{1}{1- \Sigma(t)} \exp(- \, i \, \Q \, L \, \Q \, \tau) \, L_\ed(t - \tau)\, \Q \, \rho_\TE \nonumber \\
    & + \int_0^t d\tau \, \tr_\R\, L_\SR\, \frac{1}{1 - \Sigma(t)} \exp(- \, i \, \Q \, L \, \Q \, \tau) \, \Sigma(t - \tau) \, L_\ed(t - \tau) \, \rho_\TE \,, \qquad
\label{DTCL}
\end{align}
with $\Sigma(t)$ defined by
\begin{equation}
\Sigma(t) = 1 - \theta^{-1}(t) = - \, i \int_0^t d\tau \exp(- \, i \, \Q \, L \, \Q \, \tau) \, 
\Q \, L_\SR\, \P \exp(i \, L \, \tau). \qquad
\label{Sigma-t}
\end{equation}
The first term of $D(t)$ given by (\ref{DTCL}) is called the ``first interference term" and represents the effects of initial correlation of the system and reservoir, because if the initial state $\rho_\TE$ [(\ref{rhoTE})] is approximated to the decoupled one $\rho_0 \rho_\R$, this term vanishes. 
But, the first term of $D(t)$ [(\ref{DTCL})] includes the memory effects and is different from $\bar{D}(t)$ given by (\ref{DbarTC}). 
The second term of $D(t)$ is called the ``second interference term" and represents the memory effects, which are the effects of collision of the system with the heat reservoir, because this term is derived by renormalizing the time-convolution in (\ref{QrhoTC}). 
We assume that the correlation times of the heat reservoir are finite and that Eq. (\ref{rhoTCL}) has a stationary solution of the form (\ref{rho1-inf}). Then, $\rho_1[\omega]$ satisfies the equation
\begin{equation}
- \, i \, \omega \, \rho_1[\omega] = - \, i \, L_\SS \, \rho_1[\omega] + C \, \rho_1[\omega] - i \, L_\ed[\omega] \, \rho_0 + D[\omega] \,. \qquad 
\label{rho1TCLeq}
\end{equation}
On the right-hand side of Eq. (\ref{rho1TCLeq}), $C$ is the collision operator given by 
\begin{equation}
C = C(\infty) = - \, i \cdot \tr_\R \, L_\SR \, \Sigma \, (1 - \Sigma)^{-1} \rho_\R 
=  - \, i \cdot \langle L_\SR \, \Sigma \, (1 - \Sigma)^{-1}\, \rangle_\R \,, \qquad
\label{C-inf}
\end{equation}
and $D[\omega]$ is the interference term defined through the relation
\begin{equation}
D(t) = \sum_\omega D[\omega] \, e^{- \, i \,\omega \, t}, \qquad \qquad \qquad \qquad 
(t \to \infty) \qquad \ 
\label{D}
\end{equation}
and is shown in Appendix B to take the expression
\begin{align}
D[\omega] = & - \int_0^\infty d\tau \, \tr_\R \, L_\SR\, \frac{1}{1- \Sigma} \exp(- \, i \, \Q \, L \, \Q \, \tau) \, L_\ed[\omega] \, \Q \, \rho_\TE \exp(i \, \omega \, \tau) \nonumber \\
    & - i \int_0^\infty d\tau \int_0^\tau ds \, \tr_\R\, L_\SR\, \frac{1}{1 - \Sigma} \exp(- \, i \, \Q \, L \, \Q \, \tau) \, \Q \, L_\SR \, \P \exp\{i \, L \, (\tau - s)\} L_\ed[\omega] \, \rho_\TE \exp(i \, \omega \, s), \quad 
\label{D-omega}
\end{align}
where $\Sigma$ is given by
\begin{equation}
\Sigma = \Sigma(\infty) = 1 - \theta^{-1}(\infty) = - \, i \int_0^\infty d\tau 
\exp(- \, i \, \Q \, L \, \Q \, \tau) \, \Q \, L_\SR\, \P \exp(i \, L \, \tau). \quad
\label{Sigma}
\end{equation}
The summation $\sum_\omega$ in (\ref{D}) is over the same frequencies as those in (\ref{edf}). By substituting the formal solution of Eq. (\ref{rho1TCLeq})\,:
\begin{equation}
\rho_1[\omega] = \big(\, i \, (L_\SS - \omega) - C \,\big)^{-1} \big\{\! - i \, L_\ed[\omega]\, \rho_0 + D[\omega] \,\big\} \qquad \ 
\label{rho1TCL}
\end{equation}
into (\ref{rho1-inf}) and by using the definition (\ref{Eadmit}) of the admittance for the interaction (\ref{edf}) of the system with the external driving field, which is turned on adiabatically at initial time $t$\,=\,0, we can obtain
\begin{equation}
\chi_{ij}^{\TCLE}(\omega) = \tr \, A_i \, \frac{1}{\, i \, (L_\SS - \omega) - C \,} 
\Big\{ \frac{i}{\hbar}\, [\, A_j \,,\, \rho_0 \,] + D_j[\omega] \Big\}, \qquad \ 
\label{TCLEadmit}
\end{equation}
which is a general form of the admittance obtained using the TCLE method, 
where the interference term $D_j[\omega]$ is given by
\begin{align}
D_j[\omega] = & \, \frac{1}{\hbar} \int_0^\infty d\tau \, \tr_\R \, L_\SR\, \frac{1}{1- \Sigma} \exp(- \, i \, \Q \, L \, \Q \, \tau) \, \Q \, [\, A_j \,,\, \rho_\TE \,] \exp(i \, \omega \, \tau) \nonumber \\
    & + \, \frac{i}{\hbar} \int_0^\infty d\tau \int_0^\tau ds \, \tr_\R\, L_\SR\, \frac{1}{1 - \Sigma} \exp(- \, i \, \Q \, L \, \Q \, \tau)\, \Q \, L_\SR \, \P \exp\{i \, L \, (\tau - s)\}\, [\, A_j \,,\, \rho_\TE \,] \exp(i \, \omega \, s).
\label{Dj-omega}
\end{align}
The first term of $D_j[\omega]$ represents the effects of initial correlation of the system and reservoir, and its second term represents the memory effects.

Expanding Eq. (\ref{rho1TCLeq}) up to second order in powers of the system-reservoir interaction $\H_\SR$, it reduces to
\begin{equation}
- \, i \, \omega \, \rho_1[\omega] = - \, i \, L_\SS\, \rho_1[\omega] 
+ C^{(2)} \, \rho_1[\omega] - i \, L_\ed[\omega] \, \big( \rho_\SS + \rho_0^{(2)} \big) + D^{(2)}[\omega] \,, \qquad
\label{rho1TCL2eq}
\end{equation}
where the collision operator $C^{(2)}$ and the interference term $D^{(2)}[\omega]$ 
are given by
\begin{align}
    & C^{(2)} = C^{(2)}(\infty) = - \int_0^\infty d\tau \, \big\langle L_\SR \exp(- \, i \, L_0 \, \tau) \, L_\SR \exp(i \, L_0 \, \tau) \big\rangle_\R = - \int_0^\infty d\tau \, \big\langle L_\SR \, L_\SR(- \, \tau) \big\rangle_\R \,, 
\label{C2-inf} \\
    & D^{(2)}[\omega] = \bar{D}^{(2)}[\omega] - i \int_0^\infty d\tau \int_0^\tau ds \, \tr_\R\, L_\SR \exp(- \, i \, L_0 \, \tau) \, L_\SR \exp\{ i \, L_0 \, (\tau - s) \} \, L_\ed[\omega] \, \rho_\SS \, \rho_\R \exp(i \, \omega \, s), \ 
\label{D2-omega}
\end{align}
with $\bar{D}^{(2)}[\omega]$ given by (\ref{bar-D2b-omega}) or (\ref{bar-D2t-omega}). 
The first term of $D^{(2)}[\omega]$, which is called the ``first interference term", is equal to the interference term $\bar{D}^{(2)}[\omega]$ in the TC equation and represents the effects of initial correlation of the system and reservoir. 
Its second term, which is called the ``second interference term", represents the memory effects. 
Substituting the formal solution of Eq. (\ref{rho1TCL2eq})\,:
\begin{equation}
\rho_1[\omega] = \big(\, i \,(L_\SS - \omega) - C^{(2)} \, \big)^{-1} \big\{\! - i \, L_\ed[\omega] \, \big( \rho_\SS + \rho_0^{(2)} \big) + D^{(2)}[\omega] \,\big\} \qquad \ 
\label{rho1TCL2}
\end{equation}
into (\ref{rho1-inf}) and using the definition (\ref{Eadmit}) of the admittance, the admittance 
takes the form \cite{S4,S6}
\begin{equation}
\chi_{ij}^{\TCLE}(\omega) = \tr \, A_i \, \frac{1}{\, i \, (L_\SS - \omega) - C^{(2)} \,} \Big\{ \frac{i}{\hbar}\, \big[ A_j \,,\, \rho_\SS + \rho_0^{(2)} \,\big] + D_j^{(2)}[\omega] \Big\}, \qquad \ \label{TCLE2admit}
\end{equation}
which is a general form of the admittance obtained using the TCLE method 
in the lowest Born approximation for the system-reservoir interaction. 
Here, the interference term $D_j^{(2)}[\omega]$ is given by
\begin{equation}
D_j^{(2)}[\omega] = \bar{D}_j^{(2)}[\omega] + \frac{i}{\hbar} \int_0^\infty d\tau \int_0^\tau ds \, \tr_\R \, L_\SR \exp(- \, i \, L_0 \, \tau) \, L_\SR \exp\{i \, L_0 \, (\tau - s) \} \big[ A_j \,,\, \rho_\SS \,\big] \, \rho_\R \exp(i \, \omega \, s), \ 
\label{Dj2-omega}
\end{equation}
with $\bar{D}_j^{(2)}[\omega]$ given by (\ref{barDj2b-omega}) or (\ref{barDj2t-omega}). 
The first term of $D_j^{(2)}[\omega]$ is equal to $\bar{D}_j^{(2)}[\omega]$ and represents the effects of initial correlation of the system and reservoir. Its second term represents the memory effects. The above result (\ref{TCLE2admit}) can be obtained by expanding the TCL equation (\ref{rhoTCL}) up to second order in powers of the system-reservoir interaction too \cite{S4,S6}.
%
\section{Relations among admittances}

In this section, we examine the relations among the forms of admittance derived in Sections 2 and 3 analytically. 
As found in the previous section, the admittance (\ref{RTCadmit}) obtained using the relaxation TC method has the same form as the admittance (\ref{TCEadmit}) obtained using the TCE method, i.e.,
\begin{equation}
\chi^{\R\TC}_{ij}(\omega) = \chi^{\TCE}_{ij}(\omega). \qquad \qquad 
\label{RTC-TCE}
\end{equation}
We examine analytically the relations among the admittances (\ref{RTC2admit}) [or (\ref{TCE2admit})], (\ref{RTCL2admit}) and (\ref{TCLE2admit}), which are obtained using the relaxation TC method [or the TCE method], the relaxation TCL method and the TCLE method, respectively, in the lowest Born approximation for the system-reservoir interaction.

One of the authors has investigated the relation between the admittance (\ref{RTCL2admit}) obtained using the relaxation TCL method and the admittance (\ref{TCLE2admit}) obtained using the TCLE method \cite{S4,S5,S6}. We first survey the relation between the admittances (\ref{RTCL2admit}) and (\ref{TCLE2admit}) in a compact way. The first term of the admittance $\chi^{\R\TCL}_{ij}(\omega)$ [(\ref{RTCL2admit})] can be rewritten as
\begin{align}
(\mathrm{first}\ \mathrm{term}\ \mathrm{of}\ \chi^{\R\TCL}_{ij}(\omega)\ [(\ref{RTCL2admit})]) & = \frac{i}{\hbar}\int_0^\infty dt \, \tr \, A_i \exp_\gets \!\Big\{ i \, (\omega - L_\SS) \, t + C^{(2)}\, t + \int_0^t d\tau \, (C^{(2)}(\tau) - C^{(2)}) \Big\} \big[ A_j \,,\, \rho_\SS + \rho_0^{(2)} \,\big] , \nonumber \\
    & = \frac{i}{\hbar}\int_0^\infty dt \, \tr \, A_i \, U(\omega , t) \exp_\gets \! \Big\{ \int_0^t d\tau \, U^{-1}(\omega , \tau) \, \big(C^{(2)}(\tau) - C^{(2)}\big) \, U(\omega , \tau) \Big\} \nonumber \\
    & \qquad \qquad \qquad \times \big[ A_j \,,\, \rho_\SS + \rho_0^{(2)} \,\big], 
\label{RTCL2admit-f1}
\end{align}
with $U(\omega , t)$ defined by 
\begin{equation}
U(\omega , t) = \exp\{\, i \, (\omega - L_\SS) \, t + C^{(2)} \, t \,\}, \qquad \quad
\label{U-omega-t}
\end{equation}
and can be expanded in powers of the system-reservoir interaction as
\begin{align}
(\mathrm{first}\ \mathrm{term}\ \mathrm{of}\ \chi^{\R\TCL}_{ij}(\omega)\ [(\ref{RTCL2admit})]) & = \frac{i}{\hbar} \int_0^\infty dt \, \tr \, A_i \, U(\omega , t) \, \big[ A_j \,,\, \rho_\SS + \rho_0^{(2)} \,\big] \nonumber \\
    & \ + \frac{i}{\hbar} \int_0^\infty \! dt \int_0^t d\tau \int_\tau^\infty \! ds \, \tr \, A_i \, U(\omega , t - \tau) \, \tr_\R \, L_\SR \exp(- \, i \, L_0 \, s) L_\SR \exp(i \, L_0 \, s) \, \rho_\R \, U(\omega , \tau) \nonumber \\
    & \qquad \qquad \qquad \qquad \qquad \qquad \times \big[ A_j \,,\, \rho_\SS + \rho_0^{(2)} \,\big] + \, \cdots \,,
\label{RTCL2admit-f2}
\end{align}
where (\ref{C2-t}) and (\ref{C2-inf}) have been substituted into $C^{(2)}(t)$ and $C^{(2)}$, respectively. The first term of (\ref{RTCL2admit-f2}) can be integrated to give
\begin{equation}
\frac{i}{\hbar} \int_0^\infty dt \, \tr \, A_i \, U(\omega , t) \, \big[ A_j \,,\, \rho_\SS + \rho_0^{(2)} \,\big] = \frac{i}{\hbar} \, \tr \, A_i \, \big(\, i \, (L_\SS - \omega) - C^{(2)} \,\big)^{-1} \big[ A_j \,,\, \rho_\SS + \rho_0^{(2)} \,\big]. \ 
\label{RTCL2admit-f3}
\end{equation}
Considering that in the second term of (\ref{RTCL2admit-f2}), 
\begin{align}
& \int_0^\infty dt \int_0^t d\tau \int_\tau^\infty ds \, \tr \, A_i \, U(\omega , t - \tau) = \int_0^\infty d\tau \int_\tau^\infty dt \int_\tau^\infty ds \, \tr \, A_i \, U(\omega , t - \tau), \nonumber \\
    = & \, \int_0^\infty d\tau \int_\tau^\infty ds \, \tr \, A_i \, \big(\, i \, (L_\SS - \omega) - C^{(2)} \,\big)^{-1} = \int_0^\infty ds \int_0^s d\tau \, \tr \, A_i \, \big(\, i \, (L_\SS - \omega) - C^{(2)} \,\big)^{-1}, \quad \label{RTCL2admit-f4}
\end{align}
the first term of the admittance $\chi^{\R\TCL}_{ij}(\omega)$ [(\ref{RTCL2admit})]) can be expressed as
\begin{align}
(\mathrm{first}\ \mathrm{term}\ \mathrm{of}\ \chi^{\R\TCL}_{ij}(\omega)\ [(\ref{RTCL2admit})]) = & \, \tr \, A_i \, \big(\, i \, (L_\SS - \omega) - C^{(2)} \,\big)^{-1} \, \big\{ (i / \hbar) \big[ A_j \,,\, \rho_\SS + \rho_0^{(2)} \,\big] + \big( \mathrm{second}\ \mathrm{term}\ \mathrm{of}\ D_j^{(2)}[\omega] \big) \big\} \nonumber \\
    & + \HOT \ (\mathrm{higher}\ \mathrm{order}\ \mathrm{terms}),
\label{RTCL2admit-f5}
\end{align}
where $D_j^{(2)}[\omega]$ is given by (\ref{Dj2-omega}). The second term of the admittance (\ref{RTCL2admit}) can be rewritten by substituting (\ref{I2-t}), as 
\begin{align}
(\mathrm{second}\ \mathrm{term}\ \mathrm{of}\ \chi^{\R\TCL}_{ij}(\omega)\ [\ref{RTCL2admit}]) & = - \, \frac{1}{\hbar} \int_0^\infty \! dt \int_0^t d\tau \, \tr \, A_i \exp_\gets \! \Big\{ (i \, (\omega - L_\SS) + C^{(2)})(t - \tau) + \! \int_\tau^t ds \, (C^{(2)}(s) - C^{(2)}) \Big\} \nonumber \\
    & \qquad \qquad \qquad \times \int_0^\beta d\beta' \, \tr_\R \, L_\SR \exp(- \, i \, L_0 \, \tau) \big[ A_j \,,\, \rho_\SS \, \rho_\R \, \H_\SR(- \, i \, \hbar \, \beta') \big] \exp(i \, \omega \, \tau), 
\label{RTCL2admit-s1}
\end{align}
and can be expanded in powers of the system-reservoir interaction as
\begin{align}
(\mathrm{second}\ \mathrm{term}\ \mathrm{of}\ \chi^{\R\TCL}_{ij}(\omega)\ [\ref{RTCL2admit}]) & = - \, \frac{1}{\hbar} \int_0^\infty d\tau \int_\tau^\infty dt \int_0^\beta d\beta' \, \tr \, A_i \exp \big\{ \big(i \, (\omega - L_\SS) + C^{(2)} \big) (t - \tau) \big\} \nonumber \\
    & \qquad \qquad \qquad \qquad \qquad \qquad { } \times \tr_\R \, L_\SR \exp(- \, i \, L_0 \, \tau) \, \big[ A_j \,,\, \rho_\SS \, \rho_\R \, \H_\SR(- \, i \, \hbar \, \beta') \big] \exp(i \, \omega \, \tau) \nonumber \\
    & \quad + \cdots \,,
\label{RTCL2admit-s2}
\end{align}
which can be expressed by integrating with respect to $t$, as
\begin{equation}
(\mathrm{second}\ \mathrm{term}\ \mathrm{of}\ \chi^{\R\TCL}_{ij}(\omega)\ [(\ref{RTCL2admit})]) = \tr \, A_i \, \big(\, i \, (L_\SS - \omega) - C^{(2)} \,\big)^{-1} \, \big( \bar{D}_j^{(2)}[\omega]\ \mathrm{or}\ \mathrm{first}\ \mathrm{term}\ \mathrm{of}\ D_j^{(2)}[\omega] \big) + \HOT, \ 
\label{RTCL2admit-s3}
\end{equation}
where $\bar{D}_j^{(2)}[\omega]$ is given by (\ref{barDj2b-omega}) and coincides with the first term of $D_j^{(2)}[\omega]$. Expansions (\ref{RTCL2admit-f5}) and (\ref{RTCL2admit-s3}) show that the admittance obtained using the relaxation TCL method, coincides with the admittance obtained using the TCLE method in the lowest Born approximation for the system-reservoir interaction \cite{S5,S6}, i.e.,
\begin{equation}
\chi^{\R\TCL}_{ij}(\omega)\ [(\ref{RTCL2admit})] = \chi^{\TCLE}_{ij}(\omega)\ [(\ref{TCLE2admit})] + \HOT \ (\mathrm{higher}\ \mathrm{order}\ \mathrm{terms}). \qquad \ \label{RTCL2-TCLE2}
\end{equation}

We next examine the relation between the admittance (\ref{RTCL2admit}) obtained using the relaxation TCL method and the admittance (\ref{RTC2admit}) obtained using the relaxation TC method [or the admittance (\ref{TCE2admit}) obtained using the TCE method]. The collision operator $\bar{C}^{(2)}[\omega]$ given by (\ref{bar-C2-omega}) can be rewritten as
\begin{equation}
\bar{C}^{(2)}[\omega] = - \int_0^\infty d \tau \, \tr_\R \, L_\SR \exp(- \, i \, L_0 \, \tau) \, L_\SR \, \rho_\R \exp(i \, \omega \, \tau) = C^{(2)} + \Delta C^{(2)}[\omega], \qquad \label{C2-bar-omega}
\end{equation}
with $C^{(2)}$ given by (\ref{C2-inf}), where $\Delta C^{(2)}[\omega]$ is defined by
\begin{equation}
\Delta C^{(2)}[\omega] = - \int_0^\infty d \tau \, \tr_\R \, L_\SR \exp(- \, i \, L_0 \, \tau) \, L_\SR \, \big\{ \exp(i \, \omega \, \tau) - \exp(i \, L_0 \, \tau) \big\} \, \rho_\R \,. \qquad 
\label{DeltaC2-omega}
\end{equation}
Then, the first term of the admittance $\chi^{\R\TCL}_{ij}(\omega)$ given by (\ref{RTCL2admit}) can be rewritten as
\begin{align}
& \, (\mathrm{first}\ \mathrm{term}\ \mathrm{of}\ \chi^{\R\TCL}_{ij}(\omega)\ [(\ref{RTCL2admit})]), \nonumber \\
= & \, \frac{i}{\hbar}\int_0^\infty dt \, \tr \, A_i \exp_\gets \! \Big\{ i \, (\omega - L_\SS)\, t + \bar{C}^{(2)}[\omega]\, t + \int_0^t d\tau \, \big(\, C^{(2)}(\tau) - \bar{C}^{(2)}[\omega] \,\big) \Big\} \big[ A_j \,,\, \rho_\SS + \rho_0^{(2)} \,\big], \nonumber \\
= & \, \frac{i}{\hbar}\int_0^\infty dt \, \tr \, A_i \, \bar{U}(\omega , t) \exp_\gets \! \Big\{ \int_0^t d\tau \, \bar{U}^{-1}(\omega , \tau) \, \big(\, C^{(2)}(\tau) - C^{(2)} - \Delta C^{(2)}[\omega] \,\big) \, \bar{U}(\omega , \tau) \Big\} \big[ A_j \,,\, \rho_\SS + \rho_0^{(2)} \,\big], \ \label{RTC2admit-f1}
\end{align}
with $\bar{U}(\omega , t)$ defined by
\begin{equation}
\bar{U}(\omega , t) = \exp\big\{ i \, (\omega - L_\SS) \, t + \bar{C}^{(2)}[\omega] \, t \big\}, \qquad \quad
\label{Ubar-omega-t}
\end{equation}
and can be expanded in powers of the system-reservoir interaction as
\begin{align}
(\mathrm{first}\ \mathrm{term}\ \mathrm{of}\ \chi^{\R\TCL}_{ij}(\omega)\ [(\ref{RTCL2admit})]) & = \frac{i}{\hbar} \int_0^\infty dt \, \tr \, A_i \, \bar{U}(\omega , t) \, \big[ A_j \,,\, \rho_\SS + \rho_0^{(2)} \,\big] \nonumber \\
    & \,\,\ + \frac{i}{\hbar} \int_0^\infty dt \int_0^t d\tau \, \tr \, A_i \, \bar{U}(\omega , t - \tau) \, \Big\{ \int_\tau^\infty ds \, \tr_\R \, L_\SR \exp(- \, i \, L_0 \, s) \, L_\SR \exp(i \, L_0 \, s) \, \rho_\R \nonumber \\
    & \qquad \qquad \qquad \qquad \ + \int_0^\infty ds \, \tr_\R \, L_\SR \exp(- \, i \, L_0 \, s) \, L_\SR \, \big\{\! \exp(i \, \omega \, s) - \exp(i \, L_0 \, s) \big\}\, \rho_\R \Big\} \ \nonumber \\
    & \qquad \qquad \qquad \qquad \qquad \qquad \quad \ { } \times \bar{U}(\omega , \tau) \, \big[ A_j \,,\, \rho_\SS + \rho_0^{(2)} \,\big] + \, \cdots \,, 
\label{RTC2admit-f2}
\end{align}
where (\ref{C2-t}), (\ref{C2-inf}) and (\ref{bar-C2-omega}) have been substituted into $C^{(2)}(t)$, $C^{(2)}$ and $\bar{C}^{(2)}[\omega]$, respectively. The first term of (\ref{RTC2admit-f2}) can be integrated to give
\begin{equation}
\frac{i}{\hbar} \int_0^\infty dt \, \tr \, A_i \, \bar{U}(\omega , t) \, \big[ A_j \,,\, \rho_\SS + \rho_0^{(2)} \,\big] = \frac{i}{\hbar} \, \tr \, A_i \, \big(\, i \, (L_\SS - \omega) - \bar{C}^{(2)}[\omega] \,\big)^{-1} \, \big[ A_j \,,\, \rho_\SS + \rho_0^{(2)} \,\big]. \qquad
\label{RTC2admit-f3}
\end{equation}
The second term of (\ref{RTC2admit-f2}) becomes, by integrating in the same procedure as in (\ref{RTCL2admit-f4}), as follows 
\begin{align}
    & \, \frac{i}{\hbar} \int_0^\infty d\tau \, \tr \, A_i \, \big(\, i \, (L_\SS - \omega) - \bar{C}^{(2)}[\omega] \,\big)^{-1} \Big\{ \int_0^\tau ds \, \tr_\R \, L_\SR \exp(- \, i \, L_0 \, \tau) \, L_\SR \exp(i \, L_0 \, \tau) \, \bar{U}(\omega , s) \, \rho_\R \nonumber \\
    & \qquad \qquad \qquad \qquad \qquad \quad + \int_0^\infty ds \, \tr_\R \, L_\SR \exp(- \, i \, L_0 \, \tau) \, L_\SR \, \big\{\! \exp(i \, \omega \, \tau) - \exp(i \, L_0 \, \tau) \big\} \, \bar{U}(\omega , s) \, \rho_\R \Big\} \big[ A_j \,,\, \rho_\SS + \rho_0^{(2)} \,\big] \nonumber \\
    = & \, \frac{i}{\hbar} \int_0^\infty d\tau \, \tr \, A_i \, \big(\, i \, (L_\SS - \omega) - \bar{C}^{(2)}[\omega] \,\big)^{-1} \, \tr_\R \, L_\SR \exp(- \, i \, L_0 \, \tau) \, L_\SR \, \big\{ \exp(i \, L_0 \, \tau) \exp\big(\, i \, (\omega - L_\SS) \, \tau + \bar{C}^{(2)}[\omega] \, \tau \,\big) \nonumber \\
    & \qquad \qquad \qquad \qquad - \exp(i \, L_0 \, \tau) - \big(\! \exp(i \, \omega \, \tau) - \exp(i \, L_0 \, \tau) \big) \big\} \big(\, i \, (\omega - L_\SS) + \bar{C}^{(2)}[\omega] \,\big)^{-1} \rho_\R \, \big[ A_j \,,\, \rho_\SS + \rho_0^{(2)} \,\big],
\label{RTC2admit-f4}
\end{align}
which is the higher order term in powers of the system-reservoir interaction. Thus, the first term of the admittance $\chi^{\R\TCL}_{ij}(\omega)$ [(\ref{RTCL2admit})] can be expressed as
\begin{equation}
(\mathrm{first}\ \mathrm{term}\ \mathrm{of}\ \chi^{\R\TCL}_{ij}(\omega)\ [(\ref{RTCL2admit})]) = \tr \, A_i \, \big( i \, (L_\SS - \omega) - \bar{C}^{(2)}[\omega] \,\big)^{-1} \, (i / \hbar) \, \big[ A_j \,,\, \rho_\SS + \rho_0^{(2)} \,\big] + \HOT. \quad
\label{RTC2admit-f5}
\end{equation}
The second term of the admittance $\chi^{\R\TCL}_{ij}(\omega)$ given by (\ref{RTCL2admit}) can be rewritten by substituting (\ref{I2-t}), as 
\begin{align}
& \, (\mathrm{second}\ \mathrm{term}\ \mathrm{of}\ \chi^{\R\TCL}_{ij}(\omega)\ [(\ref{RTCL2admit})]), \nonumber \\
= & \, \frac{- 1}{\hbar} \int_0^\infty dt \int_0^t d\tau \int_0^\beta d\beta' \, \tr \, A_i \exp_\gets \! \Big\{ \big(\, i \, (\omega - L_\SS) + \bar{C}^{(2)}[\omega] \,\big) (t - \tau) + \int_\tau^t ds \, \big(C^{(2)}(s) - \bar{C}^{(2)}[\omega] \big) \Big\} \quad \nonumber \\
    & \qquad \qquad \qquad \qquad \qquad \qquad \times \tr_\R \, L_\SR \exp(- \, i \, L_0 \, \tau) \big[ A_j \,,\, \rho_\SS \, \rho_\R \, \H_\SR(- \, i \, \hbar \, \beta') \, \big] \exp(i \, \omega \, \tau),
\label{RTC2admit-s1}
\end{align}
and can be expanded in powers of the system-reservoir interaction as
\begin{align}
(\mathrm{second}\ \mathrm{term}\ \mathrm{of}\ \chi^{\R\TCL}_{ij}(\omega)\ [\ref{RTCL2admit}]) & = - \, \frac{1}{\hbar} \int_0^\infty d\tau \int_\tau^\infty dt \int_0^\beta d\beta' \, \tr \, A_i \exp \big\{ \big(\, i \, (\omega - L_\SS) + \bar{C}^{(2)}[\omega] \, \big) (t - \tau) \big\} \quad \nonumber \\
    & \qquad \qquad \qquad { } \times \tr_\R \, L_\SR \exp(- \, i \, L_0 \, \tau) \big[ A_j \,,\, \rho_\SS \, \rho_\R \, \H_\SR(- \, i \, \hbar \, \beta') \,\big] \exp(i \, \omega \, \tau) + \cdots \,,
\label{RTC2admit-s2}
\end{align}
which can be expressed by integrating with respect to $t$, as
\begin{equation}
(\mathrm{second}\ \mathrm{term}\ \mathrm{of}\ \chi^{\R\TCL}_{ij}(\omega)\ [(\ref{RTCL2admit})]) = \tr \, A_i \, \big(i \, (L_\SS - \omega) - \bar{C}^{(2)}[\omega] \big)^{-1} \, \bar{D}_j^{(2)}[\omega] + \HOT, \quad
\label{RTC2admit-s3}
\end{equation}
with $\bar{D}_j^{(2)}[\omega]$ given by (\ref{barDj2b-omega}). Expansions (\ref{RTC2admit-f5}) and (\ref{RTC2admit-s3}) show that the admittance obtained using the relaxation TCL method, coincides with the admittance obtained using the relaxation TC method [or the admittance obtained using the TCE method] in the lowest Born approximation for the system-reservoir interaction, i.e.,
\begin{equation}
\chi^{\R\TCL}_{ij}(\omega)\ [(\ref{RTCL2admit})] = \chi^{\R\TC}_{ij}(\omega)\ [(\ref{RTC2admit})]\ \big(\mathrm{or}\ \chi_{ij}^{\TCE}(\omega)\ [(\ref{TCE2admit})] \big) + \HOT \ (\mathrm{higher}\ \mathrm{order}\ \mathrm{terms}). \quad
\label{RTCL2-RTCE2}
\end{equation}

We also examine the relation between the admittance (\ref{RTC2admit}) obtained using the relaxation TC method [or the admittance (\ref{TCE2admit}) obtained using the TCE method] and the admittance (\ref{TCLE2admit}) obtained using the TCLE method. The operator $\big( i \, (L_\SS - \omega) - \bar{C}^{(2)}[\omega] \big)^{-1}$ can be rewritten by using $\Delta C^{(2)}[\omega]$ given by (\ref{DeltaC2-omega}), as 
\begin{align}
& \big(\, i \, (L_\SS - \omega) - \bar{C}^{(2)}[\omega] \,\big)^{-1} = \big\{ i \, (L_\SS - \omega) - C^{(2)} - \Delta C^{(2)}[\omega] \big\}^{-1}, \nonumber \\
    = & \, \big\{ ( i \, (L_\SS - \omega) - C^{(2)} ) \, \{1 + (i \, (L_\SS - \omega) - C^{(2)})^{-1} \, (- \, \Delta C^{(2)}[\omega]) \} \big\}^{-1}, \nonumber \\
    = & \, \big\{1 + (i \, (L_\SS - \omega) - C^{(2)})^{-1} \, (- \, \Delta C^{(2)}[\omega]) \big\}^{-1}\, \big(\, i \, (L_\SS - \omega) - C^{(2)} \,\big)^{-1}, \qquad
\label{TCLE2admit-1}
\end{align}
and can be expanded in powers of the system-reservoir interaction as
\begin{equation}
\big(\, i \, (L_\SS - \omega) - \bar{C}^{(2)}[\omega] \,\big)^{-1} = \big(\, i \, (L_\SS - \omega) - C^{(2)} \,\big)^{-1} \, \big\{ 1 + \Delta C^{(2)}[\omega] \, \big(\, i \, (L_\SS - \omega) - C^{(2)} \,\big)^{-1} \big\} + \, \cdots \, . \quad
\label{TCLE2admit-2}
\end{equation}
Substituting the above expansion into $\chi^{\R\TC}_{ij}(\omega)$ given by (\ref{RTC2admit}) [or $\chi^{\TCE}_{ij}(\omega)$ given by (\ref{TCE2admit})], we have
\begin{align}
\chi^{\R\TC}_{ij}(\omega) = & \, \tr \, A_i \, \big(i \, (L_\SS - \omega) - C^{(2)} \big)^{-1}\, \big\{ (i / \hbar)\, \big[ A_j \,,\, \rho_\SS + \rho_0^{(2)} \,\big] + \bar{D}^{(2)}_j[\omega] \big\} \nonumber \\
    & + \tr \, A_i \, \big( i \, (L_\SS - \omega) - C^{(2)} \big)^{-1} \, \Delta C^{(2)}[\omega] \, \big( i \, (L_\SS - \omega) - C^{(2)} \big)^{-1} \, (i / \hbar) \, \big[ A_j \,,\, \rho_\SS + \rho_0^{(2)} \,\big] + \HOT.
\label{TCLE2admit-3}
\end{align}
where $\bar{D}_j^{(2)}[\omega]$ is given by (\ref{barDj2b-omega}) or (\ref{barDj2t-omega}), and is equal to the first term of the interference term $D_j^{(2)}[\omega]$ given by (\ref{Dj2-omega}). Substituting (\ref{DeltaC2-omega}) into the second term of the above expansion and considering that
\begin{align}
& \, \Delta C^{(2)}[\omega] \, \big( i \, (L_\SS - \omega) - C^{(2)} \big)^{-1} \, (i/ \hbar)\, \big[ A_j \,,\, \rho_\SS + \rho_0^{(2)} \,\big] \,, \nonumber \\
    = & \, - \int_0^\infty d \tau \, \tr_\R \, L_\SR \exp(- \, i \, L_0 \, \tau) \, L_\SR \, \big\{\! \exp(i \, \omega \, \tau) - \exp(i \, L_0 \, \tau) \big\} \, \rho_\R \, \big(i \, (L_\SS - \omega) - C^{(2)} \big)^{-1} \, \frac{i}{\hbar} \, \big[ A_j \,,\, \rho_\SS + \rho_0^{(2)} \,\big] \,, \ \nonumber \\
    = & \, \frac{i}{\hbar} \int_0^\infty d \tau \int_0^\tau d s \, \tr_\R \, L_\SR \exp(- \, i \, L_0 \, \tau) \, L_\SR \exp\{i \, L_0 \, (\tau - s)\} \, \rho_\R \, \big[ A_j \,,\, \rho_\SS \big] \exp(i \, \omega \, s) + \HOT, \nonumber \\
    = & \, (\mathrm{second}\ \mathrm{term}\ \mathrm{of}\ D_j^{(2)}[\omega]) + \HOT,
\label{TCLE2admit-4}
\end{align}
we find that the admittance obtained using the relaxation TC method [or the admittance obtained using the TCE method], coincides with the admittance obtained using the TCLE method in the lowest Born approximation for the system-reservoir interaction, i.e.,
\begin{equation}
\chi^{\R\TC}_{ij}(\omega)\ [(\ref{RTC2admit})] = \chi^{\TCE}_{ij}(\omega)\ [(\ref{TCE2admit})] = \chi^{\TCLE}_{ij}(\omega)\ [(\ref{TCLE2admit})] + \HOT \ (\mathrm{higher}\ \mathrm{order}\ \mathrm{terms}). \qquad
\label{RTCE2-TCLE2}
\end{equation}

According to the relations (\ref{RTC-TCE}), (\ref{RTCL2-TCLE2}), (\ref{RTCL2-RTCE2}) and (\ref{RTCE2-TCLE2}), the admittance obtained using the relaxation TC method [or the TCE method], the admittance obtained using the relaxation TCL method and the admittance obtained using the TCLE method, coincide with each other in the lowest Born approximation for the system-reservoir interaction $\H_\SR$, i.e.,
\begin{equation}
\chi^{\R\TC}_{ij}(\omega) = \chi^{\TCE}_{ij}(\omega) = \chi^{\R\TCL}_{ij}(\omega) = \chi^{\TCLE}_{ij}(\omega), \qquad \qquad \qquad \qquad \big( \mathrm{up}\ \mathrm{to}\ O(\H_\SR^2) \big) \qquad
\label{Relations}
\end{equation}
though the exact admittances (\ref{RTCadmit}) [or (\ref{TCEadmit})], (\ref{RTCLadmit}) and (\ref{TCLEadmit}), which are obtained using the relaxation TC method [or the TCE method], the relaxation TCL method and the TCLE method, respectively, take the forms different from each other. The admittances that are obtained using the relaxation TC method [or the TCE method], the relaxation TCL method and the TCLE method, respectively, in the lowest Born approximation for the system-reservoir interaction, have the same second order term and the higher order terms different from each other.
%
\section{Higher order expansions}

In this section, we derive the formulae necessary for the higher order expansions in powers of the system-reservoir interaction $\H_\SR$, and give the forms of the admittances in the $n$-th order approximation for $\H_\SR$.
%
\subsection{Expansions of the collision operators $C(t)$, $C$ and $\bar{C}[\omega]$}

The collision operator $C(t)$ given by (\ref{C-t}) can be rewritten as
\begin{equation}
C(t) = - \, i \, \big\langle L_\SR \, \Sigma(t)\, \{1 - \Sigma(t) \}^{-1} \big\rangle _\R = - \, i \sum_{n=1}^\infty \big\langle L_\SR \, \Sigma(t)^n \big\rangle _\R \,, \qquad 
\label{C-t-Sigma}
\end{equation}
with $\Sigma(t)$ defined by (\ref{Sigma-t}). Defining the unperturbed evolution operator $U_0(t)$ as
\begin{equation}
U_0(t) = \exp(- \, i \, L_0 \, t) = \exp\{- \, i \, (L_\SS + L_\R) \, t \}, \qquad \qquad
\label{U0-t}
\end{equation}
we have 
\begin{equation}
L_\SR(- \, t) = U_0(t) \, L_\SR \, U_0^{-1}(t), \qquad \qquad \qquad \H_\SR(t) = \big( U_0^{-1}(t) \, \H_\SR \big) = \exp(i \, \H_0 \, t / \hbar) \, \H_\SR \, \exp(- \, i \, \H_0 \, t / \hbar),
\end{equation}
and
\begin{align}
& \exp(i \, L \, t) = U_0^{-1}(t) \exp_\gets \! \Big\{\, i \int_0^t d \tau \, L_\SR(- \, \tau) \Big\}, 
\label{exp-iLt} \\
& \exp(- \, i \, \Q \, L \, \Q \, t)\, \Q = \exp_\to \! \Big\{\! - i \int_0^t d \tau \, \Q \, L_\SR(- \, \tau) \, \Q \Big\}\, U_0(t)\, \Q \,.\qquad \qquad
\label{exp-iQLQt}
\end{align}
Then, $\Sigma(t)$ defined by (\ref{Sigma-t}) can be rewritten as
\begin{equation}
\Sigma(t) = - \, i \int_0^t d\tau \exp_\to \!\Big\{\! - i \int_0^\tau d s_1 \, \Q \, L_\SR(- \, s_1) \, \Q \Big\}\, \Q \, L_\SR(- \, \tau) \, \P \exp_\gets \!\Big\{\, i \int_0^\tau d s_2 \, L_\SR(- \, s_2) \Big\}, \quad 
\end{equation}
which can be expanded in powers of the system-reservoir interaction $\H_\SR$ as
\begin{equation}
\Sigma(t) = \Sigma^{(1)}(t) + \Sigma^{(2)}(t) + \Sigma^{(3)}(t) + \cdots \,, \qquad \quad
\end{equation}
where $\Sigma^{(n)}(t)$ is the $n$-th order part in powers of $\H_\SR$. The first-order, second-order and third-order parts of $\Sigma(t)$ take the forms, respectively, 
\begin{align}
\Sigma^{(1)}(t) & = - \, i \int_0^t d\tau \, \Q \, L_\SR(- \, \tau) \, \P , \\
\Sigma^{(2)}(t) & = (- \, i)^2 \int_0^t d\tau_1 \int_0^{\tau_1} d\tau_2 \, \big\{ \Q \, L_\SR(- \, \tau_2) \, \Q \, L_\SR(- \, \tau_1) \, \P - \Q \, L_\SR(- \, \tau_1) \, \P \, L_\SR(- \, \tau_2) \big\}, \\
\Sigma^{(3)}(t) & = (- \, i)^3 \! \int_0^t d\tau_1 \int_0^{\tau_1} \! d\tau_2 \int_0^{\tau_2} \! d\tau_3 \, \big\{ \Q \, L_\SR(- \, \tau_3) \, \Q \, L_\SR(- \, \tau_2) \, \Q \, L_\SR(- \, \tau_1) \, \P + \Q \, L_\SR(- \, \tau_1) \, \P \, L_\SR(- \, \tau_2) \, L_\SR(- \, \tau_3) \nonumber \\
    & \qquad \qquad \qquad \qquad \qquad \, { } - \Q \, L_\SR(- \, \tau_2) \, \Q \, L_\SR(- \, \tau_1) \, \P \, L_\SR(- \, \tau_3) - \Q \, L_\SR(- \, \tau_3) \, \Q \, L_\SR(- \, \tau_1) \, \P \, L_\SR(- \, \tau_2) \big\}.
\end{align}
In the derivation of $\Sigma^{(3)}(t)$, the following integral transformation has been used
\begin{equation}
\int_0^{\tau_1} d\tau_2 \int_0^{\tau_1} d\tau_3 = \int_0^{\tau_1} d\tau_2 \int_0^{\tau_2} d\tau_3 + \int_0^{\tau_1} d\tau_2 \int_{\tau_2}^{\tau_1} d\tau_3 = \int_0^{\tau_1} d\tau_2 \int_0^{\tau_2} d\tau_3 + \int_0^{\tau_1} d\tau_3 \int_0^{\tau_3} d\tau_2 \,. \quad \label{integral-trans-1}
\end{equation}
Expanding the operator $C(t)$ in powers of the system-reservoir interaction $\H_\SR$ as
\begin{equation}
C(t) = C^{(2)}(t) + C^{(3)}(t) + C^{(4)}(t) + \cdots \,, \qquad \quad
\end{equation}
where $C^{(n)}(t)$ is the $n$-th order part in powers of $\H_\SR$, the second-order part of $C(t)$ is given by (\ref{C2-t}) and the third-order and fourth-order parts can be obtained as
\begin{align}
C^{(3)}(t) & = (- \, i)^3 \int_0^t d\tau_1 \int_0^{\tau_1} d\tau_2 \, \big\langle L_\SR \, L_\SR(- \, \tau_2) \, L_\SR(- \, \tau_1) \big\rangle_\R \,, 
\label{C-t-3} \\
C^{(4)}(t) & = (- \, i)^4 \! \int_0^t d\tau_1 \int_0^{\tau_1} \! d\tau_2 \int_0^{\tau_2} \! d\tau_3 \, \big\{ \big\langle L_\SR \, L_\SR(- \, \tau_3) L_\SR(- \, \tau_2) L_\SR(- \, \tau_1) \big\rangle_\R - \big\langle L_\SR \, L_\SR(- \, \tau_3) \big\rangle_\R \big\langle L_\SR(- \, \tau_2) L_\SR(- \, \tau_1) \big\rangle_\R \nonumber \\
    & \qquad \qquad \qquad \qquad \qquad \quad - \big\langle L_\SR \, L_\SR(- \, \tau_2) \big\rangle_\R \big\langle L_\SR(- \, \tau_3) L_\SR(- \, \tau_1) \big\rangle_\R - \big\langle L_\SR \, L_\SR(- \, \tau_1) \big\rangle_\R \big\langle L_\SR(- \, \tau_3) L_\SR(- \, \tau_2) \big\rangle_\R \, \big\},
\label{C-t-4}
\end{align}
where the following integral transformation has been performed
\begin{align}
    & \int_0^t d\tau_1 \int_0^{\tau_1} d\tau_2 \int_0^t d\tau = \int_0^t d\tau_1 \int_0^{\tau_1} d\tau_2 \, \Big\{ \int_0^{\tau_2} d\tau + \int_{\tau_2}^{\tau_1} d\tau + \int_{\tau_1}^t d\tau \Big\}, \nonumber \\
    = & \int_0^t d\tau_1 \int_0^{\tau_1} d\tau_2 \int_0^{\tau_2} d\tau + \int_0^t d\tau_1 \int_0^{\tau_1} d\tau \int_0^\tau d\tau_2 + \int_0^t d\tau \int_0^\tau d\tau_1 \int_0^{\tau_1} d\tau_2 \,. \qquad \label{integral-trans-2}
\end{align}
In the derivations of (\ref{C-t-3}) and (\ref{C-t-4}), we have considered that $ \langle \H_\SR(- \, t) \rangle_\R $\,=\,0 by virtue of the renormalization (\ref{renormalize}). The $n$-th order part $C^{(n)}(t)$ in the expansion of the collision operator $C(t)$ has the following structure\,:
\begin{equation}
C^{(n)}(t) = (- \, i)^n \int_0^t d\tau_1 \int_0^{\tau_1} d\tau_2 \int_0^{\tau_2} d\tau_3 \cdots \int_0^{\tau_{n-2}} d\tau_{n-1} \big\langle L_\SR(0) \, L_\SR(- \, \tau_{n-1}) \, L_\SR(- \, \tau_{n-2}) \cdots L_\SR(- \, \tau_2) \, L_\SR(- \, \tau_1) \big\rangle_\oc \,,
\label{C-t-n}
\end{equation}
for $n$\,$ \ge $\,2, where $\langle \cdots \rangle_\oc $ denotes the ``ordered cumulants" \cite{van-Kampen,Chaturvedi-Shibata,Shibata-Arimitsu}.

The expansion formulae of the collision operator $C$ can be obtained by expanding $C$ [=\,$ C(\infty)$] in powers of $\H_\SR$ as
\begin{equation}
C = C(\infty) = C^{(2)} + C^{(3)} + C^{(4)} + \cdots \,, \qquad \qquad \qquad \quad [\, C^{(n)} = C^{(n)}(\infty) \,]. \quad
\end{equation}

The collision operator $\bar{C}[\omega]$ given by (\ref{bar-C-omega}) can be rewritten by considering that $ \langle \H_\SR \rangle_\R $\,=\,0, as
\begin{equation}
\bar{C}[\omega] = - \int_0^\infty dt \, \big\langle L_\SR \exp_\to \!\big\{\! - i \int_0^t d \tau \, \Q \, L_\SR(- \, \tau) \, \Q \big\}\, U_0(t) \, L_\SR \big\rangle_\R \exp(i \, \omega \, t). \qquad
\label{C-bar-omega}
\end{equation}
Expanding the collision operator $\bar{C}[\omega]$ in powers of the system-reservoir interaction $\H_\SR$ as
\begin{equation}
\bar{C}[\omega] = \bar{C}^{(2)}[\omega] + \bar{C}^{(3)}[\omega] + \bar{C}^{(4)}[\omega] + \cdots \,, \qquad \quad
\end{equation}
where $\bar{C}^{(n)}[\omega]$ is the $n$-th order part in powers of $\H_\SR$, the second-order part of $\bar{C}[\omega]$ is given by (\ref{bar-C2-omega}) and the third-order and fourth-order parts can be obtained as
\begin{align}
\bar{C}^{(3)}[\omega] & = (- \, i)^3 \int_0^\infty \!\! d\tau_1 \! \int_0^{\tau_1} \!\! d \tau_2 \, \big\langle L_\SR \, L_\SR(- \, \tau_2) \, L_\SR(- \, \tau_1) \big\rangle_\R \exp\{ i \, (\omega - L_\SS) \, \tau_1 \}\,, 
\label{C3-bar-omega} \\
\bar{C}^{(4)}[\omega] & = (- \, i)^4 \int_0^\infty d\tau_1 \int_0^{\tau_1} d \tau_2 \int_0^{\tau_2} d \tau_3 \, \big\{ \big\langle L_\SR \, L_\SR(- \, \tau_3) \, L_\SR(- \, \tau_2) \, L_\SR(- \, \tau_1) \big\rangle_\R \nonumber \\
    & \qquad \qquad \qquad \qquad \qquad \qquad \qquad \quad { } - \big\langle L_\SR \, L_\SR(- \, \tau_3) \big\rangle_\R \big\langle L_\SR(- \, \tau_2) \, L_\SR(- \, \tau_1) \big\rangle_\R \, \big\} \exp\{ i \, (\omega - L_\SS) \, \tau_1 \}\,.
\label{C4-bar-omega}
\end{align}
The $n$-th order part $\bar{C}^{(n)}[\omega]$ in the expansion of the collision operator $\bar{C}[\omega]$ has the following structure\,:
\begin{align}
\bar{C}^{(n)}[\omega] & = (- \, i)^n \int_0^t d\tau_1 \int_0^{\tau_1} d\tau_2 \int_0^{\tau_2} d\tau_3 \cdots \int_0^{\tau_{n-2}} d\tau_{n-1} \big\langle L_\SR(0) \, L_\SR(- \, \tau_{n-1}) \, L_\SR(- \, \tau_{n-2}) \cdots L_\SR(- \, \tau_2) \, L_\SR(- \, \tau_1) \big\rangle_\pc \nonumber \\
    & \qquad \qquad \qquad \qquad \qquad \qquad \qquad \qquad \qquad \qquad \quad { } \times \exp\{ i \, (\omega - L_\SS) \, \tau_1 \} \,, \label{Cn-bar-omega}
\end{align}
for $n$\,$ \ge $\,2, where $\langle \cdots \rangle_\pc $ denotes the ``partial cumulants" \cite{van-Kampen,Chaturvedi-Shibata,Shibata-Arimitsu}.
%
\subsection{Expansions of the inhomogeneous terms and interference terms}

We first expand the thermal equilibrium state $\rho_\TE$ given by (\ref{rhoTE}) in powers of the system-reservoir interaction $\H_\SR$. Writing the expansion of $ \exp(- \, \beta \, \H) $\,=\,$ \exp\{- \, \beta \, (\H_0 $\,+\,$ \H_\SR)\}$ in powers of $ \H_\SR $ as
\begin{equation}
\exp(- \, \beta \, \H) = \exp(- \, \beta \, \H_0) \exp_\gets \!\Big\{\! - \int_0^\beta d\beta\,' \, \exp(\beta\,' \, \H_0) \, \H_\SR \, \exp(- \, \beta\,' \, \H_0) \Big\} = \exp(- \, \beta \, \H_0) \Big\{ 1 + \sum_{n = 1}^\infty \Xi_n \Big\}, 
\label{exp-betaH}
\end{equation}
with $ \Xi_n $ defined by
\begin{equation}
\Xi_n = (-)^n \int_0^\beta d\beta_1 \int_0^{\beta_1} d\beta_2 \,\cdots \, \int_0^{\beta_{n-1}} d\beta_n \, \H_\SR(- \, i \, \hbar \, \beta_1) \, \H_\SR(- \, i \, \hbar \, \beta_2) \, \cdots \, \H_\SR(- \, i \, \hbar \, \beta_n), \quad
\label{Xi-n}
\end{equation}
we have
\begin{equation}
\Tr \exp(- \, \beta \, \H) = \{ \, \Tr \exp(- \, \beta \, \H_0) \} \! \cdot \! (1 + \xi) , \qquad \qquad \qquad \qquad \xi = \xi_2 + \xi_3 + \xi_4 + \cdots \,, \quad
\label{Tr-exp-betaH}
\end{equation}
where we have defined
\begin{align}
\xi_n & = \Tr \, \{ \exp(- \, \beta \, \H_0)\, \Xi_n \} / (\, \Tr \exp(- \, \beta \, \H_0)) = \Tr \, \rho_\SS \, \rho_\R \, \Xi_n \,, \nonumber \\
    & = (-)^n \int_0^\beta d\beta_1 \int_0^{\beta_1} d\beta_2 \, \cdots \int_0^{\beta_{n-1}} d\beta_n \, \big\langle \, \H_\SR(- \, i \, \hbar \, \beta_1) \, \H_\SR(- \, i \, \hbar \, \beta_2) \, \cdots \, \H_\SR(- \, i \, \hbar \, \beta_n) \big\rangle_{\SS\R}\,, \quad \label{smallXi-n}
\end{align}
with the notation $ \langle \cdots \rangle_{\SS\R}$ defined by $ \langle \cdots \rangle_{\SS\R}$\,=\,$ \Tr \, \rho_\SS \, \rho_\R \cdots $, where $\rho_\SS$ and $\rho_\R$ are given by (\ref{rhoS}) and (\ref{rhoR}), respectively. Here, we have considered that $ \langle \H_\SR(- i \hbar \beta) \rangle_\R $\,=\,0. Then, $\rho_\TE$ can be expanded as
\begin{align}
\rho_\TE & = \exp(- \, \beta \, \H_0)\, (1 + \Xi_1 + \Xi_2 + \Xi_3 + \cdots )\, / \, \{ (\, \Tr \exp(- \, \beta \, \H_0)) \! \cdot \! (1 + \xi) \}, \qquad \nonumber \\
    & = \rho_\SS \, \rho_\R \, (1 + \Xi_1 + \Xi_2 + \Xi_3 + \cdots )\, (1 - \xi + \xi^2 - \xi^3 + \cdots ). 
\end{align}
Expanding $\rho_\TE$ in powers of the system-reservoir interaction $\H_\SR$ as
\begin{equation}
\rho_\TE = \rho_\TE^{(0)} + \rho_\TE^{(1)} + \rho_\TE^{(2)} + \rho_\TE^{(3)} + \cdots \,, \qquad \quad
\label{rhoTE-expand}
\end{equation}
where $\rho_\TE^{(n)}$ is the $n$-th order part in powers of $\H_\SR$, we have
\begin{align}
\rho_\TE^{(0)} = & \, \rho_\SS \, \rho_\R \,, \qquad \qquad \qquad \qquad \qquad \rho_\TE^{(1)} = \rho_\SS \, \rho_\R \, \Xi_1 = - \, \rho_\SS \, \rho_\R \int_0^\beta d\beta_1 \, \H_\SR(- \, i \, \hbar \, \beta_1),
\label{rhoTE-01} \\
\rho_\TE^{(2)} = & \, \rho_\SS \, \rho_\R \, (\Xi_2 - \xi_2) = \rho_\SS \, \rho_\R \int_0^\beta d\beta_1 \int_0^{\beta_1} d\beta_2 \, \big\{ \, \H_\SR(- \, i \, \hbar \, \beta_1) \, \H_\SR(- \, i \, \hbar \, \beta_2) - \big\langle \H_\SR(- \, i \, \hbar \, \beta_1) \, \H_\SR(- \, i \, \hbar \, \beta_2) \big\rangle_{\SS\R} \, \big\},
\label{rhoTE-2} \\
\rho_\TE^{(3)} = & \, \rho_\SS \, \rho_\R \, (\Xi_3 - \xi_3 - \Xi_1 \, \xi_2), \nonumber \\
    = & - \rho_\SS \, \rho_\R \int_0^\beta d\beta_1 \int_0^{\beta_1} d\beta_2 \int_0^{\beta_2} d\beta_3 \, \big\{ \H_\SR(- \, i \, \hbar \, \beta_1) \, \H_\SR(- \, i \, \hbar \, \beta_2) \, \H_\SR(- \, i \, \hbar \, \beta_3) \nonumber \\
      & \qquad \qquad \qquad \qquad \qquad \qquad \qquad \ - \big\langle \H_\SR(- \, i \, \hbar \, \beta_1) \, \H_\SR(- \, i \, \hbar \, \beta_2) \, \H_\SR(- \, i \, \hbar \, \beta_3) \big\rangle_{\SS\R} \, \big\} \nonumber \\
      & + \rho_\SS \, \rho_\R \int_0^\beta d\beta\,' \, \H_\SR(- \, i \, \hbar \, \beta\,' ) \int_0^\beta d\beta_1 \int_0^{\beta_1} d\beta_2 \, \big\langle \H_\SR(- \, i \, \hbar \, \beta_1) \, \H_\SR(- \, i \, \hbar \, \beta_2) \big\rangle_{\SS\R}\,,
\label{rhoTE-3} \\
\rho_\TE^{(4)} = & \, \rho_\SS \, \rho_\R \, \{\, \Xi_4 - \xi_4 - (\Xi_2 - \xi_2) \, \xi_2 - \Xi_1 \, \xi_3 \}, \nonumber \\
    = & \, \rho_\SS \, \rho_\R \int_0^\beta d\beta_1 \int_0^{\beta_1} d\beta_2 \int_0^{\beta_2} d\beta_3 \int_0^{\beta_3} d\beta_4 \, \big\{ \H_\SR(- \, i \, \hbar \, \beta_1) \, \H_\SR(- \, i \, \hbar \, \beta_2) \, \H_\SR(- \, i \, \hbar \, \beta_3) \, \H_\SR(- \, i \, \hbar \, \beta_4) \nonumber \\
      & \qquad \qquad \qquad \qquad \qquad \qquad \qquad \qquad \quad - \big\langle \H_\SR(- \, i \, \hbar \, \beta_1) \, \H_\SR(- \, i \, \hbar \, \beta_2) \, \H_\SR(- \, i \, \hbar \, \beta_3) \, \H_\SR(- \, i \, \hbar \, \beta_4) \big\rangle_{\SS\R} \big\} \nonumber \\
    & - \rho_\SS \, \rho_\R \int_0^\beta d\beta_1 \int_0^{\beta_1} d\beta_2 \big\{ \H_\SR(- \, i \, \hbar \, \beta_1) \, \H_\SR(- \, i \, \hbar \, \beta_2) - \big\langle \H_\SR(- \, i \, \hbar \, \beta_1) \, \H_\SR(- \, i \, \hbar \, \beta_2) \big\rangle_{\SS\R} \big\} \nonumber \\
      & \qquad \qquad \qquad \quad { } \times \int_0^\beta d\beta\,' \int_0^{\beta\,'} d\beta\,'' \, \big\langle \H_\SR(- \, i \, \hbar \, \beta\,') \, \H_\SR(- \, i \, \hbar \, \beta\,'') \big\rangle_{\SS\R} \nonumber \\
    & - \rho_\SS \, \rho_\R \int_0^\beta d\beta\,' \, \H_\SR(- \, i \, \hbar \, \beta\,' ) \int_0^\beta d\beta_1 \int_0^{\beta_1} d\beta_2 \int_0^{\beta_2} d\beta_3 \, \big\langle \H_\SR(- \, i \, \hbar \, \beta_1) \, \H_\SR(- \, i \, \hbar \, \beta_2) \, \H_\SR(- \, i \, \hbar \, \beta_3) \big\rangle_{\SS\R} \, .
\label{rhoTE-4}
\end{align}

The inhomogeneous terms $\bar{I}(t)$ and $I(t)$ which are given by (\ref{bar-I-t}) and (\ref{I-t}), can be rewritten using (\ref{exp-iQLQt}) as
\begin{align}
\bar{I}(t) & = - \, i \! \cdot \! \tr_\R \, L_\SR \exp_\to \!\Big\{\! - i \int_0^t d \tau \, \Q \, L_\SR(- \, \tau) \, \Q \Big\}\, U_0(t)\, \Q \, \big[ A \,,\, \rho_\TE \,\big], 
\label{I-bar-expand} \\
I(t) & = - \, i \! \cdot \! \tr_\R \, L_\SR \, \{1 - \Sigma(t)\}^{- 1} \exp_\to \!\Big\{\! - i \int_0^t d \tau \, \Q \, L_\SR(- \, \tau) \, \Q \Big\}\, U_0(t)\, \Q \, \big[ A \,,\, \rho_\TE \,\big], \qquad 
\label{I-expand}
\end{align}
which can be expanded in powers of the system-reservoir interaction $\H_\SR$, respectively, as
\begin{equation}
\bar{I}(t) = \bar{I}^{(2)}(t) + \bar{I}^{(3)}(t) + \bar{I}^{(4)}(t) + \cdots \,, \qquad \qquad \qquad \quad I(t) = I^{(2)}(t) + I^{(3)}(t) + I^{(4)}(t) + \cdots \,,\ 
\end{equation}
because $ \bar{I}^{(1)}(t) $\,=\,$ I^{(1)}(t) $\,=\,0 according to $ \Q \, \rho_\TE^{(0)} $\,=\,$ \Q \, \rho_\SS \, \rho_\R $\,=\,0, where $\bar{I}^{(n)}(t)$ and $I^{(n)}(t)$ are the $n$-th order parts in powers of $\H_\SR$. The second-order parts are given by (\ref{bar-I2-t}) and (\ref{I2-t}), and the third-order and fourth-order parts can be obtained as
\begin{align}
\bar{I}^{(3)}(t) = & \, I^{(3)}(t) = \int_0^t d \tau \int_0^\beta d\beta_1 \, \tr_\R \, L_\SR \, L_\SR(- \, \tau) \, U_0(t) \big[ A \,,\, \rho_\SS \, \rho_\R \, \H_\SR(- \, i \, \hbar \, \beta_1) \big] \nonumber \\
    & \qquad \quad \ - i \int_0^\beta d\beta_1 \int_0^{\beta_1} d\beta_2 \, \tr_\R \, L_\SR \, U_0(t) \big[ A \,,\, \rho_\SS \, \rho_\R \, \H_\SR(- \, i \, \hbar \, \beta_1) \, \H_\SR(- \, i \, \hbar \, \beta_2) \big], 
\label{I-bar-I-3} \\
\bar{I}^{(4)}(t) = & - i \int_0^t d\tau_1 \int_0^{\tau_1} d\tau_2 \int_0^\beta d\beta_1 \, \tr_\R \, L_\SR \, L_\SR(- \, \tau_2) \, \Q \, L_\SR(- \, \tau_1) \, U_0(t) \big[ A \,,\, \rho_\SS \, \rho_\R \, \H_\SR(- \, i \, \hbar \, \beta_1) \big] \nonumber \\
    & - \int_0^t d \tau \int_0^\beta d\beta_1 \int_0^{\beta_1} d\beta_2 \, \tr_\R \, L_\SR \, L_\SR(- \, \tau) \nonumber \\
    & \qquad \qquad \qquad \qquad \qquad \quad \times U_0(t) \big[ A \,, \rho_\SS \, \rho_\R \, \big\{ \H_\SR(- \, i \, \hbar \, \beta_1) \, \H_\SR(- \, i \, \hbar \, \beta_2) - \big\langle \H_\SR(- \, i \, \hbar \, \beta_1) \, \H_\SR(- \, i \, \hbar \, \beta_2) \big\rangle_{\R}  \big\} \big] \nonumber \\
    & + i \int_0^\beta d\beta_1 \int_0^{\beta_1} d\beta_2 \int_0^{\beta_2} d\beta_3 \, \tr_\R \, L_\SR \, U_0(t) \big[ A \,,\, \rho_\SS \, \rho_\R \, \H_\SR(- \, i \, \hbar \, \beta_1) \, \H_\SR(- \, i \, \hbar \, \beta_2) \, \H_\SR(- \, i \, \hbar \, \beta_3) \big] \nonumber \\
    & - i \int_0^\beta d\beta\,' \int_0^\beta d\beta_1 \int_0^{\beta_1} d\beta_2 \, \tr_\R \, L_\SR \, U_0(t) \big[ A \,,\, \rho_\SS \, \rho_\R \, \H_\SR(- \, i \, \hbar \, \beta\,' ) \big] \big\langle \H_\SR(- \, i \, \hbar \, \beta_1) \, \H_\SR(- \, i \, \hbar \, \beta_2) \big\rangle_{\SS\R} \,,
\label{I-bar-4}
\end{align}
\begin{equation}
I^{(4)}(t) = \bar{I}^{(4)}(t) + i \int_0^t d\tau_1 \int_0^{\tau_1} d\tau_2 \int_0^\beta d \beta_1 \, \tr_\R \, L_\SR \, L_\SR(- \, \tau_1) \, \P \, L_\SR(- \, \tau_2) \, U_0(t) \big[ A \,,\, \rho_\SS \, \rho_\R \, \H_\SR(- \, i \, \hbar \, \beta_1) \big], \quad 
\label{I-4}
\end{equation}
where we have considered that $ \langle \H_\SR(- \, t) \rangle_\R $\,=\,$ \langle \H_\SR(- i \hbar \beta) \rangle_\R $\,=\,$ \langle \H_\SR \rangle_\R $\,=\,0 and $ \Q \, \rho_\TE^{(1)} $\,=\,$ \rho_\TE^{(1)} $. The second term of (\ref{I-4}) comes from $ \Sigma(t) $\,=\,1\,$ - $\,$ \theta^{-1}(t) $ which produces by renormalizing the time-convolution and represents the memory effect. Thus, the fourth-order part $I^{(4)}(t)$ of $I(t)$ includes the memory effect.

The interference terms $\bar{D}[\omega]$ and $D[\omega]$ which are given by (\ref{bar-D-omega}) and (\ref{D-omega}), can be rewritten using (\ref{exp-iLt}) and (\ref{exp-iQLQt}) as
\begin{align}
\bar{D}[\omega] = & - \int_0^\infty d\tau \, \tr_\R \, L_\SR \exp_\to \! \Big\{\! - i \int_0^\tau d s \, \Q \, L_\SR(- \, s) \, \Q \Big\} \, U_0(\tau)\, L_\ed[\omega] \, \Q \, \rho_\TE \, e^{i \, \omega \, \tau}, \\
D[\omega] = & - \int_0^\infty d\tau \, \tr_\R \, L_\SR \Big( \sum_{n = 0}^\infty \Sigma^n \Big) \exp_\to \!\Big\{\! - i \int_0^\tau d s \, \Q \, L_\SR(- \, s) \, \Q \Big\} \, U_0(\tau)\, L_\ed[\omega] \, \Q \, \rho_\TE \, e^{i \, \omega \, \tau}, \nonumber \\
    & - i \int_0^\infty d\tau_1 \int_0^{\tau_1} d\tau_2 \, \tr_\R \, L_\SR \Big( \sum_{n = 0}^\infty \Sigma^n \Big) \exp_\to \! \Big\{\! - i \int_0^{\tau_1} d s_1 \, \Q \, L_\SR(- \, s_1) \, \Q \Big\} \, U_0(\tau_1) \, \Q \, L_\SR \nonumber \\
    & \qquad \qquad \qquad \qquad \quad \times \P \, U_0^{- 1}(\tau_1 - \tau_2) \exp_\gets \!\Big\{ i \int_0^{\tau_1 - \tau_2} d s_2 \, L_\SR(- \, s_2) \Big\}\, L_\ed[\omega] \, \rho_\TE \exp(i \, \omega \, \tau_2),
\end{align}
which can be expanded in powers of the system-reservoir interaction $\H_\SR$, respectively, as
\begin{equation}
\bar{D}[\omega] = \bar{D}^{(2)}[\omega] + \bar{D}^{(3)}[\omega] + \bar{D}^{(4)}[\omega] + \cdots \,, \qquad \qquad \qquad \quad D[\omega] = D^{(2)}[\omega] + D^{(3)}[\omega] + D^{(4)}[\omega] + \cdots \,, \ 
\end{equation}
because $ \bar{D}^{(1)}[\omega] $\,=\,$ D^{(1)}[\omega] $\,=\,0 according to $ \Q \, \rho_\TE^{(0)} $\,=\,$ \Q \, \rho_\SS \rho_\R $\,=\,0, where $\bar{D}^{(n)}[\omega]$ and $D^{(n)}[\omega]$ are the $n$-th order parts in powers of $\H_\SR$. Here, $\Sigma$ is given by (\ref{Sigma}) and can be expanded in powers of $\H_\SR$ as 
$ \Sigma$\,=\,$ \Sigma(\infty) $\,=\,$ \Sigma^{(1)} $\,+\,$ \Sigma^{(2)} $\,+\,$ \Sigma^{(3)} $\,+\,$ \cdots $ with $ \Sigma^{(n)} $\,=\,$ \Sigma^{(n)}(\infty) $. 
The second-order parts are given by (\ref{bar-D2b-omega}) [or (\ref{bar-D2t-omega})] and (\ref{D2-omega}), and the third-order and fourth-order parts can be obtained as
\begin{align}
\bar{D}^{(3)}[\omega] = & - \, i \int_0^\infty d\tau_1 \int_0^{\tau_1} d \tau_2 \int_0^\beta d\beta_1 \, \tr_\R \, L_\SR \, L_\SR(- \, \tau_2) \, U_0(\tau_1)\, L_\ed[\omega] \, \rho_\SS \, \rho_\R \, \H_\SR(- \, i \, \hbar \, \beta_1) \exp(i \, \omega \, \tau_1) \qquad \qquad \nonumber \\
    & - \int_0^\infty d\tau \int_0^\beta d\beta_1 \int_0^{\beta_1} d\beta_2 \, \tr_\R \, L_\SR \, U_0(\tau) \, L_\ed[\omega] \, \rho_\SS \, \rho_\R \, \H_\SR(- \, i \, \hbar \, \beta_1) \, \H_\SR(- \, i \, \hbar \, \beta_2) \exp(i \, \omega \, \tau),
\label{barD3b-omega} \\
D^{(3)}[\omega] = & \, \bar{D}^{(3)}[\omega] - \int_0^\infty d\tau_1 \int_0^{\tau_1} d \tau_2 \int_0^{\tau_2} d \tau_3 \, \tr_\R \, L_\SR \, \big\{ L_\SR(- \, \tau_3) \, L_\SR(- \, \tau_1) \, U_0(\tau_2) \, L_\ed[\omega] \, \rho_\SS \, \rho_\R \exp(i \, \omega \, \tau_2) \nonumber \\
    & \qquad \qquad \qquad \qquad \qquad \qquad \qquad \qquad \qquad + L_\SR(- \, \tau_2) \, L_\SR(- \, \tau_1) \, U_0(\tau_3) \, L_\ed[\omega] \, \rho_\SS \, \rho_\R \exp(i \, \omega \, \tau_3) \big\},
\label{D3-omega} \\
\bar{D}^{(4)}[\omega] = & - \int_0^\infty \! d\tau_1 \int_0^{\tau_1} d\tau_2 \int_0^{\tau_2} d\tau_3 \int_0^\beta d\beta_1 \, \tr_\R \, L_\SR \, L_\SR(- \, \tau_3) \, \Q \, L_\SR(- \, \tau_2) \, U_0(\tau_1)\, L_\ed[\omega] \, \rho_\SS \, \rho_\R \, \H_\SR(- i \, \hbar \, \beta_1) \exp(i \, \omega \, \tau_1) \nonumber \\
    & + i \int_0^\infty d\tau_1 \int_0^{\tau_1} d \tau_2 \int_0^\beta d \beta_1 \int_0^{\beta_1} d \beta_2 \, \tr_\R \, L_\SR \, L_\SR(- \, \tau_2) \, U_0(\tau_1) L_\ed[\omega] \, \rho_\SS \, \rho_\R \exp(i \, \omega \, \tau_1) \nonumber \\
    & \qquad \qquad \qquad \qquad \qquad \qquad \qquad \qquad \times \big\{ \H_\SR(- \, i \, \hbar \, \beta_1) \, \H_\SR(- \, i \, \hbar \, \beta_2) - \big\langle \H_\SR(- \, i \, \hbar \, \beta_1) \, \H_\SR(- \, i \, \hbar \, \beta_2) \big\rangle_\R \, \big\} \nonumber \\
    & + \int_0^\infty \! d\tau \int_0^\beta d\beta_1 \int_0^{\beta_1} \! d\beta_2 \int_0^{\beta_2} \! d\beta_3 \, \tr_\R \, L_\SR \, U_0(\tau) L_\ed[\omega] \, \rho_\SS \, \rho_\R \, \H_\SR(- i \, \hbar \, \beta_1) \H_\SR(- i \, \hbar \, \beta_2) \H_\SR(- i \, \hbar \, \beta_3) \exp(i \, \omega \, \tau) \nonumber \\
    & - \int_0^\infty d\tau \int_0^\beta d\beta\,' \int_0^\beta d\beta_1 \int_0^{\beta_1} d \beta_2 \, \tr_\R \, L_\SR \, U_0(\tau)\, L_\ed[\omega] \, \rho_\SS \, \rho_\R \, \H_\SR(- \, i \, \hbar \, \beta\,' ) \, \big\langle \H_\SR(- \, i \, \hbar \, \beta_1) \, \H_\SR(- \, i \, \hbar \, \beta_2) \big\rangle_{\SS \R} \nonumber \\
    & \qquad \qquad \qquad \qquad \qquad \qquad \qquad \qquad \qquad \qquad \qquad \qquad \qquad \times \exp(i \, \omega \, \tau),
\label{barD4b-omega} 
\end{align}
\begin{align}
D^{(4)}[\omega] = & \, \bar{D}^{(4)}[\omega] \nonumber \\
    & + \int_0^\infty d\tau \int_0^\infty d\tau_1 \int_0^{\tau_1} d\tau_2 \int_0^\beta d\beta_1 \, \tr_\R \, L_\SR \, L_\SR(- \, \tau_1) \, \P \, L_\SR(- \, \tau_2) \, U_0(\tau) \, L_\ed[\omega] \, \rho_\SS \, \rho_\R \, \H_\SR(- \, i \, \hbar \, \beta_1) \exp(i \, \omega \, \tau) \nonumber \\
    & + i \int_0^\infty d\tau_1 \int_0^{\tau_1} d\tau_2 \int_0^{\tau_1} d s_1 \int_0^{s_1} d s_2 \, \tr_\R \, L_\SR \, L_\SR(- \, s_2) \, \Q \, L_\SR(- \, s_1) \, L_\SR(- \, \tau_1) \, U_0(\tau_2) \, L_\ed[\omega] \, \rho_\SS \, \rho_\R \exp(i \, \omega \, \tau_2) \nonumber \\
    & - i \int_0^\infty d\tau_1 \int_0^{\tau_1} d\tau_2 \int_0^\infty d s_1 \int_0^{s_1} ds_2 \, \tr_\R \, L_\SR \, L_\SR(- \, s_1) \, \P \, L_\SR(- \, s_2) \, L_\SR(- \, \tau_1) \, U_0(\tau_2) \, L_\ed[\omega] \, \rho_\SS \, \rho_\R \exp(i \, \omega \, \tau_2) \nonumber \\
    & + i \int_0^\infty d\tau_1 \int_0^{\tau_1} d\tau_2 \int_0^{\tau_1 - \tau_2} d \tau_3 \int_0^{\tau_3} d \tau_4 \, \tr_\R \, L_\SR \, U_0(\tau_1) \, L_\SR \, \P \, U_0^{-1}(\tau_1 - \tau_2) \, L_\SR(- \, \tau_3) \, L_\SR(- \, \tau_4) \, L_\ed[\omega] \, \rho_\SS \, \rho_\R \nonumber \\
    & \qquad \qquad \qquad \qquad \qquad \qquad \qquad \qquad \qquad \qquad \qquad \qquad \qquad \times \exp(i \, \omega \, \tau_2) \nonumber \\
    & - \int_0^\infty d\tau_1 \int_0^{\tau_1} d\tau_2 \int_0^{\tau_1 - \tau_2} d \tau_3 \int_0^\beta d\beta_1 \, \tr_\R \, L_\SR \, U_0(\tau_1) \, L_\SR \, \P \, U_0^{-1}(\tau_1 - \tau_2) \, L_\SR(- \, \tau_3) \, L_\ed[\omega] \, \rho_\SS \, \rho_\R \, \H_\SR(- \, i \, \hbar \, \beta_1) \nonumber \\
    & \qquad \qquad \qquad \qquad \qquad \qquad \qquad \qquad \qquad \qquad \qquad \qquad \qquad \times \exp(i \, \omega \, \tau_2) \nonumber \\
    & - i \int_0^\infty d\tau_1 \int_0^{\tau_1} d\tau_2 \int_0^\beta d\beta_1 \int_0^{\beta_1} d\beta_2 \, \tr_\R \, L_\SR \, L_\SR(- \, \tau_1) \, \P \, U_0(\tau_2) \, L_\ed[\omega] \, \rho_\SS \, \rho_\R \exp(i \, \omega \, \tau_2) \nonumber \\
    & \qquad \qquad \qquad \qquad \qquad \qquad \qquad \qquad \times \big\{ \H_\SR(- i \, \hbar \, \beta_1) \H_\SR(- i \, \hbar \, \beta_2) - \big\langle \H_\SR(- i \, \hbar \, \beta_1) \H_\SR(- i \, \hbar \, \beta_2) \big\rangle_{\SS\R} \big\},
\label{D4b-omega}
\end{align}
where we have considered that $ \langle \H_\SR(- \, t) \rangle_\R $\,=\,$ \langle \H_\SR(- i \hbar \beta) \rangle_\R $\,=\,$ \langle \H_\SR \rangle_\R $\,=\,0 and $ \Q \, \rho_\TE^{(1)} $\,=\,$ \rho_\TE^{(1)} $. 
The interference term $D[\omega]$ consists of the interference term $\bar{D}[\omega]$ in the TC equation, which represents the effects of initial correlation of the system and reservoir, and of the terms that represent the memory effects.
\subsection{Admittance in the $n$-th order approximation for $\H_\SR$}

In this subsection, we give the forms of the admittances in the $n$-th order approximation for the system-reservoir interaction $\H_\SR$. The admittance obtained using the relaxation TCL method by expanding Eq. (\ref{tilde-a-TCL}) up to $n$-th order in powers of $\H_\SR$ and by proceeding in the same way as in Subsection III.B, takes the form
\begin{align}
\chi^{\R\TCL}_{ij}(\omega) = & \, \frac{i}{\hbar} \int_0^\infty dt \, \tr \, A_i \exp_\gets \! \Big\{ i \, (\omega - L_\SS) \, t + \int_0^t d\tau \sum_{m = 2}^n C^{(m)}(\tau) \Big\} \Big[ A_j \,,\, \rho_\SS + \sum_{m = 2}^n \rho_0^{(m)} \,\Big] \nonumber \\
    & + \frac{i}{\hbar} \int_0^\infty dt \int_0^t d\tau \, \tr \, A_i \exp_\gets \! \Big\{ i \, (\omega - L_\SS) (t - \tau) + \int_\tau^t ds \sum_{m = 2}^n C^{(m)}(s) \Big\} \sum_{m = 2}^n I_j^{(m)}(\tau) \exp( i \, \omega \, \tau ), \ 
\label{RTCL-admit-n}
\end{align}
with $ \rho_0^{(m)} $\,=\,$ \tr_\R \, \rho_\TE^{(m)} $, where $I_j^{(n)}(t)$ is equal to $I^{(n)}(t)$ with $A_j$ in place of $A$. The admittance obtained using the relaxation TC method by expanding Eq. (\ref{tilde-a-TC}) or (\ref{tilde-a-TComega}) up to $n$-th order in powers of $\H_\SR$ and by proceeding in the same way as in Subsection III.A, is equal to that obtained using the TCE method by expanding Eq. (\ref{rhoTC}) or (\ref{rho1TCeq}) up to $n$-th order in powers of $\H_\SR$ and by proceeding in the same way as in Subsection III.C, and takes the form
\begin{equation}
\chi^{\R\TC}_{ij}(\omega) = \chi^{\TCE}_{ij}(\omega) = \tr \, A_i \, \Big( i \, (L_\SS - \omega) - \sum_{m = 2}^n \bar{C}^{(m)}[\omega] \Big)^{-1} \Big\{ \frac{i}{\hbar} \, \Big[ A_j \,,\, \rho_\SS + \sum_{m = 2}^n \rho_0^{(m)} \,\Big] + \sum_{m = 2}^n \bar{D}_j^{(m)}[\omega] \Big\}\,. \ 
\label{RTCE-admit-n}
\end{equation}
The second-order part $\bar{D}_j^{(2)}[\omega]$ of $ \bar{D}_j[\omega]$ is given by (\ref{barDj2b-omega}) or (\ref{barDj2t-omega}), and the third-order and fourth-order parts are given by
\begin{align}
\bar{D}_j^{(3)}[\omega] = & \, \frac{i}{\hbar} \int_0^\infty d\tau_1 \int_0^{\tau_1} d \tau_2 \int_0^\beta d\beta_1 \, \tr_\R \, L_\SR \, L_\SR(- \, \tau_2) \, U_0(\tau_1) \big[ A_j \,,\, \rho_\SS \, \rho_\R \, \H_\SR(- \, i \, \hbar \, \beta_1) \big] \exp(i \, \omega \, \tau_1) \nonumber \\
    & + \frac{1}{\hbar} \int_0^\infty d\tau \int_0^\beta d\beta_1 \int_0^{\beta_1} d\beta_2 \, \tr_\R \, L_\SR \, U_0(\tau) \big[ A_j \,,\, \rho_\SS \, \rho_\R \, \H_\SR(- \, i \, \hbar \, \beta_1) \, \H_\SR(- \, i \, \hbar \, \beta_2) \big] \exp(i \, \omega \, \tau), \qquad \label{barDj3b-omega}
\end{align}
\begin{align}
\bar{D}_j^{(4)}[\omega] = & \, \frac{1}{\hbar} \int_0^\infty d\tau_1 \int_0^{\tau_1} d\tau_2 \int_0^{\tau_2} d\tau_3 \int_0^\beta d\beta_1 \big\{ \tr_\R \, L_\SR \, L_\SR(- \, \tau_3) \, L_\SR(- \, \tau_2) - \big\langle L_\SR \, L_\SR(- \, \tau_3) \big\rangle_\R \, \tr_\R \, L_\SR(- \, \tau_2) \big\} \nonumber \\
    & \qquad \qquad \qquad \qquad \qquad \qquad \qquad \qquad \times U_0(\tau_1) \big[ A_j \,,\, \rho_\SS \, \rho_\R \, \H_\SR(- \, i \, \hbar \, \beta_1) \big] \exp(i \, \omega \, \tau_1) \nonumber \\
    & - \frac{i}{\hbar} \int_0^\infty d\tau_1 \int_0^{\tau_1} d \tau_2 \int_0^\beta d \beta_1 \int_0^{\beta_1} d \beta_2 \, \tr_\R \, L_\SR \, L_\SR(- \, \tau_2) \, U_0(\tau_1) \exp(i \, \omega \, \tau_1) \nonumber \\
    & \qquad \qquad \qquad \qquad \qquad \qquad \qquad \times \big[ A_j \,,\, \rho_\SS \, \rho_\R \, \big\{ \H_\SR(- \, i \, \hbar \, \beta_1) \, \H_\SR(- \, i \, \hbar \, \beta_2) - \big\langle \H_\SR(- \, i \, \hbar \, \beta_1) \, \H_\SR(- \, i \, \hbar \, \beta_2) \big\rangle_\R \big\} \big] \nonumber \\
    & - \frac{1}{\hbar} \int_0^\infty d\tau \int_0^\beta d\beta_1 \int_0^{\beta_1} d\beta_2 \int_0^{\beta_2} d\beta_3 \, \tr_\R \, L_\SR \, U_0(\tau) \big[ A_j \,,\,  \rho_\SS \, \rho_\R \, \H_\SR(- \, i \, \hbar \, \beta_1) \, \H_\SR(- \, i \, \hbar \, \beta_2) \, \H_\SR(- \, i \, \hbar \, \beta_3) \big] \nonumber \\
    & \qquad \qquad \qquad \qquad \qquad \qquad \qquad \qquad \qquad \qquad \qquad \qquad \qquad \times \exp(i \, \omega \, \tau) \nonumber \\
    & + \frac{1}{\hbar} \int_0^\infty d\tau \int_0^\beta d\beta\,' \int_0^\beta d\beta_1 \int_0^{\beta_1} d \beta_2 \, \tr_\R \, L_\SR \, U_0(\tau) \big[ A_j \,,\, \rho_\SS \, \rho_\R \, \H_\SR(- \, i \, \hbar \, \beta\,' ) \big] \big\langle \H_\SR(- \, i \, \hbar \, \beta_1) \, \H_\SR(- \, i \, \hbar \, \beta_2) \big\rangle_{\SS \R} \nonumber \\
    & \qquad \qquad \qquad \qquad \qquad \qquad \qquad \qquad \qquad \qquad \qquad \qquad \qquad \times \exp(i \, \omega \, \tau).
\label{barDj4b-omega}
\end{align}
The admittance obtained using the TCLE method by expanding Eq. (\ref{rhoTCL}) or (\ref{rho1TCLeq}) up to $n$-th order in powers of $\H_\SR$ and by proceeding in the same way as in Subsection III.D, takes the form
\begin{equation}
\chi_{ij}^{\TCLE}(\omega) = \tr \, A_i \, \Big( i \, (L_\SS - \omega) - \sum_{m = 2}^n C^{(m)} \Big)^{-1} \Big\{ \frac{i}{\hbar} \, \Big[ A_j \,,\, \rho_\SS + \sum_{m = 2}^n \rho_0^{(m)} \,\Big] + \sum_{m = 2}^n D_j^{(m)}[\omega] \Big\}\,. \quad \ 
\label{TCLE-admit-n}
\end{equation}
The second-order part $D_j^{(2)}[\omega]$ of $D_j[\omega]$ is given by (\ref{Dj2-omega}), and the third-order and fourth-order parts are given by
\begin{align}
D_j^{(3)}[\omega] & = \bar{D}_j^{(3)}[\omega] + \frac{1}{\hbar} \int_0^\infty d\tau_1 \int_0^{\tau_1} d \tau_2 \int_0^{\tau_2} d \tau_3 \, \big\{ \big\langle L_\SR \, L_\SR(- \, \tau_3) \, L_\SR(- \, \tau_1) \big\rangle_\R \, \big[ A_j(- \, \tau_2) \,,\, \rho_\SS \big] \exp(i \, \omega \, \tau_2) \nonumber \\
    & \qquad \qquad \qquad \qquad \qquad \qquad \qquad \qquad + \big\langle L_\SR \, L_\SR(- \, \tau_2) \, L_\SR(- \, \tau_1) \big\rangle_\R \, \big[ A_j(- \, \tau_3) \,,\, \rho_\SS \big] \exp(i \, \omega \, \tau_3) \big\}, \label{Dj3-omega} \\
D_j^{(4)}[\omega] & = \bar{D}_j^{(4)}[\omega] \nonumber \\
    & - \frac{1}{\hbar} \int_0^\infty \! d\tau \int_0^\infty \! d\tau_1 \int_0^{\tau_1} d\tau_2 \int_0^\beta d\beta_1 \, \big\langle L_\SR \, L_\SR(- \, \tau_1) \big\rangle_\R \, \tr_\R \, L_\SR(- \, \tau_2) \, U_0(\tau) \, \big[ A_j \,,\, \rho_\SS \, \rho_\R \, \H_\SR(- \, i \, \hbar \, \beta_1) \big] \exp(i \, \omega \, \tau) \nonumber \\
    & - \frac{i}{\hbar} \int_0^\infty \! d\tau_1 \int_0^{\tau_1} d\tau_2 \int_0^{\tau_1} d \tau_3 \int_0^{\tau_3} d \tau_4 \, \big\{ \big\langle L_\SR \, L_\SR(- \, \tau_4) \, L_\SR(- \, \tau_3) \, L_\SR(- \, \tau_1) \big\rangle_\R \nonumber \\
    & \qquad \qquad \qquad \qquad \qquad \qquad \qquad \qquad { } - \big\langle L_\SR \, L_\SR(- \, \tau_4) \big\rangle_\R \big\langle L_\SR(- \, \tau_3) \, L_\SR(- \, \tau_1) \big\rangle_\R \, \big\} \big[ A_j(- \, \tau_2) \,,\, \rho_\SS \big] \exp(i \, \omega \, \tau_2) \nonumber \\
    & + \frac{i}{\hbar} \int_0^\infty d\tau_1 \int_0^{\tau_1} d\tau_2 \int_0^\infty d \tau_3 \int_0^{\tau_3} d \tau_4 \, \big\langle L_\SR \, L_\SR(- \, \tau_3) \big\rangle_\R \big\langle L_\SR(- \, \tau_4) \, L_\SR(- \, \tau_1) \big\rangle_\R \, \big[ A_j(- \, \tau_2) \,,\, \rho_\SS \big] \exp(i \, \omega \, \tau_2) \nonumber \\
    & - \frac{i}{\hbar} \int_0^\infty \! d\tau_1 \int_0^{\tau_1} d\tau_2 \int_0^{\tau_1 - \tau_2} d \tau_3 \int_0^{\tau_3} d \tau_4 \, \big\langle L_\SR \, L_\SR (- \, \tau_1) \big\rangle_\R \, U_0(\tau_2) \, \big\langle L_\SR(- \, \tau_3) \, L_\SR(- \, \tau_4) \big\rangle_\R \, \big[ A_j \,,\, \rho_\SS \big] \exp(i \, \omega \, \tau_2) \nonumber \\
    & + \frac{1}{\hbar} \int_0^\infty \! d\tau_1 \int_0^{\tau_1} \! d\tau_2 \int_0^{\tau_1 - \tau_2} \!\! d \tau_3 \int_0^\beta \! d\beta_1 \, \big\langle L_\SR \, L_\SR(- \, \tau_1) \big\rangle_\R \, U_0(\tau_2) \, \tr_\R \, L_\SR(- \, \tau_3) \big[ A_j \,,\, \rho_\SS \, \rho_\R \, \H_\SR(- i \, \hbar \, \beta_1) \big] \exp(i \, \omega \, \tau_2) \nonumber \\
    & + \frac{i}{\hbar} \int_0^\infty d\tau_1 \int_0^{\tau_1} d\tau_2 \int_0^\beta d\beta_1 \int_0^{\beta_1} d\beta_2 \, \big\langle L_\SR \, L_\SR(- \, \tau_1) \big\rangle_\R \, U_0(\tau_2) \, \big[ A_j \,,\, \rho_\SS \, \big\{ \big\langle \H_\SR(- \, i \, \hbar \, \beta_1) \, \H_\SR(- \, i \, \hbar \, \beta_2) \big\rangle_\R \nonumber \\
    & \qquad \qquad \qquad \qquad \qquad \qquad \qquad \qquad \qquad \qquad \quad - \big\langle \H_\SR(- \, i \, \hbar \, \beta_1) \, \H_\SR(- \, i \, \hbar \, \beta_2) \big\rangle_{\SS\R} \, \big\} \, \big] \exp(i \, \omega \, \tau_2),
\label{Dj4b-omega} 
\end{align}
where $A(t)$ is defined by 
\begin{equation}
A(t) = \exp(i \, L_\SS \, t)\, A = \exp(i \, \H_\SS \, t / \hbar)\, A \exp(- \, i \, \H_\SS \, t / \hbar). \qquad \qquad
\end{equation}

As expressed using time-integrals alone for the second-order interference terms $\bar{D}_j^{(2)}[\omega]$ and $D_j^{(2)}[\omega]$ in (\ref{barDj2t-omega}) and (\ref{Dj2-omega}), the inhomogeneous term $I_j(t)$ and the interference terms $\bar{D}_j[\omega]$ and $D_j[\omega]$ can be expressed using time-integrals alone. By virtue of the results of Appendixes A, C and D, the second-order, third-order and fourth-order parts $I_j^{(n)}(t)$ ($n$\,=\,2,\,3,\,4) of $I_j(t)$ take the forms using time-integrals alone\,:
%
\begin{align}
I_j^{(2)}(t) = & - \int_t^\infty d \tau \, \tr_\R \, L_\SR \, \big[ A_j(- \, t) \,,\, L_\SR(- \, \tau) \, \rho_\SS \, \rho_\R \, \big]\,, \\
I_j^{(3)}(t) = & \,\, i \int_t^\infty d\tau_1 \int_0^t d \tau_2 \, \tr_\R \, L_\SR \, L_\SR(- \, \tau_2) \, \big[ A_j(- \, t) \,,\, L_\SR(- \, \tau_1) \, \rho_\SS \, \rho_\R \, \big] \nonumber \\
    & + i \int_t^\infty d\tau_1 \int_t^{\tau_1} d\tau_2 \, \tr_\R \, L_\SR \, \big[ A_j(- \, t) \,,\, L_\SR(- \, \tau_2) \, L_\SR(- \, \tau_1) \, \rho_\SS \, \rho_\R \, \big] \,, \\
I_j^{(4)}(t) = & \int_t^\infty d\tau_1 \int_0^t d\tau_2 \int_0^{\tau_2} d\tau_3 \, \big\{ \, \tr_\R \, L_\SR \, L_\SR(- \, \tau_3) \, L_\SR(- \, \tau_2) \, \big[ A_j(- \, t) \,,\, L_\SR(- \, \tau_1) \, \rho_\SS \, \rho_\R \, \big] \nonumber \\
    & \qquad \qquad \qquad \qquad \qquad \quad - \big\langle L_\SR \, L_\SR(- \, \tau_3) \big\rangle_\R \, \tr_\R \, L_\SR(- \, \tau_2) \, \big[ A_j(- \, t) \,,\, L_\SR(- \, \tau_1) \, \rho_\SS \, \rho_\R \, \big] \big\} \nonumber \\
    & + \int_t^\infty d\tau_1 \int_t^{\tau_1} d\tau_2 \int_0^t d \tau_3 \, \big\{ \, \tr_\R \, L_\SR \, L_\SR(- \, \tau_3) \, \big[ A_j(- \, t) \,,\, L_\SR(- \, \tau_2) \, L_\SR(- \, \tau_1) \, \rho_\SS \, \rho_\R \, \big] \nonumber \\
    & \qquad \qquad \qquad \qquad \qquad \qquad - \big\langle L_\SR \, L_\SR(- \, \tau_3) \big\rangle_\R \, \tr_\R \, \big[ A_j(- \, t) \,,\, L_\SR(- \, \tau_2) \, L_\SR(- \, \tau_1) \, \rho_\SS \, \rho_\R \, \big] \big\} \qquad \nonumber \\
    & + \int_t^\infty d\tau_1 \int_t^{\tau_1} d\tau_2 \int_t^{\tau_2} d\tau_3 \, \tr_\R \, L_\SR \, \big[ A_j(- \, t) \,,\, L_\SR(- \, \tau_3) \, L_\SR(- \, \tau_2) \, L_\SR(- \, \tau_1) \, \rho_\SS \, \rho_\R \, \big] \nonumber \\
    & - \int_t^\infty d \tau_1 \int_0^t d\tau_2 \int_0^{\tau_2} d\tau_3 \, \big\langle L_\SR \, L_\SR(- \, \tau_2) \big\rangle_\R \, \tr_\R \, L_\SR(- \, \tau_3) \, \big[ A_j(- \, t) \,,\, L_\SR(- \, \tau_1) \, \rho_\SS \, \rho_\R \, \big]\,.
\end{align}
The second-order interference terms $\bar{D}_j^{(2)}[\omega]$ and $D_j^{(2)}[\omega]$ given by (\ref{barDj2t-omega}) and (\ref{Dj2-omega}) can be, respectively, expressed in the compact forms
\begin{align}
\bar{D}_j^{(2)}[\omega] & = - \, \frac{i}{\hbar} \int_0^\infty d\tau_1 \int_0^{\tau_1} d\tau_2 \, \tr_\R \, L_\SR \, \big[\, A_j(- \, \tau_2) \,,\, L_\SR(- \, \tau_1) \, \rho_\SS \, \rho_\R \,\big] \exp(i \, \omega \, \tau_2),
\label{barDj2t-omega-1} \\
D_j^{(2)}[\omega] & = \bar{D}_j^{(2)}[\omega] + \frac{i}{\hbar} \int_0^\infty d\tau_1 \int_0^{\tau_1} d\tau_2 \, \big\langle L_\SR \, L_\SR(- \, \tau_1) \big\rangle_\R \, \big[ A_j(- \, \tau_2) \,,\, \rho_\SS \,\big] \exp(i \, \omega \, \tau_2). \qquad 
\label{Dj2-omega-1}
\end{align}
By virtue of the results of Appendixes A and C, the third-order interference term $\bar{D}_j^{(3)}[\omega]$ given by (\ref{barDj3b-omega}) takes the form using time-integrals alone\,:
\begin{align}
\bar{D}_j^{(3)}[\omega] = & - \frac{1}{\hbar} \int_0^\infty d\tau_1 \int_0^{\tau_1} d \tau_2 \int_0^{\tau_2} d\tau_3 \, \big\{\, \tr_\R \, L_\SR \, L_\SR(- \, \tau_3) \, \big[ A_j(- \, \tau_2) \,,\, L_\SR(- \, \tau_1) \, \rho_\SS \, \rho_\R \, \big] \exp(i \, \omega \, \tau_2) \nonumber \\
    & \qquad \qquad \qquad \qquad \qquad \qquad \quad + \tr_\R \, L_\SR \, \big[ A(- \, \tau_3) \,,\, L_\SR(- \, \tau_2) \, L_\SR(- \, \tau_1) \, \rho_\SS \, \rho_\R \, \big] \exp(i \, \omega \, \tau_3) \,\big\}. \quad \label{barDj3t-omega}
\end{align}
The third-order interference term $D_j^{(3)}[\omega]$ is given by (\ref{Dj3-omega}) with the above $\bar{D}_j^{(3)}[\omega]$. By virtue of the results of Appendixes A, C and D, the fourth-order interference terms $\bar{D}_j^{(4)}[\omega]$ and $D_j^{(4)}[\omega]$ given by (\ref{barD4b-omega}) and (\ref{D4b-omega}) take the forms using time-integrals alone\,:
\begin{align}
\bar{D}_j^{(4)}[\omega] = 
    & \, \frac{i}{\hbar} \int_0^\infty d\tau_1 \int_0^{\tau_1} d\tau_2 \int_0^{\tau_2} d\tau_3 \int_0^{\tau_3} d\tau_4 \, \big\{ \, \tr_\R \, L_\SR \, L_\SR(- \, \tau_4) \, L_\SR(- \, \tau_3) \, \big[ A_j(- \, \tau_2) \,,\, L_\SR(- \, \tau_1) \, \rho_\SS \, \rho_\R \, \big] \nonumber \\
    & \qquad \qquad \qquad \qquad \qquad \qquad \qquad \quad - \big\langle L_\SR \, L_\SR(- \, \tau_4) \big\rangle_\R \, \tr_\R \, L_\SR(- \, \tau_3) \, \big[ A_j(- \, \tau_2) \,,\, L_\SR(- \, \tau_1) \, \rho_\SS \, \rho_\R \, \big] \big\} \exp(i \, \omega \, \tau_2) \nonumber \\
    & + \frac{i}{\hbar} \int_0^\infty d\tau_1 \int_0^{\tau_1} d\tau_2 \int_0^{\tau_2} d\tau_3 \int_0^{\tau_3} d \tau_4 \, \big\{\, \tr_\R \, L_\SR \, L_\SR(- \, \tau_4) \, \big[ A_j(- \, \tau_3) \,,\, L_\SR(- \, \tau_2) \, L_\SR(- \, \tau_1) \, \rho_\SS \, \rho_\R \, \big] \nonumber \\
    & \qquad \qquad \qquad \qquad \qquad \qquad \qquad \qquad - \big\langle L_\SR \, L_\SR(- \, \tau_4) \big\rangle_\R \, \big[ A_j(- \, \tau_3) \,, \big\langle L_\SR(- \, \tau_2) \, L_\SR(- \, \tau_1) \big\rangle_\R \, \rho_\SS \, \big] \big\} \exp(i \, \omega \, \tau_3) \nonumber \\
    & + \frac{i}{\hbar} \int_0^\infty d\tau_1 \int_0^{\tau_1} d\tau_2 \int_0^{\tau_2} d\tau_3 \int_0^{\tau_3} d\tau_4 \, \tr_\R \, L_\SR \, \big[ A_j(- \, \tau_4) \,,\, L_\SR(- \, \tau_3) \, L_\SR(- \, \tau_2) \, L_\SR(- \, \tau_1) \, \rho_\SS \, \rho_\R \, \big] \exp(i \, \omega \, \tau_4),
\end{align}
\begin{align}
D_j^{(4)}[\omega] = & \,\, \bar{D}_j^{(4)}[\omega] \nonumber \\
%
    & - \frac{i}{\hbar} \int_0^\infty d\tau_1 \int_0^{\tau_1} d\tau_2 \int_0^\infty d\tau_3 \int_0^{\tau_3} d\tau_4 \, \big\langle L_\SR \, L_\SR(- \, \tau_3) \big\rangle_\R \, \tr_\R \, L_\SR(- \, \tau_4) \, \big[ A_j(- \, \tau_2) \,,\, L_\SR(- \, \tau_1) \, \rho_\SS \, \rho_\R \, \big] \exp(i \, \omega \, \tau_2) \nonumber \\
    & - \frac{i}{\hbar} \int_0^\infty d\tau_1 \int_0^{\tau_1} d\tau_2 \int_0^{\tau_1} d \tau_3 \int_0^{\tau_3} d \tau_4 \, \big\{ \big\langle L_\SR \, L_\SR(- \, \tau_4) \, L_\SR(- \, \tau_3) \, L_\SR(- \, \tau_1) \big\rangle_\R \nonumber \\
    & \qquad \qquad \qquad \qquad \qquad \qquad \qquad \qquad - \big\langle L_\SR \, L_\SR(- \, \tau_4) \big\rangle_\R \big\langle L_\SR(- \, \tau_3) \, L_\SR(- \, \tau_1) \big\rangle_\R \big\} \, \big[ A_j(- \, \tau_2) \,,\, \rho_\SS \, \big] \exp(i \, \omega \, \tau_2) \nonumber \\
    & + \frac{i}{\hbar} \int_0^\infty d\tau_1 \int_0^{\tau_1} d\tau_2 \int_0^\infty d \tau_3 \int_0^{\tau_3} d \tau_4 \, \big\langle L_\SR \, L_\SR(- \, \tau_3) \big\rangle_\R \big\langle L_\SR(- \, \tau_4) \, L_\SR(- \, \tau_1) \big\rangle_\R \, \big[ A_j(- \, \tau_2) \,,\, \rho_\SS \, \big] \exp(i \, \omega \, \tau_2) \nonumber \\
%
    & - \frac{i}{\hbar} \int_0^\infty d\tau_1 \int_0^{\tau_1} d\tau_2 \int_0^{\tau_2} d\tau_3 \int_0^{\tau_3} d\tau_4 \, \big\langle L_\SR \, L_\SR (- \, \tau_1) \big\rangle_\R \, \big\langle L_\SR(- \, \tau_2) \, L_\SR(- \, \tau_3) \big\rangle_\R \, \big[ A_j(- \, \tau_4) \,,\, \rho_\SS \, \big] \exp(i \, \omega \, \tau_4) \nonumber \\
%
    & + \frac{i}{\hbar} \int_0^\infty d\tau_1 \int_0^{\tau_1} d\tau_2 \int_0^{\tau_2} d \tau_3 \int_{\tau_3}^\infty d\tau_4 \, \big\langle L_\SR \, L_\SR(- \, \tau_1) \big\rangle_\R \, \tr_\R \, L_\SR(- \, \tau_2) \big[ A_j(- \, \tau_3) \,,\, L_\SR(- \, \tau_4) \, \rho_\SS \, \rho_\R \, \big] \exp(i \, \omega \, \tau_3) \nonumber \\
    & - \frac{i}{\hbar} \int_0^\infty d\tau_1 \int_0^{\tau_1} d\tau_2 \int_{\tau_2}^\infty d\tau_3 \int_{\tau_2}^{\tau_3} d\tau_4 \, \big\langle L_\SR \, L_\SR(- \, \tau_1) \big\rangle_\R \, \big[ A_j(- \, \tau_2) \,,\, \big\langle L_\SR(- \, \tau_4) \, L_\SR(- \, \tau_3) \big\rangle_\R \, \rho_\SS \, \big] \exp(i \, \omega \, \tau_2),
\end{align}
where we have performed some integral transformations as (\ref{integral-trans}). By virtue of the results of Appendixes C, E and D, $\rho_0^{(n)}$ [=\,$ \tr_\R \rho_\TE^{(n)}$] ($n$\,=\,2,\,3) can be expressed using time-integrals as
\begin{align}
\rho_0^{(2)} = & \,\, - \int_0^\infty d\tau_1 \int_0^{\tau_1} d\tau_2 \, \tr_\R \, L_\SR(- \, \tau_2) \, L_\SR(- \, \tau_1) \, \rho_\SS \, \rho_\R \exp(- \, \epsilon \, \tau_1) \, \big|_{\epsilon \to +0}\,, 
\label{rho0-t-2} \\
\rho_0^{(3)} = & \,\, i \int_0^\infty d\tau_1 \int_0^{\tau_1} d\tau_2 \int_0^{\tau_2} d\tau_3 \, \tr_\R \, L_\SR(- \, \tau_3) \, L_\SR(- \, \tau_2) \, L_\SR(- \, \tau_1) \, \rho_\SS \, \rho_\R \exp(- \, \epsilon \, \tau_1) \, \big|_{\epsilon \to +0}\,, 
\label{rho0-t-3} \\
\rho_0^{(4)} = & \int_0^\infty d\tau_1 \int_0^{\tau_1} d\tau_2 \int_0^{\tau_2} d\tau_3 \int_0^{\tau_3} d\tau_4 \, \tr_\R \, L_\SR(- \, \tau_4) \, L_\SR(- \, \tau_3) \, L_\SR(- \, \tau_2) \, L_\SR(- \, \tau_1) \, \rho_\SS \, \rho_\R \exp(- \, \epsilon \, \tau_1) \, \big|_{\epsilon \to +0}\,, 
\label{rho0-t-4}
\end{align}
where we have considered that $ \langle \H_\SR(- i \hbar \beta) \rangle_\R $\,=\,0. 
Thus, the admittances have been expressed using the time-integrals alone in the fourth-order approximation for the system-reservoir interaction. In the higher order approximation, the admittances can be expressed using the time-integrals alone by the same way. This facilitates the calculations of the admittances, because when the time-correlation functions of the heat reservoir are given, the admittances can be calculated.
\end{widetext}
%
%
\section{Comparison with analytically solvable models}

In the previous sections, we have studied expressions of the susceptibility in various methods. 
We may think some of them is better than others. 
There have been discussed comparative merits and demerits of the TC and TCL methods. 
Both of them are correct expression, and thus, if we take all the terms in expansion, they must be the same. 
When we truncate the perturbation at a finite term, they are different with each other. 
Then, one of them gives the better results than the other. 
In this section, we consider the two exactly solvable models, and compare each exact form of admittance 
with the results obtained by using the methods derived in Section III in the lowest Born approximation. \\
%
\quad \\
(1) {\it A quantum oscillator model} \\

We first consider the model of a system of quantum oscillator interacting with a heat reservoir composed of many quantum oscillators. 
We study the susceptibility of the system for an external driving field which is a periodic function of the frequency $\omega$. 
We adopt the following Hamiltonians:
\begin{align}
    & \H_\SS = \hbar \, \omega_0 \, b^\dagger b \,, \qquad \qquad \H_\R = \sum_\alpha \hbar \, \omega_\alpha \, b_\alpha^\dagger b_\alpha \,, \quad \label{H-OR}\\
    & \H_\SR = \hbar \, b \, R^\dagger + \hbar \, b^\dagger R \,, \qquad \quad (R = \!\sum_\alpha g_\alpha b_\alpha) \label{H-ORint}
\end{align}
where $b$ is the boson operator of the quantum oscillator system, $b_\alpha$ represents the boson operator of the mode $\alpha$ which composes the heat reservoir, 
$\omega_0$ and $\omega_\alpha$ are the characteristic frequencies of the oscillators, 
and $g_\alpha$ is the coupling constant between the boson system and the boson of the mode $\alpha$. 
The interaction of the quantum oscillator system system with the external driving field is given by
\begin{equation}
\H_\ed(t) = - \, \hbar \, b \, F^* e^{i \, \omega \, t} - \hbar \, b^\dagger F \, e^{- \, i \, \omega \, t}, \label{H-Oexternal}
\end{equation}
where $F$ is the magnitude of the external driving force. 
Then, $\H_\SR(t)$ defined by the interaction representation (\ref{H-SR-t}) takes the form
\begin{align}
\H_\SR(t) & = \hbar \sum_\alpha \big\{ g_\alpha^* \, b \, b_\alpha^\dagger \exp\{- \, i \, (\omega_0 - \omega_\alpha) \, t \} \nonumber \\
    & \qquad \quad \ + g_\alpha \, b^\dagger \, b_\alpha \exp\{i \, (\omega_0 - \omega_\alpha) \, t \} \big\}. \quad \label{Hboson-SR-t}
\end{align}
The admittance $\chi_{b b^\dagger}(\omega)$ can be written as \cite{Kubo}
\begin{align}
\chi_{b b^\dagger}(\omega) & = \frac{i}{\hbar} \int_0^\infty \! dt \, \Tr \, \hbar \, b \, e^{- \, i \, L \, t} \, [\, \hbar \, b^\dagger , \, \rho_\TE \,] \, e^{i \, \omega \, t \, - \, \epsilon \, t}, \nonumber \\
    & = i \, \hbar \int_0^\infty \! dt \, \Tr \, b \, e^{- \, i \, L_0 \, t} \exp_\gets \! \big\{\! - i \! \int_0^t \! d \tau \, L_\SR(\tau) \big\} \nonumber \\
       & \qquad \qquad \qquad \times [\, b^\dagger , \, \rho_\TE \,] \, e^{i \, \omega \, t \, - \, \epsilon \, t}, \label{chi-bbd} 
\end{align}
with $\epsilon $$ \to $$ +0$. 
As shown in Appendix F, the above admittance can be exactly calculated as
\begin{equation}
\chi_{b b^\dagger}(\omega) = \frac{i \, \hbar}{i \, (\omega_0 - \bar{\omega}) + \phi(\bar{\omega})}, \qquad \label{exact-chi-bbd}
\end{equation}
with $ \bar{\omega}$\,=\,$ \omega $\,+\,$ i \, \epsilon $ \,($ \epsilon $$ \to $+0), where $\phi(\omega)$ is defined by
\begin{subequations}
\label{phi-omega}
\begin{align}
\phi(\bar{\omega}) & = \int_0^\infty d\tau \, \langle \, [\, R(\tau)\,,\, R^\dagger \,] \, \rangle_\R \, e^{i \, \bar{\omega} \, \tau} , \label{phi-omega-1}\\
    & = i \sum_\alpha |\, g_\alpha \,|^2 / (\bar{\omega} - \omega_\alpha). \label{phi-omega-2}
\end{align}
\end{subequations}

Applying the relaxation TC method (or the TCE method) for the Hamiltonians (\ref{H-OR})\,$-$\,(\ref{H-Oexternal}), we obtain
\begin{align}
    & \tr \, b \, \bar{C}^{(2)}[\omega] \, \tilde{a}_{b^\dagger}[\omega] = - \, \phi(\bar{\omega}) \, \tr \, b \, \tilde{a}_{b^\dagger}[\omega] + \HOT, \\
    & \tr \, b \, \bar{I}^{(2)}_{b^\dagger}[\omega] = - \int_0^\infty \! d\tau \int_0^\tau \! ds \, \big\{ \langle R(\tau) R^\dagger \rangle_\R \, \tr \, [\, \hbar \, b^\dagger , b \, \rho_\SS \,] \nonumber \\
      & \qquad \qquad \qquad - \langle R^\dagger R(\tau) \rangle_\R \, \tr \, [\, \hbar \, b^\dagger , \rho_\SS \, b \,] \big\} e^{i \omega_0 \tau + i (\omega - \omega_0) s}, \nonumber \\
      & \qquad \qquad = 0,
\end{align}
with $ \bar{\omega}$\,=\,$ \omega $\,+\,$ i \, \epsilon $ \,($\epsilon$$\to$+0), where $\HOT$ denotes the higher order terms in powers of $\H_\SR$\,. 
In the lowest Born approximation for the boson-reservoir interaction $\H_\SR$, we have
\begin{equation}
\{\, i \, (\omega_0 - \bar{\omega}) + \phi(\bar{\omega}) \,\}\, \tr \, b \, \tilde{a}_{b^\dagger}[\omega] = \tr \, b \, \tilde{a}_{b^\dagger}(0), \label{RTC-bbd}
\end{equation}
which leads to the admittance
\begin{equation}
\chi_{b b^\dagger}^{\R\TC}(\omega) = \frac{i}{\hbar} \, \tr \, \hbar \, b \, \tilde{a}_{b^\dagger}[\omega] = \frac{i \, \hbar}{i \, (\omega_0 - \bar{\omega}) + \phi(\bar{\omega})},
\label{chi-RTC-bbd}
\end{equation}
with $ \bar{\omega}$\,=\,$ \omega $\,+\,$ i \, \epsilon $ \,($\epsilon$$\to$+0). This is equal to the exact admittance (\ref{exact-chi-bbd}).
Therefore, for the quantum oscillator model under consideration, 
the admittance (\ref{chi-RTC-bbd}) obtained by using the relaxation TC method (or the TCE method) 
in the lowest Born approximation for the boson-reservoir interaction, coincides with the exact one.

Applying the relaxation TCL method for the Hamiltonians (\ref{H-OR})\,$-$\,(\ref{H-Oexternal}), we obtain
\begin{align}
    & \tr \, b \, C^{(2)}(t) \, \tilde{a}_{b^\dagger}(t) = - \, \phi(\omega_0 , t) \, \tr \, b \, \tilde{a}_{b^\dagger}(t), \\
    & \tr \, b \, I^{(2)}_{b^\dagger}(t) = - \int_t^\infty \! d\tau \, \big\{ \langle R(\tau) R^\dagger \rangle_\R \, \tr \, [\, \hbar \, b^\dagger , b \, \rho_\SS \,] \nonumber \\
      & \qquad \qquad \qquad \quad - \langle R^\dagger R(\tau) \rangle_\R \, \tr \, [\, \hbar \, b^\dagger , \rho_\SS \, b \,] \big\} e^{i \, \omega_0 \, (\tau - t)}, \nonumber \\
      & \qquad \qquad = 0,
\end{align}
where $\phi(\omega , t)$ is defined by
\begin{subequations}
\label{phi-omega-t}
\begin{align}
\phi(\omega , t) & = \int_0^t d\tau \, \langle \, [\, R(\tau) \,,\, R^\dagger \,] \, \rangle_\R \, e^{i \, \omega \, \tau}, \label{phi-omega-t-1}\\
    & = i \sum_\alpha |\, g_\alpha \,|^2 \, \frac{\, 1 - \exp\{ i \, (\omega - \omega_\alpha) \, t \} \,}{\omega - \omega_\alpha}. \label{phi-omega-t-2}
\end{align}
\end{subequations}
In the lowest Born approximation for the boson-reservoir interaction $\H_\SR$, we have the TCL equation
\begin{equation}
(d/dt) \, \tr \, b \, \tilde{a}_{b^\dagger}(t) = - \, i \, \omega_0 \, \tr \, b \, \tilde{a}_{b^\dagger}(t) - \phi(\omega_0 , t) \, \tr \, b \, \tilde{a}_{b^\dagger}(t),
\label{RTCL-bbd}
\end{equation}
which leads to the admittance \cite{S4}
\begin{align}
\chi_{b b^\dagger}^{\R\TCL}(\omega) & = \frac{i}{\hbar} \int_0^\infty dt \, \tr \, \hbar \, b \, \tilde{a}_{b^\dagger}(t) \, e^{i \, \bar{\omega} \, t}, \nonumber \\
    & = i \, \hbar \int_0^\infty \! dt \exp\Big\{ i \, (\bar{\omega} - \omega_0) \, t - \! \int_0^t \! d\tau \, \phi(\omega_0 , \tau) \Big\},
\label{chi-RTCL-bbd}
\end{align}
with $ \bar{\omega}$\,=\,$ \omega $\,+\,$ i \, \epsilon $ \,($ \epsilon $$ \to $+0).

By using the TCLE method in the lowest Born approximation for the boson-reservoir interaction, the admittance was obtained by one of the authors as \cite{S4}
\begin{equation}
\chi_{b b^\dagger}^{\TCLE}(\omega) = \hbar \, \frac{i + \Phi(\bar{\omega})}{\, i \, (\omega_0 - \bar{\omega}) + \phi(\bar{\omega_0}) \,}, \quad
\label{chi-TCLE-bbd}
\end{equation}
with $ \bar{\omega}$\,=\,$ \omega $\,+\,$ i \, \epsilon $ \,($ \epsilon $$ \to $+0), 
where $\Phi(\bar{\omega})$ comes from the interference term and is given by
\begin{equation}
\Phi(\bar{\omega}) = ( \phi(\bar{\omega}) - \phi(\bar{\omega}_0) ) / (\bar{\omega} - \omega_0). \quad \ 
\end{equation}

Now, we have found that the admittance (\ref{chi-RTC-bbd}) obtained by using the relaxation TC method (or the TCE method) in the lowest Born approximation for the boson-reservoir interaction is equal to the exact one (\ref{exact-chi-bbd}). 
On the other hand, the admittance (\ref{chi-RTCL-bbd}) obtained by using the relaxation TCL method and the admittance (\ref{chi-TCLE-bbd}) obtained by using the TCLE method take the forms different from the exact one (\ref{exact-chi-bbd}). 

Let us compare these admittances analytically. 
The second-order terms of these admittances in powers of the boson-reservoir interaction coincide and are equal to 
\begin{equation}
\chi_{b b^\dagger}^{(2)}(\omega) = \frac{i \, \hbar \, \phi(\bar{\omega})}{\, (\bar{\omega} - \omega_0)^2 \,} = \sum_\alpha \frac{- \, \hbar \, |\, g_\alpha \,|^2}{(\bar{\omega} - \omega_0)^2 \, (\bar{\omega} - \omega_\alpha)}, \label{chi-bbd(2)}
\end{equation}
with $ \bar{\omega}$\,=\,$ \omega $\,+\,$ i \, \epsilon $ \,($ \epsilon $$ \to $+0). The fourth-order terms of those admittances are obtained, by expanding (\ref{exact-chi-bbd}) [or (\ref{chi-RTC-bbd})], (\ref{chi-RTCL-bbd}) and (\ref{chi-TCLE-bbd}) in powers of $\phi$, as
\begin{align}
& \chi_{b b^\dagger}^{(4)}(\omega) = \chi_{b b^\dagger}^{\R\TC(4)}(\omega) = - \, \frac{\, \hbar \, \phi(\bar{\omega})^2 \,}{(\omega_0 - \bar{\omega})^3}, \nonumber \\
    & \qquad \quad \, = \sum_{\alpha \,,\, \alpha'} \, \frac{\hbar \, |\, g_\alpha |^2 \, |\, g_{\alpha'} |^2}{(\omega_0 - \bar{\omega})^3 \, (\bar{\omega} - \omega_\alpha) \, (\bar{\omega} - \omega_{\alpha'})} \,, \label{chi-bbd(4)} \\
& \chi_{b b^\dagger}^{\TCLE(4)}(\omega) = - \, \hbar \, \frac{\, \phi(\bar{\omega}) \, \phi(\omega_0) \,}{(\omega_0 - \bar{\omega})^3} , \nonumber \\
    & \qquad \qquad = \sum_{\alpha \,,\, \alpha'} \, \frac{\hbar \, |\, g_\alpha |^2 \, |\, g_{\alpha'} |^2}{(\omega_0 - \bar{\omega})^3 \, (\bar{\omega} - \omega_\alpha) \, (\omega_0 - \omega_{\alpha'})} \,, \label{chi-bbd-TCLE(4)} \\
& \chi_{b b^\dagger}^{\R\TCL(4)}(\omega) = \sum_{\alpha \,,\, \alpha'} \, \frac{\hbar \, |\, g_\alpha |^2 \, |\, g_{\alpha'} |^2}{(\omega_0 - \bar{\omega}) \, (\omega_0 - \omega_\alpha) \, (\omega_0 - \omega_{\alpha'})} \nonumber \\
    & \qquad \qquad \!\! \times \Big\{ \frac{1}{(\omega_0 - \bar{\omega})^2} - \frac{1}{2} \Big( \frac{1}{(\bar{\omega} - \omega_\alpha)^2} + \frac{1}{(\bar{\omega} - \omega_{\alpha'})^2} \Big) \nonumber \\
    & \qquad \qquad \qquad + \frac{1}{2} \! \cdot \! \frac{2 \, \omega_0 - \omega_\alpha - \omega_{\alpha'}}{(\omega_0 - \bar{\omega}) \, (\bar{\omega} + \omega_0 - \omega_\alpha - \omega_{\alpha'})}\nonumber \\
    & \qquad \qquad \qquad \qquad \times \Big( \frac{1}{\bar{\omega} - \omega_\alpha} + \frac{1}{\bar{\omega} - \omega_{\alpha'}} \Big) \Big\}, \label{chi-bbd-RTCL(4)}
\end{align}
with $ \bar{\omega}$\,=\,$ \omega $\,+\,$ i \, \epsilon $ \,($ \epsilon $$ \to $+0). The fourth-order term $\chi_{b b^\dagger}^{\TCLE(4)}(\omega)$ takes a form similar to the exact one $\chi_{b b^\dagger}^{(4)}(\omega)$. 
The fourth-order term $\chi_{b b^\dagger}^{\R\TCL(4)}(\omega)$ takes a form different from the exact one. 
It should be noted that 
$\chi_{b b^\dagger}^{\TCLE(4)}(\omega)$ and $\chi_{b b^\dagger}^{\R\TCL(4)}(\omega)$ coincide 
with the exact one $\chi_{b b^\dagger}^{(4)}(\omega)$ in the limit $\omega $\,$\to$\,$ \omega_0$. 
Thus for the quantum oscillator model (\ref{H-OR})$-$(\ref{H-Oexternal}), the relaxation TC method (or the TCE method) in the lowest Born approximation for the boson-reservoir interaction gives the exact admittance, and in that approximation the TCLE method gives the form similar to the exact one, while the relaxation TCL method gives the form different from the exact one in the higher-order terms. \\
\quad \\
(2) {\it A quantum spin model} \\

We next consider the model of a quantum spin system of magnitude $S$\,=\,1/2 under an external static magnetic field $\vec{H}_0$ in the $z$ direction, interacting with a heat reservoir which is composed of many quantum oscillators. 
We study the susceptibility of the system for an external driving magnetic field. 
We adopt the following Hamiltonians of the quantum spin system and heat reservoir: 
\begin{align}
    & \H_\SS = - \, \hbar \, \omega_0 \, S_z \,, \qquad \qquad \H_\R = \sum_\alpha \hbar \, \omega_\alpha \, b_\alpha^\dagger b_\alpha \,, \quad \label{H-SR}\\
    & \H_\SR = - \, \hbar \, S_z \, ( R + R^\dagger ), \qquad \quad (R = \!\sum_\alpha g_\alpha b_\alpha) \label{H-SRint}
\end{align}
where $b_\alpha$ represents the boson operator of the mode $\alpha$ which composes the heat reservoir, 
$\omega_\alpha$ is the characteristic frequency of each oscillator, 
and $g_\alpha$ is the coupling constant between the spin system and the boson of the mode $\alpha$. 
Here, $\omega_0$ is the Zeeman frequency $ \omega_0 $\,=\,$ \gamma H_0 $ with the magnetomechanical ratio $\gamma$. 
The interaction of the quantum spin system with the external driving magnetic field is given by
\begin{equation}
\H_\ed(t) = - \, \hbar \, S_+ \, \gamma \, \frac{H}{2} \, e^{i \, \omega \, t} - \hbar \, S_- \, \gamma \, \frac{H}{2} \, e^{- \, i \, \omega \, t}, \label{H-Sexternal}
\end{equation}
where $H$ is the magnitude of the driving magnetic field. 
Then, $\H_\SR(t)$ defined by the interaction representation (\ref{H-SR-t}) takes the form
\begin{equation}
\H_\SR(t) = - \, \hbar \, S_z \sum_\alpha \big( g_\alpha \, b_\alpha \, e^{- \, i \, \omega_\alpha \, t} + g_\alpha^* \, b_\alpha^\dagger \, e^{i \, \omega_\alpha \, t} \big). \label{Hspin-SR-t}
\end{equation}
In the study of this spin model, we neglect the effect of initial correlation between the spin and reservoir for brevity, 
and adopt the initial state $\rho_\TE$ in the decoupled form
\begin{equation}
\rho_\TE = \rho_0 \, \rho_\R \,, \qquad
\end{equation}
with $\rho_0$ given by (\ref{rho}).

Applying the relaxation TCL method to the above spin model, we obtain
\begin{equation}
\tr \, S_+ \, C^{(2)}(t) \, \tilde{a}_-(t) = - \, \psi(0 , t) \, \tr \, S_+ \, \tilde{a}_-(t), \quad \label{S+C(2)(t)a(t)}
\end{equation}
where $\phi(0 , t)$ is defined by
\begin{subequations}
\label{phi-0-t}
\begin{align}
\psi(0 , t) & = \frac{1}{2} \int_0^t d\tau \, \big( \langle \, [\, R(\tau) \,,\, R^\dagger \,]_+ \, \rangle_\R + \langle \, [\, R^\dagger(\tau) \,,\, R \,]_+ \, \rangle_\R \big), \label{psi-0-t-1}\\
    & = \frac{1}{2} \sum_\alpha |\, g_\alpha \,|^2 \, (2 \, \bar{n}_\alpha + 1) \int_0^t d\tau \, \big( e^{i \, \omega_\alpha \tau} + e^{- i \, \omega_\alpha \tau} \big), \label{psi-0-t-2}
\end{align}
\end{subequations}
with $\bar{n}_\alpha$ defined by 
\begin{equation}
\bar{n}_\alpha = 1 / \{\exp(\beta \, \hbar \, \omega_\alpha) - 1\}. \qquad \label{bar-n}
\end{equation}
Here, we have used the relations $S_z S_+$\,=\,$S_+/2$ and $S_+ S_z$\,=\,$- S_+/2$ for spin $S$\,=\,1/2. In the lowest Born approximation for the spin-reservoir interaction $\H_\SR$, we have the TCL equation
\begin{equation}
\frac{d}{dt} \, \tr \, S_+ \, \tilde{a}_-(t) = - \, i \, \omega_0 \, \tr \, S_+ \, \tilde{a}_-(t) - \psi(0 , t) \, \tr \, S_+ \, \tilde{a}_-(t),
\label{RTCL-bbd}
\end{equation}
which leads the admittance 
\begin{align}
& \chi_{+ -}^{\R\TCL}(\omega) = \frac{i}{\hbar} \int_0^\infty dt \, \tr \, \hbar \, S_+ \, \tilde{a}_-(t) \, e^{i \, \bar{\omega} \, t}, \nonumber \\
    & = 2 \, i \, \hbar \, \langle S_z \rangle_0 \int_0^\infty \! dt \exp\Big\{ i \, (\bar{\omega} - \omega_0) \, t - \! \int_0^t \! d\tau \, \psi(0 , \tau) \Big\},
\label{chi-RTCL-S+S-}
\end{align}
with $ \bar{\omega}$\,=\,$ \omega $\,+\,$ i \, \epsilon $ \,($ \epsilon $$ \to $+0), where $\langle S_z \rangle_0$ is given by $ \langle S_z \rangle_0 $\,=\,$ \tr \, S_z \rho_0$. 
It should be noted that in the present spin model, the sum of the higher-order ordered cumulants vanishes, as shown in Appendix G. 
Therefore, the admittance obtained by using the relaxation TCL method in the lowest Born approximatin for the spin-reservoir interaction, 
is equal to the exact one except for the term of initial correlation between the spin and reservoir, i.e., 
\begin{equation}
\chi_{+ -}(\omega) = \chi_{+ -}^{\R\TCL}(\omega).
\label{chi-S+S-}
\end{equation}

Applying the relaxation TC method (or the TCE method) for the present spin model, we obtain
\begin{equation}
\tr \, S_+ \, \bar{C}^{(2)}[\omega] \, \tilde{a}_- [\omega] = - \, \psi(\bar{\omega} - \omega_0) \, \tr \, S_+ \, \tilde{a}_- [\omega] + \HOT,
\end{equation}
with $ \bar{\omega}$\,=\,$ \omega $\,+\,$ i \, \epsilon $ \,($ \epsilon $$ \to $+0), 
where $\psi(\omega)$ is defined by
\begin{subequations}
\label{psi-omega}
\begin{align}
\psi(\bar{\omega}) & = \frac{1}{2} \int_0^\infty \! d\tau \, \big( \big\langle [\, R(\tau)\,,\, R^\dagger \,]_+ \big\rangle_\R + \big\langle [\, R^\dagger(\tau) \,,\, R \,]_+ \big\rangle_\R \big) \nonumber \\
    & \qquad \qquad \qquad \times e^{i \, \bar{\omega} \, \tau}, \label{psi-omega-1}\\
    & = \frac{i}{2} \sum_\alpha |\, g_\alpha \,|^2 \, (2 \, \bar{n}_\alpha + 1) \, \Big\{ \frac{1}{\bar{\omega} + \omega_\alpha} + \frac{1}{\bar{\omega} - \omega_\alpha} \Big\}. \label{psi-omega-2}
\end{align}
\end{subequations}
In the lowest Born approximation for the spin-reservoir interaction, we have the admittance
\begin{equation}
\chi_{+-}^{\R\TC}(\omega) = \frac{i}{\hbar} \, \tr \, \hbar \, S_+ \, \tilde{a}_-[\omega] = \frac{2 \, i \, \hbar \, \langle S_z \rangle_0}{i \, (\omega_0 - \bar{\omega}) + \psi(\bar{\omega} - \omega_0)},
\label{chi-RTC-S+S-}
\end{equation}
which takes the form different from (\ref{chi-RTCL-S+S-}).

Applying the TCLE method to the spin model under consideration, we obtain
\begin{align}
    & \tr \, S_+ \, C^{(2)} \, \rho_1[\omega] = - \, \psi(\bar{0}) \, \tr \, S_+ \rho_1[\omega] \,, \label{S+C(2)rho1} \\
    & \tr \, S_+ \, D_-^{(2)}[\omega] = 2 \, \langle S_z \rangle_0 \, \big( \psi(\bar{\omega} - \omega_0) - \psi(\bar{0}) \big) / (\bar{\omega} - \omega_0),  \label{S+D(2)}
\end{align}
with $ \bar{\omega}$\,=\,$ \omega $\,+\,$ i \, \epsilon $ \,($ \epsilon $$ \to $+0), where $\psi(\bar{0})$ is given by
\begin{equation}
\psi(\bar{0}) = \pi \sum_\alpha |\, g_\alpha \,|^2 \, (2 \, \bar{n}_\alpha + 1) \, \delta(\omega_\alpha). \ \label{psi-0}
\end{equation}
In the lowest Born approximation for the spin-reservoir interaction, we have the admittance
\begin{equation}
\chi_{+-}^{\TCLE}(\omega) = 2 \, \hbar \, \langle S_z \rangle_0 \, \frac{i + \Psi(\bar{\omega} - \omega_0)}{i \, (\omega_0 - \bar{\omega}) + \psi(\bar{0})},
\label{chi-TCLE-S+S-}
\end{equation}
which is different from (\ref{chi-RTCL-S+S-}), where $\Phi(\bar{\omega} - \omega_0)$ comes from the interference term and is given by
\begin{equation}
\Psi(\bar{\omega} - \omega_0) = ( \psi(\bar{\omega} - \omega_0) - \psi(\bar{0}) ) / (\bar{\omega} - \omega_0).
\end{equation}

Thus for the quantum spin model (\ref{H-SR})$-$(\ref{H-Sexternal}), 
the admittance (\ref{chi-RTCL-S+S-}) obtained by using the relaxation TCL method in the lowest Born approximation for the spin-reservoir interaction, is equal to the exact one except for the term of initial correlation between the spin and reservoir. 
On the other hand, in the lowest Born approximation the admittance (\ref{chi-RTC-S+S-}) obtained by using the relaxation TC method (or the TCE method) 
and the admittance (\ref{chi-TCLE-S+S-}) obtained by using the TCLE method have the forms different from the exact one. 

Let us compare these admittances analytically. 
The second-order terms of these admittances in powers of the spin-reservoir interaction coincide and are equal to
\begin{align}
\chi_{+-}^{(2)}(\omega) & = 2 \, i \, \hbar \, \langle S_z \rangle_0 \, \psi(\bar{\omega} - \omega_0) / (\bar{\omega} - \omega_0)^2 , \nonumber \\
    & = - \sum_\alpha \frac{ 2 \, \hbar \, \langle S_z \rangle_0 \, |\, g_\alpha \,|^2 \, (2 \, \bar{n}_\alpha + 1)}{\, (\bar{\omega} - \omega_0) \,\{\, (\bar{\omega} - \omega_0)^2 - \omega_\alpha^2 \,\} \,}. \label{chi-S+S-(2)}
\end{align}
The fourth-order terms of those admittances are obtained, by expanding (\ref{chi-RTCL-S+S-}), (\ref{chi-RTC-S+S-}) and (\ref{chi-TCLE-S+S-}) in powers of $\psi$, as
\begin{align}
& \chi_{+ -}^{\R\TCL(4)}(\omega) = i \, \hbar \, \langle S_z \rangle_0 \int_0^\infty dt \int_0^t d\tau_1 \int_0^t d\tau_2 \nonumber \\
  & \qquad \qquad \qquad \ \times \psi(0, \tau_1) \, \psi(0, \tau_2) \exp\{ i \, (\bar{\omega} - \omega_0) \, t \}, \nonumber \\
    & = \sum_{\alpha ,\, \alpha'} \frac{ - \, \hbar \, \langle S_z \rangle_0 \, |\, g_\alpha \,|^2 \!\cdot\! |\, g_{\alpha'} \,|^2 \, (2 \, \bar{n}_\alpha + 1) (2 \, \bar{n}_{\alpha'} + 1) }{ (\bar{\omega} - \omega_0) \{ (\bar{\omega} - \omega_0)^2 - \omega_\alpha^2 \} \{ (\bar{\omega} - \omega_0)^2 - \omega_{\alpha'}^2 \} } \nonumber \\
  & \times \frac{6 \, (\bar{\omega} - \omega_0)^4 - 3 \, (\bar{\omega} - \omega_0)^2 \, (\omega_\alpha^2 + \omega_{\alpha'}^2) + (\omega_\alpha^2 - \omega_{\alpha'}^2)^2}{ \{(\bar{\omega} - \omega_0)^2 - (\omega_\alpha + \omega_{\alpha'})^2 \} \{(\bar{\omega} - \omega_0)^2 - (\omega_\alpha - \omega_{\alpha'})^2 \}},
\label{chi-RTCL-S+S-(4)} \\
& \chi_{+-}^{\R\TC(4)}(\omega) = 2 \, \hbar \, \langle S_z \rangle_0 \, \psi(\bar{\omega} - \omega_0)^2 / (\bar{\omega} - \omega_0)^3 , \nonumber \\
    & \quad = \sum_{\alpha ,\, \alpha'} \frac{ - \, 2 \, \hbar \, \langle S_z \rangle_0 \, |\, g_\alpha \,|^2 \!\cdot\! |\, g_{\alpha'} \,|^2 \, (2 \, \bar{n}_\alpha + 1) (2 \, \bar{n}_{\alpha'} + 1) }{ (\bar{\omega} - \omega_0) \{ (\bar{\omega} - \omega_0)^2 - \omega_\alpha^2 \} \{ (\bar{\omega} - \omega_0)^2 - \omega_{\alpha'}^2 \} },
\label{chi-RTC-S+S-(4)} \\
    & \chi_{+-}^{\TCLE(4)}(\omega) = 2 \, \hbar \, \langle S_z \rangle_0 \, \psi(\bar{\omega} - \omega_0)\, \psi(\bar{0}) / (\bar{\omega} - \omega_0)^3 , \nonumber \\
    & \quad = \sum_{\alpha ,\, \alpha'} \frac{\, 2 \, i \, \pi \, \hbar \, \langle S_z \rangle_0 \, |\, g_\alpha \,|^2 \!\cdot\! |\, g_{\alpha'} \,|^2 \, (2 \, \bar{n}_\alpha + 1) (2 \, \bar{n}_{\alpha'} + 1) }{ (\bar{\omega} - \omega_0)^2 \{ (\bar{\omega} - \omega_0)^2 - \omega_\alpha^2 \} } \nonumber \\
    & \qquad \qquad \times \delta(\omega_{\alpha'}),
\label{chi-TCLE-S+S-(4)}
\end{align}
with $ \bar{\omega}$\,=\,$ \omega $\,+\,$ i \, \epsilon $ \,($ \epsilon $$ \to $+0). These fourth-order terms of admittances take the forms different from each other. 
They do not coincide even in the limit $\omega $\,$ \to $\,$ \omega_0$. 
For this quantum spin model, the relaxation TC method (or the TCE method) and the TCLE method in the lowest Born approximation for the spin-reservoir interaction give the forms of admittance different from the exact one in the higher-order terms.


For the quantum oscillator model (\ref{H-OR})$-$(\ref{H-Oexternal}), the relaxation TC method (or the TCE method) in the lowest Born approximation for the boson-reservoir interaction gives the exact admittance, and in that approximation the relaxation TCL method and the TCLE method give the forms of admittance different from the exact one in the higher-order terms. On the other hand, for the quantum spin model (\ref{H-SR})$-$(\ref{H-Sexternal}), the relaxation TCL method in the lowest Born approximation for the spin-reservoir interaction gives the exact admittance except for the term of initial correlation between the spin and reservoir, 
and in that approximation the relaxation TC method (or the TCE method) and the TCLE method give the forms of admittance different from the exact one in the higher-order terms. 
This means that accuracy of the admittances obtained by using the relaxation TC method (or the TCE method), 
the relaxation TCL method and the TCLE method in the lowest Born approximation for the system-reservoir interaction, 
depends on the model of physical system under consideration.
%
%
\section{Summary and concluding remarks}

We have derived the exact expressions (\ref{RTCadmit}), (\ref{RTCLadmit}), (\ref{TCEadmit}) and (\ref{TCLEadmit}) of the complex admittance for quantum system in contact with heat reservoir using the relaxation method and the external-field method, respectively, for the two types of equation of motion, which are the time-convolution (TC) equation and time-convolutionless (TCL) equation. We have shown that the expression of the complex admittance obtained using the relaxation method with the TC equation (the relaxation TC method), coincides exactly with that obtained using the external-field method with the TC equation (the TCE method), i.e., (\ref{RTC-TCE}). We have also shown that the three expressions of the complex admittance coincide with each other in the lowest Born approximation for the system-reservoir interaction, i.e., (\ref{Relations}), though the three expressions of the exact admittances are different from each other. We have besides given the expressions of the admittance in the $n$-th order approximation for the system-reservoir interaction, and have expressed the admittance in the fourth order approximation using time-integrals alone by transforming inverse-temperature-integrals into time-integrals. Furthermore, we have compared the admittances for the two exactly solvable models analytically and have found that for the quantum oscillator model (\ref{H-OR})$-$(\ref{H-Oexternal}), the relaxation TC method (or the TCE method) in the lowest Born approximation for the boson-reservoir interaction gives the exact admittance, that for the quantum spin model (\ref{H-SR})$-$(\ref{H-Sexternal}), the relaxation TCL method in the lowest Born approximation for the spin-reservoir interaction gives the exact admittance except for the term of initial correlation between the spin and reservoir, and thus that accuracy of the admittances depends on the model of physical system under consideration.

We here compare the admittances (\ref{RTC2admit}) [or (\ref{TCE2admit})], (\ref{RTCL2admit}) and (\ref{TCLE2admit}), which are, respectively, obtained using the relaxation TC method [or the TCE method], the relaxation TCL method and the TCLE method in the lowest Born approximation for the system-reservoir interaction, with the results in the conventional Markovian approximation \cite{Louisell,KTH,QME5,Redfield,STM2000} or in the van Hove limit \cite{vanHove}. 
In the conventional Markovian approximation, the motion of the system is determined in the van Hove limit \cite{vanHove} or in the narrowing limit \cite{Louisell,KTH} in which the heat reservoir is damped very rapidly, i.e., the reservoir correlation time $\tau_c$\,$\to$\,0. 
In this limit, the second-order TC equation (\ref{tilde-a-TC2}) and the second-order TCL equation (\ref{tilde-a-TCL2}) 
become \cite{Louisell,KTH,QME5,S4,S6}
\begin{equation}
(d/dt)\, \tilde{a}(t) = - \, i \, L_\SS \, \tilde{a}(t) + C^{(2)} \, \tilde{a}(t), 
\label{tilde-a-vanHove-eq}
\end{equation}
where $C^{(2)}$ is given by (\ref{C2-inf}). 
Substituting the formal solution of the above equation\,:
\begin{equation}
\tilde{a}(t) = \exp\{- \, i \, L_\SS \, t + C^{(2)} \, t \}\, \tilde{a}(0) \,\,\quad 
\label{tilde-a-vanHove}
\end{equation}
into the Kubo formula (\ref{KF}), the admittance takes the form \cite{S4,S5,S6}
\begin{equation}
\chi^{\v}_{ij}(\omega) = (i / \hbar) \, \tr \, A_i \big\{ i \, (L_\SS - \omega) - C^{(2)} \big\}^{- 1} \big[ A_j , \rho_\SS + \rho_0^{(2)} \big], 
\label{vlimit-admit}
\end{equation}
which is valid only in the van Hove limit \cite{vanHove} or in the narrowing limit \cite{Louisell,KTH,QME5,Redfield,STM2000}, where $\rho_0^{(2)}$ is given by (\ref{rho-0-2}) or (\ref{rho0-t-2}). 
Comparing the above admittance (\ref{vlimit-admit}) with the admittance (\ref{TCLE2admit}) obtained using the TCLE method in the lowest Born approximation for the system-reservoir interaction, the latter admittance includes the interference term $D_j^{(2)}[\omega]$ which is not included in the former admittance and represents the effects of the initial correlation and memory for the system and reservoir. 
The effects of the initial correlation and memory are the effects of the deviation from the van Hove limit \cite{vanHove} or the narrowing limit, because the effects of initial correlation can be neglected in that limit as shown in Ref. \cite{S5} and the memory effects are the effects of collision of the system with the heat reservoir. 
These effects are neglected in the conventional Markovian approximation \cite{Louisell,KTH,QME5,Redfield,STM2000}. 
As examined in Section IV, the admittance obtained using the relaxation TC method [or the TCE method], the admittance obtained using the relaxation TCL method and the admittance obtained using the TCLE method, coincide with each other in the lowest Born approximation for the system-reservoir interaction. 
Therefore, the admittances (\ref{RTC2admit}) [or (\ref{TCE2admit})], (\ref{RTCL2admit}) and (\ref{TCLE2admit}) include the effects of the initial correlation and memory for the system and reservoir.

The relaxation method, in which the Kubo formula is calculated for systems with no external driving fields, is essentially equal with the external-field method in which the admittance is directly calculated from equations of motion with external driving terms. 
Because, when the TC equation of motion is used to calculate the admittance, the expression of the complex admittance obtained using the relaxation TC method, coincides exactly with that obtained using the TCE method (the external-field method with the TC equation). 
When the TCL equation of motion is used to calculate the admittance, the two expressions of the complex admittance obtained using the relaxation TCL method and TCLE method (the external-field method with the TCL equation) are different to each other, and also they are different from the expression of the admittance obtained using the relaxation TC method (or the TCE method). 
The difference of the expressions occurs by renormalizing the memory terms in the process of derivation of the TCL equations.

In Section IV, we have examined analytically the relations among the admittances (\ref{RTC2admit}) [or (\ref{TCE2admit})], (\ref{RTCL2admit}) and (\ref{TCLE2admit}), which are obtained using the relaxation TC method [or the TCE method], the relaxation TCL method and the TCLE method, respectively, in the lowest Born approximation for the system-reservoir interaction. 
The fact that the three expressions (\ref{RTC2admit}) [or (\ref{TCE2admit})], (\ref{RTCL2admit}) and (\ref{TCLE2admit}) of the complex admittance coincide with each other in the lowest Born approximation for the system-reservoir interaction, means that the three expressions of the admittance are respectively valid in that approximation. 
The differences among the three expressions come from the fact that the high-order terms differ from each other. 
These three expressions include the high-order terms that are dominant in the resonance region. 
Even if the expression of the complex admittance obtained in the lowest Born approximation for the system-reservoir interaction includes many more high-order terms, it is not always the more exact expression. 
Because, that expression of the complex admittance does not include part of the third-order, fourth-order and high-order terms in powers of the system-reservoir interaction. 
It is also possible to derive the admittance (\ref{RTCL2admit}) obtained using the relaxation TCL method from the admittance (\ref{TCLE2admit}) obtained using the TCLE method by proceeding in the same way as in Section IV. 
Moreover, it should be noticed that as mentioned in Section III, the admittance (\ref{RTC2admit}) obtained using the relaxation TC method by the lowest Born approximation of the equation (\ref{tilde-a-TC}) including no external driving terms, coincides with the admittance (\ref{TCE2admit}) obtained using the TCE method by the lowest Born approximation of the equation (\ref{rhoTC}) or (\ref{rho1TCeq}) including external driving terms. Thus in the relaxation TC method and TCE method, although the equations in which the perturbation is truncated are different from each other, the expressions of the obtained admittances coincides with each other.

The three expressions (\ref{RTC2admit}) [or (\ref{TCE2admit})], (\ref{RTCL2admit}) and (\ref{TCLE2admit}) of the complex admittance are respectively valid in the lowest Born approximation for the system-reservoir interaction as mentioned in the above paragraph, but the usability is different from each other. 
The expression (\ref{RTCL2admit}), which is obtained using the relaxation TCL method, 
takes the form that is not easy to use, because it is difficult in general to calculate the time integral of the ordered exponential for many-body systems or for complicated systems. 
The expression (\ref{TCLE2admit}), which is obtained using the TCLE method, 
has been formulated in terms of thermo-field dynamics (TFD) using the method of nonequilibrium thermo-field dynamics proposed by Arimitsu and Umezawa \cite{Arimitsu-Umezawa} in order to study the linear response of many-body systems interacting with heat reservoir \cite{S6,S7,S8}. It has been applied to an weakly-interacting boson system \cite{S9} and to a ferromagnetic system interacting with a phonon reservoir in the spin-wave approximation \cite{S10}, by using the generalized thermo-field dynamics \cite{S6,S7,S8}. 
Recently, fluctuation-dissipation theorem for the TCLE method has been examined in the lowest Born approximation for the system-reservoir interaction \cite{S8,S9}. The TCLE method is a method for improving the difficulty of the relaxation TCL method. 
The expression (\ref{RTC2admit}) [or (\ref{TCE2admit})], which is obtained using the relaxation TC method [or the TCE method], can be applied to a interacting spin system by making a unfied numerical method \cite{Uchiyama}. 
In practical applications, the relaxation TC method, the TCE method and the TCLE method may be the useful methods to study the linear response of many-body systems interacting with heat reservoir.

We have derived the formulae necessary for the higher order expansions in powers of the system-reservoir interaction in Section V. 
When we proceed to the high order in powers of the system-reservoir interaction, the formulae derived in Section V will be useful. 
We have there given the expressions of the admittance in the $n$-th order approximation, and have also given the forms of the admittances in the fourth-order approximation in the expression using time-integrals alone by transforming inverse-temperature-integrals into time-integrals. 
This facilitates the calculations of the admittances, because when the time-correlation functions of the heat reservoir are given, the admittances can be calculated. 

As examined in Section VI for the two exactly solvable models analytically, the relaxation TC method (or the TCE method) in the lowest Born approximation in powers of the system-reservoir interaction gives the exact admittance for the quantum oscillator model (\ref{H-OR})$-$(\ref{H-Oexternal}), and the relaxation TCL method in the lowest Born approximation in powers of the system-reservoir interaction gives the exact admittance for the quantum spin model (\ref{H-SR})$-$(\ref{H-Sexternal}) except for the term of initial correlation between the spin and reservoir. 
This means that accuracy of the admittances obtained using the relaxation TC method (or the TCE method), the relaxation TCL method and the TCLE method in the lowest Born approximation for the system-reservoir interaction, depends on the model of physical system under consideration. For the case of a non-adiabatic interaction as (\ref{H-ORint}), in which the Hamiltonian $\H_\SS$ of the quantum system is non-commutable to the system-reservoir interaction $\H_\SR$, the relaxation TC method (or the TCE method) may give a better result. For the case of an adiabatic interaction as (\ref{H-SRint}), in which the Hamiltonian $\H_\SS$ of the quantum system is commutable to the system-reservoir interaction $\H_\SR$, the relaxation TCL method may give a better result. Accuracy of the three admittances is an interesting subject of future study.
\begin{acknowledgments}
The authors are very grateful to the Miyashita research group for stimulating and valuable discussions. The present work was supported by Grant-in-Aid for Scientific Research on Priority Areas, and also the Next Generation Super Computer Project, Nanoscience Program from MEXT of Japan.
\end{acknowledgments}
\begin{widetext}
\appendix
\renewcommand{\thesection}{\Alph{section}}
\section{Transformation of the $\beta'$-integral into a time-integral}

We show that the expression (\ref{bar-D2b-omega}) for the interference term $\bar{D}^{(2)}[\omega]$ in the TC equation, which is equal to the first term of the interference term $D^{(2)}[\omega]$ given by (\ref{D2-omega}), coincides with the expression (\ref{bar-D2t-omega}). The inverse-temperature-integral ($\beta'$-integral) in the expression (\ref{bar-D2b-omega}) can be integrated and can be rewritten as follows,
\begin{align}
& \int_0^\beta d\beta' \exp(- \, i \, L_0 \, \tau) L_\ed[\omega] \, \rho_\SS \, \rho_\R \, \H_\SR(- \, i \, \hbar \, \beta' ) \, e^{i \, \omega \, \tau} = \int_0^\beta d\beta' \exp(- \, i \, L_0 \, \tau) L_\ed[\omega] \, \rho_\SS \, \rho_\R \exp(\beta' \, \hbar \, L_0)\, \H_\SR \, e^{i \, \omega \, \tau}, \ \\
    & = \exp(- \, i \, L_0 \, \tau) \, L_\ed[\omega] \, \rho_\SS \, \rho_\R \, \frac{\, \exp(\beta \, \hbar \, L_0) - 1 \,}{ \hbar \, L_0 } \, \H_\SR \, e^{i \, \omega \, \tau} = \exp(- \, i \, L_0 \, \tau) \, L_\ed[\omega] \, \frac{1}{L_0} \, L_\SR \, \rho_\SS \, \rho_\R \, e^{i \, \omega \, \tau}, \ 
\label{A-3}
\end{align}
which can be expressed using a time-integrals as
\begin{equation}
\int_0^\beta d\beta' \, e^{- \, i \, L_0 \, \tau} L_\ed[\omega]\, \rho_\SS \, \rho_\R \, \H_\SR(- \, i \, \hbar \, \beta' ) \, e^{i \, \omega \, \tau} = \lim_{\epsilon \, \to \, +0} \, i \int_\tau^\infty ds \, e^{- \, i \, L_0 \, \tau} \, L_\ed[\omega] \, e^{i \, L_0 \, (\tau \, - \, s)} \, L_\SR \, \rho_\SS \, \rho_\R \, e^{i \, \omega \, \tau \, - \, \epsilon \, s}. \ 
\label{A-3}
\end{equation}
Using the integral transformation
\begin{equation}
\int_0^\infty d\tau \int_\tau^\infty ds = \int_0^\infty ds \int_0^s d\tau = \int_0^\infty d\tau \int_0^\tau ds \, \big(\, \tau \! \iff \! s \,\big), \qquad \qquad
\label{integral-trans}
\end{equation}
the expression (\ref{bar-D2b-omega}) for $\bar{D}^{(2)}[\omega]$ coincides with the expression (\ref{bar-D2t-omega}). Similarly, the expression (\ref{barDj2b-omega}) for $\bar{D}_j^{(2)}[\omega]$, which is equal to the first term of $D_j^{(2)}[\omega]$ given by (\ref{Dj2-omega}), can be shown to coincide with the expression (\ref{barDj2t-omega}).
%
\section{Derivation of the interference term \emph{$D[\omega]$}}

The interference term $D(t)$ given by (\ref{DTCL}), in the limit $t$\,$ \to $\,$ \infty $, becomes
\begin{align}
D(t) = & \int_0^t d\tau \, \tr_\R \, L_\SR \, \frac{1}{1- \Sigma(t)} \, e^{- \, i \, \Q \, L \, \Q \,\tau} \Big\{- L_\ed(t - \tau) \, \Q \, \rho_\TE + \Sigma(t - \tau) \, L_\ed(t - \tau) \, \rho_\TE \, \Big\} \, \Big|_{t \to \infty}\,, \qquad \\
    = & - \int_0^t d\tau \sum_\omega \tr_\R \, L_\SR\, \frac{1}{1- \Sigma(t)} \, e^{- \, i \, \Q \, L \, \Q \,\tau} \, L_\ed[\omega]\, \Q \, \rho_\TE \, e^{- \, i \, \omega \, (t \, - \, \tau)} \Big|_{t \to \infty} \nonumber \\
    & - i \int_0^t d\tau \int_\tau^t ds \sum_\omega \tr_\R\, L_\SR\, \frac{1}{1 - \Sigma(t)} \, e^{- \, i \, \Q \, L \, \Q \, \tau} \, e^{- \, i \, \Q \, L \, \Q \, (s \, - \, \tau)} \Q \, L_\SR \, \P \, e^{ i \, L \, (s \, - \, \tau)} \, L_\ed[\omega] \, \rho_\TE \, e^{- \, i \, \omega \, (t \, - \, \tau)} \Big|_{t \to \infty},
\end{align}
where we have substituted (\ref{edf}) and $\Sigma(t - \tau)$, which is given by
\begin{equation}
\Sigma(t - \tau) = - \, i \int_\tau^t ds \exp\{- \, i \, \Q \, L \, \Q \, (s - \tau)\} \, \Q \, L_\SR \, \P \exp\{i \, L \, (s - \tau)\}. \qquad
\end{equation}
Then, the interference term $D[\omega]$ takes, according to the definition (\ref{D}), the form
\begin{align}
D[\omega] = & - \int_0^\infty d\tau \, \tr_\R \, L_\SR \, \frac{1}{\, 1- \Sigma \,} \exp(- \, i \, \Q \, L \, \Q \, \tau) \, L_\ed[\omega] \, \Q \, \rho_\TE \exp(i \, \omega \, \tau) \nonumber \\
    & - i \int_0^\infty d\tau \int_\tau^\infty ds \, \tr_\R\, L_\SR\, \frac{1}{\, 1 - \Sigma \,} \exp(- \, i \, \Q \, L \, \Q \, s) \, \Q \, L_\SR \, \P \exp\{i \, L \, (s - \tau) \} L_\ed[\omega] \, \rho_\TE \exp(i \, \omega \, \tau), \ 
\label{B-4}
\end{align}
which becomes equal the expression (\ref{D-omega}) by the integral transformation (\ref{integral-trans}), where $\Sigma$ is given by (\ref{Sigma}).
%
\section{Transformation of the $\beta_1$ and $\beta_2$ integrals into two time-integrals}

The $\beta_1$ and $\beta_2$ integrals in Subsection V.C can be integrated and can be rewritten as follows,
\begin{align}
    & \int_0^\beta d \beta_1 \int_0^{\beta_1} d \beta_2 \, e^{- \, i \, L_0 \, \tau} \, \big[ A \,,\, \rho_\SS \, \rho_\R \, \H_\SR(- \, i \, \hbar \, \beta_1) \, \H_\SR(- \, i \, \hbar \, \beta_2) \, \big],
\label{C-1} \\
    & = \int_0^\beta d \beta_1 \int_0^{\beta_1} d \beta_2 \, e^{- \, i \, L_0 \, \tau} \, \big[ A \,,\, \rho_\SS \, \rho_\R \, \big( e^{\beta_1 \, \hbar \, L_0}\, \H_\SR \big) \big( e^{\beta_2 \, \hbar \, L_0}\, \H_\SR \big) \big], \\
    & = \int_0^\beta d \beta_1 \, e^{- \, i \, L_0 \, \tau} \Big[ A \,,\, \rho_\SS \, \rho_\R \, \big\{\, e^{\beta_1 \, \hbar \, L_0}\, \H_\SR \, \frac{1}{\hbar \, L_0}\, \H_\SR - \big( e^{\beta_1 \, \hbar \, L_0}\, \H_\SR \big) \frac{1}{\hbar \, L_0}\, \H_\SR \,\big\} \Big], \\
    & = e^{- \, i \, L_0 \, \tau} \Big[ A \,,\, \rho_\SS \, \rho_\R \, \frac{\, e^{\beta \, \hbar \, L_0} - 1 \,}{\hbar \, L_0} \, \H_\SR \, \frac{1}{\, \hbar \, L_0}\, \H_\SR - \rho_\SS \, \rho_\R \Big( \frac{\, e^{\beta \, \hbar \, L_0} - 1 \,}{\hbar \, L_0}\, \H_\SR \Big) \frac{1}{\, \hbar \, L_0}\, \H_\SR \, \Big], \qquad \\
    & = e^{- \, i \, L_0 \, \tau} \Big[ A \,,\, \frac{1}{\hbar \, L_0}\, \big[\, \H_\SR \, \frac{1}{\hbar \, L_0}\, \H_\SR \,,\, \rho_\SS \, \rho_\R \,\big] - \big( \frac{1}{\hbar \, L_0}\, \big[\, \H_\SR \,,\, \rho_\SS \, \rho_\R \,\big] \big)\, \frac{1}{\hbar \, L_0}\, \H_\SR \, \Big],
\end{align}
which can be expressed using the two time-integrals as
\begin{align}
    & \int_0^\beta d \beta_1 \int_0^{\beta_1} d \beta_2 \, e^{- \, i \, L_0 \, \tau} \, \big[ A \,,\, \rho_\SS \, \rho_\R \, \H_\SR(- \, i \, \hbar \, \beta_1) \, \H_\SR(- \, i \, \hbar \, \beta_2) \, \big], \nonumber \\
    & = \frac{i^2}{\hbar^2} \, \Big\{ \int_\tau^\infty d\tau_1 \int_{\tau_1}^\infty d\tau_2 \, e^{- \, i \, L_0 \, \tau} \, \big[ A \,,\, e^{- \, i \, L_0 \, (\tau_1 \, - \, \tau) }\, \big[\, \H_\SR \, e^{- \, i \, L_0 \, (\tau_2 \, - \, \tau_1) }\, \H_\SR \,,\, \rho_\SS \, \rho_\R \,\big] \, \big] \nonumber \\
    & \qquad \ - \int_\tau^\infty d\tau_1 \int_\tau^\infty d\tau_2 \, e^{- \, i \, L_0 \, \tau} \, \big[ A \,,\, \big( e^{- \, i \, L_0 \, (\tau_1 \, - \, \tau) }\, [\, \H_\SR \,,\, \rho_\SS \, \rho_\R \,] \, \big)\, e^{- \, i \, L_0 \, (\tau_2 \, - \, \tau) }\, \H_\SR \, \big] \Big\} \exp(- \, \epsilon \, \tau_1 - \epsilon \, \tau_2 ) \big|_{\epsilon \to +0}\,, \\
    & = \frac{i^2}{\hbar^2} \, \Big\{ \int_\tau^\infty d\tau_1 \int_{\tau_1}^\infty d\tau_2 \, \big[ A(- \, \tau) \,, \big[\, \H_\SR(- \, \tau_1) \, \H_\SR(- \, \tau_2) \,,\, \rho_\SS \, \rho_\R \, \big] \, \big] \nonumber \\
    & \qquad \ - \int_\tau^\infty d\tau_1 \int_\tau^\infty d\tau_2 \, \big[ A(- \, \tau) \,, \big[\, \H_\SR(- \, \tau_1) \,,\, \rho_\SS \, \rho_\R \,\big] \, \H_\SR(- \, \tau_2) \, \big] \Big\} \exp(- \, \epsilon \, \tau_1 - \epsilon \, \tau_2 ) \big|_{\epsilon \to +0}\,,
\label{C-7} \\
    & = \frac{i^2}{\hbar^2} \int_\tau^\infty d\tau_1 \int_\tau^{\tau_1} d\tau_2 \, \big[ A(- \, \tau) \,,\, \H_\SR(- \, \tau_2) \big[\, \H_\SR(- \, \tau_1) \,,\, \rho_\SS \, \rho_\R \, \big] - [\, \H_\SR(- \, \tau_1) \,,\, \rho_\SS \, \rho_\R \,] \, \H_\SR(- \, \tau_2) \, \big] \, \big] \exp(- \, \epsilon \, \tau_1 ) \big|_{\epsilon \to +0}\,, \nonumber \\
    & = \frac{i^2}{\hbar^2} \int_\tau^\infty d\tau_1 \int_\tau^{\tau_1} d\tau_2 \, \big[ A(- \, \tau) \,, \big[\, \H_\SR(- \, \tau_2) \,,\, \big[\, \H_\SR(- \, \tau_1) \,,\, \rho_\SS \, \rho_\R \, \big] \, \big] \, \big] \exp(- \, \epsilon \, \tau_1 ) \big|_{\epsilon \to +0}\,, \\
    & = i^2 \int_\tau^\infty d\tau_1 \int_\tau^{\tau_1} d\tau_2 \, \big[ A(- \, \tau) \,,\, L_\SR(- \, \tau_2) \, L_\SR(- \, \tau_1) \, \rho_\SS \, \rho_\R \, \big] \exp(- \, \epsilon \, \tau_1 ) \big|_{\epsilon \to +0}\,.
\label{C-9}
\end{align}
\section{Transformation of the $\beta_1$, $\beta_2$ and $\beta_3$ integrals into three time-integrals}

The $\beta_1$, $\beta_2$ and $\beta_3$ integrals in Subsection V.C can be integrated and can be rewritten as follows,
\begin{align}
    & \int_0^\beta d \beta_1 \int_0^{\beta_1} d \beta_2 \int_0^{\beta_2} d \beta_3 \, e^{- \, i \, L_0 \, \tau} \big[ A \,,\, \rho_\SS \, \rho_\R \, \H_\SR(- \, i \, \hbar \, \beta_1) \, \H_\SR(- \, i \, \hbar \, \beta_2) \, \H_\SR(- \, i \, \hbar \, \beta_3) \, \big], 
\label{D-1} \\
    & = \int_0^\beta d \beta_1 \, e^{- \, i \, L_0 \, \tau} \, \Big[ A \,,\, \rho_\SS \, \rho_\R \, \big( e^{\beta_1 \, \hbar \, L_0}\, \H_\SR \big) \, \Big\{ \frac{\, e^{\beta_1 \, \hbar \, L_0} - 1 \,}{\hbar \, L_0}\, \H_\SR \, \frac{1}{\, \hbar \, L_0}\, \H_\SR - \big( \frac{\, e^{\beta_1 \, \hbar \, L_0} - 1 \,}{\hbar \, L_0}\, \H_\SR \big) \frac{1}{\, \hbar \, L_0}\, \H_\SR \Big\} \Big], \\
    & = \frac{1}{\hbar^3} \, e^{- \, i \, L_0 \, \tau} \, \Big[ A \,,\, \rho_\SS \, \rho_\R \Big\{ \frac{1}{L_0} \, \big( e^{\beta \, \hbar \, L_0} - 1 \big) \, \H_\SR \, \frac{1}{L_0} \, \H_\SR \, \frac{1}{L_0} \, \H_\SR - \Big( \frac{1}{L_0} \, \big( e^{\beta \, \hbar \, L_0} - 1 \big) \, \H_\SR \Big) \, \frac{1}{L_0} \, \H_\SR \, \frac{1}{L_0} \, \H_\SR \nonumber \\
    & \qquad \qquad \qquad \qquad \qquad - \Big(\frac{1}{L_0} \, \big( e^{\beta \, \hbar \, L_0} - 1 \big) \, \H_\SR \, \frac{1}{L_0} \, \H_\SR \Big) \, \frac{1}{L_0} \, \H_\SR + \Big( \frac{1}{L_0} \, \big( e^{\beta \, \hbar \, L_0} - 1 \big) \, \H_\SR \Big) \Big( \frac{1}{L_0} \, \H_\SR \Big) \, \frac{1}{L_0} \, \H_\SR \Big\} \Big], \\
    & = \frac{1}{\hbar^3} \, e^{- \, i \, L_0 \, \tau} \, \Big[ A \,,\, \frac{1}{L_0} \, \Big[\, \H_\SR \, \frac{1}{L_0} \, \H_\SR \, \frac{1}{L_0} \, \H_\SR \,,\, \rho_\SS \, \rho_\R \, \Big] - \Big( \frac{1}{L_0} \, \big[\, \H_\SR \,,\, \rho_\SS \, \rho_\R \, \big] \Big) \, \frac{1}{L_0} \, \H_\SR \, \frac{1}{L_0} \, \H_\SR \nonumber \\
    & \qquad \qquad \qquad \qquad - \Big( \frac{1}{L_0} \, \big[\, \H_\SR \, \frac{1}{L_0} \, \H_\SR \,,\, \rho_\SS \, \rho_\R \, \big] \Big) \, \frac{1}{L_0} \, \H_\SR + \Big( \frac{1}{L_0} \, \big[\, \H_\SR \,,\, \rho_\SS \, \rho_\R \, \big] \Big) \Big( \frac{1}{L_0} \, \H_\SR \Big) \, \frac{1}{L_0} \, \H_\SR \, \Big],
\end{align}
which can be expressed using the three time-integrals, by performing some integral transformations as (\ref{integral-trans}), (\ref{integral-trans-1}) and (\ref{integral-trans-2}), as
\begin{align}
    & \int_0^\beta d \beta_1 \int_0^{\beta_1} d \beta_2 \int_0^{\beta_2} d \beta_3 \, e^{- \, i \, L_0 \, \tau} \, \big[ A \,,\, \rho_\SS \, \rho_\R \, \H_\SR(- \, i \, \hbar \, \beta_1) \, \H_\SR(- \, i \, \hbar \, \beta_2) \, \H_\SR(- \, i \, \hbar \, \beta_3) \, \big], \nonumber \\
    & \! = \frac{i^3}{\hbar^3} \, \Big\{ \int_\tau^\infty \! d\tau_1 \int_{\tau_1}^\infty d\tau_2 \int_{\tau_2}^\infty d\tau_3 \, e^{- \, i \, L_0 \, \tau} \, \big[ A \,,\, e^{- \, i \, L_0 \, (\tau_1 \, - \, \tau) } \, \big[\, \H_\SR \, e^{- \, i \, L_0 \, (\tau_2 \, - \, \tau_1) } \, \H_\SR \, e^{- \, i \, L_0 \, (\tau_3 \, - \, \tau_2) } \, \H_\SR \,,\, \rho_\SS \, \rho_\R \, \big] \, \big] \nonumber \\
    & \qquad - \int_\tau^\infty d\tau_1 \int_\tau^\infty d\tau_2 \int_{\tau_2}^\infty d\tau_3 \, e^{- \, i \, L_0 \, \tau} \, \big[ A \,,\, \big( e^{- \, i \, L_0 \, (\tau_1 \, - \, \tau) }\, \big[ \, \H_\SR \,,\, \rho_\SS \, \rho_\R \, \big] \, \big)\, e^{- \, i \, L_0 \, (\tau_2 \, - \, \tau) }\, \H_\SR \, e^{- \, i \, L_0 \, (\tau_3 \, - \, \tau_2) }\, \H_\SR \, \big] \nonumber \\
    & \qquad - \int_\tau^\infty d\tau_1 \int_{\tau_1}^\infty d\tau_2 \int_{\tau}^\infty d\tau_3 \, e^{- \, i \, L_0 \, \tau} \, \big[ A \,,\, \big( e^{- \, i \, L_0 \, (\tau_1 \, - \, \tau) }\, \big[ \, \H_\SR \, e^{- \, i \, L_0 \, (\tau_2 \, - \, \tau_1) }\, \H_\SR \,,\, \rho_\SS \, \rho_\R \, \big] \, \big) \, e^{- \, i \, L_0 \, (\tau_3 \, - \, \tau) } \, \H_\SR \, \big] \nonumber \\
    & \qquad + \int_\tau^\infty d\tau_1 \int_\tau^\infty d\tau_2 \int_\tau^\infty d\tau_3 \, e^{- \, i \, L_0 \, \tau} \, \big[ A \,,\, \big( e^{- \, i \, L_0 \, (\tau_1 \, - \, \tau) }\, \big[ \, \H_\SR \,,\, \rho_\SS \, \rho_\R \, \big] \, \big) \, \big( e^{- \, i \, L_0 \, (\tau_2 \, - \, \tau) }\, \H_\SR \, \big)\, e^{- \, i \, L_0 \, (\tau_3 \, - \, \tau) }\, \H_\SR \, \big] \Big\} \nonumber \\
    & \qquad \qquad \qquad \qquad \qquad \qquad \qquad \qquad \qquad { } \times \exp(- \, \epsilon \, \tau_1 - \epsilon \, \tau_2 - \epsilon \, \tau_3 ) \big|_{\epsilon \to +0}\,, \\
    & \! = \frac{i^3}{\hbar^3} \, \Big\{ \int_\tau^\infty d\tau_1 \int_{\tau_1}^\infty d\tau_2 \int_{\tau_2}^\infty d\tau_3 \, \big[ A(- \, \tau) \,, \big[\, \H_\SR(- \, \tau_1) \, \H_\SR(- \, \tau_2) \, \H_\SR(- \, \tau_3) \,,\, \rho_\SS \, \rho_\R \, \big] \, \big] \nonumber \\
    & \qquad - \int_\tau^\infty d\tau_1 \int_\tau^\infty d\tau_2 \int_{\tau_2}^\infty d\tau_3 \, \big[ A(- \, \tau) \,, \big[\, \H_\SR(- \, \tau_1) \,,\, \rho_\SS \, \rho_\R \, \big] \, \H_\SR(- \, \tau_2) \, \H_\SR(- \, \tau_3) \, \big] \nonumber \\
    & \qquad - \int_\tau^\infty d\tau_1 \int_{\tau_1}^\infty d\tau_2 \int_{\tau}^\infty d\tau_3 \, \big[ A(- \, \tau) \,, \big[\, \H_\SR(- \, \tau_1) \, \H_\SR(- \, \tau_2) \,,\, \rho_\SS \, \rho_\R \,\big] \, \H_\SR(- \, \tau_3) \,\big] \nonumber \\
    & \qquad + \int_\tau^\infty d\tau_1 \int_\tau^\infty d\tau_2 \int_\tau^\infty d\tau_3 \, \big[ A(- \, \tau) \,, \big[\, \H_\SR(- \, \tau_1) \,,\, \rho_\SS \, \rho_\R \,\big] \, \H_\SR(- \, \tau_2) \, \H_\SR(- \, \tau_3) \,\big] \Big\}\, e^{- \, \epsilon \, \tau_1 - \epsilon \, \tau_2 - \epsilon \, \tau_3 } \big|_{\epsilon \to +0}\,, \label{D-6} \\
    & = \frac{i^3}{\hbar^3} \int_\tau^\infty d\tau_1 \int_\tau^{\tau_1} d\tau_2 \int_\tau^{\tau_2} d\tau_3 \, \big[ A(- \, \tau) \,, \big[\, \H_\SR(- \, \tau_3) \,, \big[\, \H_\SR(- \, \tau_2) \,, \big[\, \H_\SR(- \, \tau_1) \,,\, \rho_\SS \, \rho_\R \, \big] \, \big] \, \big] \, \big] \exp(- \, \epsilon \, \tau_1) \big|_{\epsilon \to +0}\,, \quad 
\label{D-7} \\
    & = \, i^3 \int_\tau^\infty d\tau_1 \int_\tau^{\tau_1} d\tau_2 \int_\tau^{\tau_2} d\tau_3 \, \big[ A(- \, \tau) \,,\, L_\SR(- \, \tau_3) \, L_\SR(- \, \tau_2) \, L_\SR(- \, \tau_1) \, \rho_\SS \, \rho_\R \, \big] \exp(- \, \epsilon \, \tau_1) \big|_{\epsilon \to +0}\,. \label{D-8}
\end{align}
\section{Transformation of the $\beta_1$, $\beta_2$, $\beta_3$ and $\beta_4$ integrals into four time-integrals}

The $\beta_1$, $\beta_2$, $\beta_3$ and $\beta_4$ integrals in the expression (\ref{rhoTE-4}) of $\rho_\TE^{(4)}$ can be integrated and can be expressed using the four time-integrals by proceeding in the same way as in Appendixes C and D as follows,
\begin{align}
    & \int_0^\beta d \beta_1 \int_0^{\beta_1} d \beta_2 \int_0^{\beta_2} d \beta_3 \int_0^{\beta_3} d \beta_4 \, \rho_\SS \, \rho_\R \, \H_\SR(- \, i \, \hbar \, \beta_1) \, \H_\SR(- \, i \, \hbar \, \beta_2) \, \H_\SR(- \, i \, \hbar \, \beta_3) \, \H_\SR(- \, i \, \hbar \, \beta_4), \qquad \label{E-1} \\
    = & \, \frac{1}{\hbar^3} \int_0^\beta d \beta_1 \, \rho_\SS \, \rho_\R \, \big( e^{\beta_1 \, \hbar \, L_0} \, \H_\SR \, \big) \Big\{ \frac{1}{L_0} \, \big( e^{\beta_1 \, \hbar \, L_0} - 1 \big) \, \H_\SR \, \frac{1}{L_0} \, \H_\SR \, \frac{1}{L_0} \, \H_\SR - \Big( \frac{1}{L_0} \, \big( e^{\beta_1 \, \hbar \, L_0} - 1 \big) \, \H_\SR \Big) \, \frac{1}{L_0} \, \H_\SR \, \frac{1}{L_0} \, \H_\SR \nonumber \\
    & \qquad \qquad \qquad \qquad \ - \Big( \frac{1}{L_0} \, \big( e^{\beta_1 \, \hbar \, L_0} - 1 \big) \, \H_\SR \, \frac{1}{L_0} \, \H_\SR \Big) \, \frac{1}{L_0} \, \H_\SR + \Big( \frac{1}{L_0} \, \big( e^{\beta_1 \, \hbar \, L_0} - 1 \big) \, \H_\SR \, \Big) \Big( \frac{1}{L_0} \, \H_\SR \Big) \, \frac{1}{L_0} \, \H_\SR \Big\}, \label{E-2} \\
    = & \, \frac{1}{\hbar^4} \, \rho_\SS \, \rho_\R \, \Big\{ \frac{1}{L_0} \, \big( e^{\beta \, \hbar \, L_0} - 1 \big) \, \H_\SR \, \frac{1}{L_0} \, \H_\SR \, \frac{1}{L_0} \, \H_\SR \, \frac{1}{L_0} \, \H_\SR - \Big( \frac{1}{L_0} \, \big( e^{\beta \, \hbar \, L_0} - 1 \big) \, \H_\SR \Big) \, \frac{1}{L_0} \, \H_\SR \, \frac{1}{L_0} \, \H_\SR \, \frac{1}{L_0} \, \H_\SR \nonumber \\
    & \quad - \Big( \frac{1}{L_0} \, \big( e^{\beta \, \hbar \, L_0} - 1 \big) \, \H_\SR \, \frac{1}{L_0} \, \H_\SR \Big) \, \frac{1}{L_0} \, \H_\SR \, \frac{1}{L_0} \, \H_\SR + \Big( \frac{1}{L_0} \, \big( e^{\beta \, \hbar \, L_0} - 1 \big) \, \H_\SR \Big) \Big( \frac{1}{L_0} \, \H_\SR \Big) \, \frac{1}{L_0} \, \H_\SR \, \frac{1}{L_0} \, \H_\SR \nonumber \\
    & \quad - \Big( \frac{1}{L_0} \, \big( e^{\beta \, \hbar \, L_0} - 1 \big) \, \H_\SR \, \frac{1}{L_0} \, \H_\SR \, \frac{1}{L_0} \, \H_\SR \Big) \, \frac{1}{L_0} \, \H_\SR + \Big( \frac{1}{L_0} \, \big( e^{\beta \, \hbar \, L_0} - 1 \big) \, \H_\SR \Big) \Big( \frac{1}{L_0} \, \H_\SR \, \frac{1}{L_0} \, \H_\SR \Big) \, \frac{1}{L_0} \, \H_\SR \nonumber \\
    & \quad + \Big( \frac{1}{L_0} \, \big( e^{\beta \, \hbar \, L_0} - 1 \big) \, \H_\SR \, \frac{1}{L_0} \, \H_\SR \Big) \Big( \frac{1}{L_0} \, \H_\SR \Big) \, \frac{1}{L_0} \, \H_\SR - \Big( \frac{1}{L_0} \big( e^{\beta \, \hbar \, L_0} - 1 \big) \, \H_\SR \Big) \Big( \frac{1}{L_0} \, \H_\SR \Big) \Big( \frac{1}{L_0} \, \H_\SR \Big) \, \frac{1}{L_0} \, \H_\SR \Big\}, \label{E-3}
\end{align}
\begin{align}
    = & \, \frac{i^4}{\hbar^4} \, \Big\{ \int_0^\infty d\tau_1 \int_{\tau_1}^\infty d\tau_2 \int_{\tau_2}^\infty d\tau_3 \int_{\tau_3}^\infty d\tau_4 \, \big[ \, \H_\SR(- \, \tau_1) \, \H_\SR(- \, \tau_2) \, \H_\SR(- \, \tau_3) \, \H_\SR(- \, \tau_4) \,,\, \rho_\SS \, \rho_\R \, \big] \nonumber \\
    & \ - \int_0^\infty d\tau_1 \int_0^\infty d\tau_2 \int_{\tau_2}^\infty d\tau_3 \int_{\tau_3}^\infty d\tau_4 \, \big[ \, \H_\SR(- \, \tau_1) \,,\, \rho_\SS \, \rho_\R \, \big] \, \H_\SR(- \, \tau_2) \, \H_\SR(- \, \tau_3) \, \H_\SR(- \, \tau_4) \nonumber \\
    & \ - \int_0^\infty d\tau_1 \int_{\tau_1}^\infty d\tau_2 \int_0^\infty d\tau_3 \int_{\tau_3}^\infty d\tau_4 \, \big[ \, \H_\SR(- \, \tau_1) \, \H_\SR(- \, \tau_2) \,,\, \rho_\SS \, \rho_\R \, \big] \, \H_\SR(- \, \tau_3) \, \H_\SR(- \, \tau_4) \nonumber \\
    & \ + \int_0^\infty d\tau_1 \int_0^\infty d\tau_2 \int_0^\infty d\tau_3 \int_{\tau_3}^\infty d\tau_4 \, \big[ \, \H_\SR(- \, \tau_1) \,,\, \rho_\SS \, \rho_\R \, \big] \, \H_\SR(- \, \tau_2) \, \H_\SR(- \, \tau_3) \, \H_\SR(- \, \tau_4) \nonumber \\
    & \ - \int_0^\infty d\tau_1 \int_{\tau_1}^\infty d\tau_2 \int_{\tau_2}^\infty d\tau_3 \int_0^\infty d\tau_4 \, \big[ \, \H_\SR(- \, \tau_1) \, \H_\SR(- \, \tau_2) \, \H_\SR(- \, \tau_3) \,,\, \rho_\SS \, \rho_\R \, \big] \, \H_\SR(- \, \tau_4) \nonumber \\
    & \ + \int_0^\infty d\tau_1 \int_0^\infty d\tau_2 \int_{\tau_2}^\infty d\tau_3 \int_0^\infty d\tau_4 \, \big[ \, \H_\SR(- \, \tau_1) \,,\, \rho_\SS \, \rho_\R \, \big] \, \H_\SR(- \, \tau_2) \, \H_\SR(- \, \tau_3) \, \H_\SR(- \, \tau_4) \nonumber \\
    & \ + \int_0^\infty d\tau_1 \int_{\tau_1}^\infty d\tau_2 \int_0^\infty d\tau_3 \int_0^\infty d\tau_4 \, \big[ \, \H_\SR(- \, \tau_1) \, \H_\SR(- \, \tau_2) \,,\, \rho_\SS \, \rho_\R \, \big] \, \H_\SR(- \, \tau_3) \, \H_\SR(- \, \tau_4) \nonumber \\
    & \ - \int_0^\infty d\tau_1 \int_0^\infty d\tau_2 \int_0^\infty d\tau_3 \int_0^\infty d\tau_4 \, \big[ \, \H_\SR(- \, \tau_1) \,,\, \rho_\SS \, \rho_\R \, \big] \, \H_\SR(- \, \tau_2) \, \H_\SR(- \, \tau_3) \, \H_\SR(- \, \tau_4) \Big\}\, e^{- \, \epsilon \, \tau_1 - \epsilon \, \tau_2 - \epsilon \, \tau_3 - \epsilon \, \tau_4 }\big|_{\epsilon \to +0}\,,
\label{E-4} \\
    = & \, \frac{i^4}{\hbar^4} \, \Big\{ \int_0^\infty d\tau_1 \int_{\tau_1}^\infty d\tau_2 \int_{\tau_2}^\infty d\tau_3 \int_{\tau_3}^\infty d\tau_4 \, \big[ \, \H_\SR(- \, \tau_1) \, \H_\SR(- \, \tau_2) \, \H_\SR(- \, \tau_3) \, \H_\SR(- \, \tau_4) \,,\, \rho_\SS \, \rho_\R \, \big] \nonumber \\
    & \quad - \int_0^\infty d\tau_1 \int_{\tau_1}^\infty d\tau_2 \int_{\tau_2}^\infty d\tau_3 \int_0^\infty d\tau_4 \, \big[ \, \H_\SR(- \, \tau_1) \, \H_\SR(- \, \tau_2) \, \H_\SR(- \, \tau_3) \,,\, \rho_\SS \, \rho_\R \, \big] \, \H_\SR(- \, \tau_4) \nonumber \\
    & \quad + \int_0^\infty d\tau_1 \int_{\tau_1}^\infty d\tau_2 \int_0^\infty d\tau_3 \int_0^{\tau_3} d\tau_4 \, \big[ \, \H_\SR(- \, \tau_1) \, \H_\SR(- \, \tau_2) \,,\, \rho_\SS \, \rho_\R \, \big] \, \H_\SR(- \, \tau_3) \, \H_\SR(- \, \tau_4) \nonumber \\
    & \quad - \int_0^\infty d\tau_1 \int_0^\infty d\tau_2 \int_0^{\tau_2} d\tau_3 \int_0^{\tau_3} d\tau_4 \, \big[ \, \H_\SR(- \, \tau_1) \,,\, \rho_\SS \, \rho_\R \, \big] \, \H_\SR(- \, \tau_2) \, \H_\SR(- \, \tau_3) \, \H_\SR(- \, \tau_4) \Big\} \qquad \qquad \qquad \qquad \qquad \nonumber \\
    & \qquad \qquad \qquad \qquad \qquad \qquad \qquad \qquad \quad \times \exp\{ - \, \epsilon \, \tau_1 - \epsilon \, \tau_2 - \epsilon \, \tau_3 - \epsilon \, \tau_4 \, \}\big|_{\epsilon \to +0}\,, \label{E-5}
\end{align}
which can be expressed, by performing some integral transformations as (\ref{integral-trans}), (\ref{integral-trans-1}) and (\ref{integral-trans-2}), in the following compact form
\begin{align}
    & \int_0^\beta d \beta_1 \int_0^{\beta_1} d \beta_2 \int_0^{\beta_2} d \beta_3 \int_0^{\beta_3} d \beta_4 \, \rho_\SS \, \rho_\R \, \H_\SR(- \, i \, \hbar \, \beta_1) \, \H_\SR(- \, i \, \hbar \, \beta_2) \, \H_\SR(- \, i \, \hbar \, \beta_3) \, \H_\SR(- \, i \, \hbar \, \beta_4), \qquad \nonumber \\
    = & \, \frac{i^4}{\hbar^4} \int_0^\infty \! d\tau_1 \int_0^{\tau_1} \! d\tau_2 \int_0^{\tau_2} \! d\tau_3 \int_0^{\tau_3} \! d\tau_4 \, \big[ \, \H_\SR(- \, \tau_4) \,, \big[ \, \H_\SR(- \, \tau_3) \,, \big[ \, \H_\SR(- \, \tau_2) \,, \big[ \, \H_\SR(- \, \tau_1) \,,\, \rho_\SS \, \rho_\R \, \big] \, \big] \, \big] \, \big] \, e^{- \, \epsilon \, \tau_1}|_{\epsilon \to +0}\, , \\
    = & \,\, i^4 \int_0^\infty d\tau_1 \int_0^{\tau_1} d\tau_2 \int_0^{\tau_2} d\tau_3 \int_0^{\tau_3} d\tau_4 \, L_\SR(- \, \tau_4) \, L_\SR(- \, \tau_3) \, L_\SR(- \, \tau_2) \, L_\SR(- \, \tau_1) \, \rho_\SS \, \rho_\R \exp(- \, \epsilon \, \tau_1) \, |_{\epsilon \to +0}\,. \label{E-7}
\end{align}
\section{Calculation of admittance $\chi_{b b^\dagger}(\omega)$ [(\ref{chi-bbd})]}

The admittance $\chi_{b b^\dagger}(\omega)$ given by (\ref{chi-bbd}) can be rewritten as
\begin{align}
\chi_{b b^\dagger}(\omega) & = i \, \hbar \int_0^\infty dt \, \Tr \, b(t) \Big\{1 + \sum_{n = 1}^\infty (- \, i)^n \int_0^t d\tau_1 \int_0^{\tau_1} d\tau_2 \, \cdots \int_0^{\tau_{n - 1}} d\tau_n \, L_\SR(\tau_1) \, L_\SR(\tau_2) \, \cdots L_\SR(\tau_n) \Big\} \qquad \nonumber \\
    & \qquad \qquad \qquad \qquad \qquad \qquad \qquad \qquad \qquad \qquad \qquad \qquad \times [\, b^\dagger , \, \rho_\TE \,] \exp(i \, \omega \, t - \epsilon \, t) \,\big|_{\epsilon \to +0} \,, \\
    & = i \, \hbar \, \Big\{ \int_0^\infty \! dt \, \Tr \, b + \sum_{n = 1}^\infty (- \, i)^n \int_0^\infty \! d\tau_n \int_{\tau_n}^\infty \! d\tau_{n-1} \, \cdots \int_{\tau_2}^\infty \! d\tau_1 \int_{\tau_1}^\infty \! dt \, \Tr \, [\, \cdots [\, [\, b \,,\, \H_\SR(\tau_1) \,] \,,\, \H_\SR(\tau_2) \,] \,,\, \cdots \quad \nonumber \\
    & \qquad \qquad \qquad \qquad \qquad \qquad \qquad \qquad \qquad \quad \cdots ,\, \H_\SR(\tau_n) \,] / \hbar^n \Big\}\, [\, b^\dagger , \, \rho_\TE \,] \exp\{ i \, (\omega - \omega_0) \, t - \epsilon \, t \} \,\big|_{\epsilon \to +0} \,,
\end{align}
which can be calculated using (\ref{Hboson-SR-t}) as
\begin{align}
\chi_{b b^\dagger}(\omega) & = \frac{\hbar}{\omega_0 - \bar{\omega}} + \frac{\hbar}{\omega_0 - \bar{\omega}} \sum_{m=1}^\infty (- \, i)^{2m} \int_0^\infty d\tau_{2m} \int_{\tau_{2m}}^\infty d\tau_{2m-1} \cdots \int_{\tau_3}^\infty d\tau_2 \int_{\tau_2}^\infty d\tau_1 \sum _{\alpha_1} \cdots \sum_{\alpha_m} |\, g_{\alpha_1} |^2 \cdots \,|\, g_{\alpha_m} |^2 \nonumber \\
    & \qquad \qquad \qquad \qquad \qquad \quad \times \Tr \, b \, [\, b^\dagger, \, \rho_\TE \,] \exp\{i \, (\bar{\omega} - \omega_0) \, \tau_1 + i \, (\omega_0 - \omega_{\alpha_1}) \, \tau_1 - i \, (\omega_0 - \omega_{\alpha_1}) \, \tau_2 + \cdots \nonumber \\
    & \qquad \qquad \qquad \qquad \qquad \qquad \qquad \qquad \qquad \qquad \cdots + i \, (\omega_0 - \omega_{\alpha_m}) \, \tau_{2m-1} - i \, (\omega_0 - \omega_{\alpha_m}) \, \tau_{2m} \} , \\
    & = \frac{\hbar}{\omega_0 - \bar{\omega}}\, \Big\{ 1 + \sum_{m = 1}^\infty \sum_{\alpha_1} \cdots \sum_{\alpha_m} \frac{(- 1)^m \, |\, g_{\alpha_1} \,|^2 \, \cdots \, |\, g_{\alpha_m} \,|^2 }{\, (\omega_0 - \bar{\omega})^m \, (\bar{\omega} - \omega_{\alpha_1}) \, \cdots \, (\bar{\omega} - \omega_{\alpha_m}) \,} , \\
    & = \frac{\hbar}{\omega_0 - \bar{\omega}}\, \Big\{ 1 + \sum_\alpha \frac{|\, g_\alpha \,|^2}{\, (\omega_0 - \bar{\omega}) \, (\bar{\omega} - \omega_\alpha) \,} \Big\}^{-1} = \frac{i \, \hbar}{\, i \, (\omega_0 - \bar{\omega}) + \phi(\bar{\omega}) \,},
\end{align}
with $ \bar{\omega}$\,=\,$ \omega $\,+\,$ i \, \epsilon $ \,($ \epsilon $$ \to $+0), where $\phi(\bar{\omega})$ is given by (\ref{phi-omega}). Thus, the exact form (\ref{exact-chi-bbd}) of $\chi_{b b^\dagger}(\omega)$ can be obtained.
\section{The ordered cumulants for the spin model (\ref{H-SR})$-$(\ref{H-Sexternal})}

We calculate the ordered cumulants for the quantum spin model (\ref{H-SR})$-$(\ref{H-Sexternal}). For the second-order ordered cumulant $\langle L_\SR L_\SR(- \, \tau) \rangle_\R$ \,$(\tau \ge 0)$, we can obtain, by using the relations $S_z S_+$\,=\,$S_+/2$ and $S_+ S_z$\,=\,$- S_+/2$,
\begin{equation}
\tr \, S_+ \, \langle L_\SR \, L_\SR(- \, \tau) \rangle_\R \, \tilde{a}_-(t)
= (1/2) \big\{ \langle \,[\, R(\tau), \, R^\dagger \,]_+ \rangle_\R + \langle \,[\, R^\dagger(\tau), \, R \,]_+ \rangle_\R \big\} \, \tr \, S_+ \, \tilde{a}_-(t). \qquad
\end{equation}
We can calculate the fourth-order ordered cumulant $\langle L_\SR L_\SR(- \, \tau_3) L_\SR(- \, \tau_2) L_\SR(- \, \tau_1) \rangle_\R$ \,$(\tau_1 \ge \tau_2 \ge \tau_3 \ge 0)$ as follows,
\begin{align}
& \tr \, S_+ \, \langle L_\SR \, L_\SR(- \, \tau_3) \, L_\SR(- \, \tau_2) \, L_\SR(- \, \tau_1) \rangle_\R \, \tilde{a}_-(t) = - \, \tr \, [\, S_+,\, S_z \,]\, \langle \, (R + R^\dagger) \, L_\SR(- \, \tau_3) \, L_\SR(- \, \tau_2) \, L_\SR(- \, \tau_1) \, \rangle_\R \, \tilde{a}_-(t), \nonumber \\
    = & - \Tr \, \big\{ S_+ \, S_z \, (R + R^\dagger) (R(- \, \tau_3) + R^\dagger(- \, \tau_3)) - S_z \, S_+ \, (R(- \, \tau_3) + R^\dagger(- \, \tau_3)) (R + R^\dagger) \big\} \, L_\SR(- \, \tau_2) \, L_\SR(- \, \tau_1) \, \rho_\R \, \tilde{a}_-(t), \nonumber \\
    = & \, (1/2) \, \Tr \, S_+ \, [\, R + R^\dagger ,\, R(- \, \tau_3) + R^\dagger(- \, \tau_3) \,]_+ \, L_\SR(- \, \tau_2) \, L_\SR(- \, \tau_1) \, \rho_\R \, \tilde{a}_-(t), \nonumber \\
    = & \, (1/ \, 2^2 \,) \, \Tr \, S_+ \, [\, [\, R + R^\dagger ,\, R(- \, \tau_3) + R^\dagger(- \, \tau_3) \,]_+ \,,\, R(- \, \tau_2) + R^\dagger(- \, \tau_2) \,]_+ \, L_\SR(- \, \tau_1) \, \rho_\R \, \tilde{a}_-(t), \nonumber \\
    = & \, (1/ \, 2^3 \,) \, \big\langle \, [\, [\, [\, R + R^\dagger ,\, R(- \, \tau_3) + R^\dagger(- \, \tau_3) \,]_+ \,,\, R(- \, \tau_2) + R^\dagger(- \, \tau_2) \,]_+ \,,\, R(- \, \tau_1) + R^\dagger(- \, \tau_1) \,]_+ \, \big\rangle_\R \, \tr \, S_+ \, \tilde{a}_-(t),
\end{align}
which can be rewritten using the second-order ordered cumulants by the Wick's theorem for finite temperature, as
\begin{align}
& \tr \, S_+ \, \langle L_\SR \, L_\SR(- \, \tau_3) \, L_\SR(- \, \tau_2) \, L_\SR(- \, \tau_1) \rangle_\R \, \tilde{a}_-(t), \nonumber \\
    = & \, (1/ \, 2^2 \,) \, \{ \langle \,[\, R + R^\dagger ,\, R(- \, \tau_3) + R^\dagger(- \, \tau_3) \,]_+ \rangle_\R \, \langle \,[\, R(- \, \tau_2) + R^\dagger(- \, \tau_2) \,,\, R(- \, \tau_1) + R^\dagger(- \, \tau_1) \,]_+ \rangle_\R \nonumber \\
    & \qquad \,\,\ + \langle \,[\, R + R^\dagger ,\, R(- \, \tau_2) + R^\dagger(- \, \tau_2) \,]_+ \rangle_\R \, \langle \,[\, R(- \, \tau_3) + R^\dagger(- \, \tau_3) \,,\, R(- \, \tau_1) + R^\dagger(- \, \tau_1) \,]_+ \rangle_\R \nonumber \\
    & \qquad \,\,\ + \langle \,[\, R + R^\dagger ,\, R(- \, \tau_1) + R^\dagger(- \, \tau_1) \,]_+ \rangle_\R \, \langle \,[\, R(- \, \tau_3) + R^\dagger(- \, \tau_3) \,,\, R(- \, \tau_2) + R^\dagger(- \, \tau_2) \,]_+ \rangle_\R \,\} \, \tr \, S_+ \, \tilde{a}_-(t). \quad
\end{align}
We can also calculate the product of the second-order ordered cumulants as
\begin{align}
& \tr \, S_+ \, \langle L_\SR \, L_\SR(- \, \tau_3) \rangle_\R \, \langle L_\SR(- \, \tau_2) \, L_\SR(- \, \tau_1) \rangle_\R \, \tilde{a}_-(t), \nonumber \\
    & = (1/2) \, \langle \, [\, R + R^\dagger ,\, R(- \, \tau_3) + R^\dagger(- \, \tau_3) \,]_+ \rangle_\R \, \tr \, S_+ \, \langle L_\SR(- \, \tau_2) \, L_\SR(- \, \tau_1) \rangle_\R \, \tilde{a}_-(t), \nonumber \\
    & = (1/ \, 2^2 \,) \, \langle \, [\, R + R^\dagger ,\, R(- \, \tau_3) + R^\dagger(- \, \tau_3) \,]_+ \rangle_\R \, \langle \,[\, R(- \, \tau_2) + R^\dagger(- \, \tau_2) \,,\, R(- \, \tau_1) + R^\dagger(- \, \tau_1) \,]_+ \rangle_\R \, \tr \, S_+ \, \tilde{a}_-(t), \\
& \tr \, S_+ \, \langle L_\SR \, L_\SR(- \, \tau_2) \rangle_\R \, \langle L_\SR(- \, \tau_3) \, L_\SR(- \, \tau_1) \rangle_\R \, \tilde{a}_-(t), \nonumber \\
    & = (1/ \, 2^2 \,) \, \langle \, [\, R + R^\dagger ,\, R(- \, \tau_2) + R^\dagger(- \, \tau_2) \,]_+ \rangle_\R \, \langle \,[\, R(- \, \tau_3) + R^\dagger(- \, \tau_3) \,,\, R(- \, \tau_1) + R^\dagger(- \, \tau_1) \,]_+ \rangle_\R \, \tr \, S_+ \, \tilde{a}_-(t), \\
& \tr \, S_+ \, \langle L_\SR \, L_\SR(- \, \tau_1) \rangle_\R \, \langle L_\SR(- \, \tau_3) \, L_\SR(- \, \tau_2) \rangle_\R \, \tilde{a}_-(t), \nonumber \\
    & = (1/ \, 2^2 \,) \, \langle \, [\, R + R^\dagger ,\, R(- \, \tau_1) + R^\dagger(- \, \tau_1) \,]_+ \rangle_\R \, \langle \,[\, R(- \, \tau_3) + R^\dagger(- \, \tau_3) \,,\, R(- \, \tau_2) + R^\dagger(- \, \tau_2) \,]_+ \rangle_\R \, \tr \, S_+ \, \tilde{a}_-(t). 
\end{align}
Therefore, the sum of the fourth-order ordered cumulants vanishes, i.e.,
\begin{align}
& \tr \, S_+ \, \big\{ \big\langle L_\SR \, L_\SR(- \, \tau_3) L_\SR(- \, \tau_2) L_\SR(- \, \tau_1) \big\rangle_\R - \big\langle L_\SR \, L_\SR(- \, \tau_3) \big\rangle_\R \big\langle L_\SR(- \, \tau_2) L_\SR(- \, \tau_1) \big\rangle_\R \nonumber \\
    & \qquad \quad - \big\langle L_\SR \, L_\SR(- \, \tau_2) \big\rangle_\R \big\langle L_\SR(- \, \tau_3) L_\SR(- \, \tau_1) \big\rangle_\R - \big\langle L_\SR \, L_\SR(- \, \tau_1) \big\rangle_\R \big\langle L_\SR(- \, \tau_3) L_\SR(- \, \tau_2) \big\rangle_\R \, \big\} \, \tilde{a}_-(t) = 0.
\end{align}
In the same way, it can be shown that the sum of the higher-order ordered cumulants vanishes.
%
\end{widetext}
%

\end{document}